# Louis de Broglie
# und die Quantenmechanik

## Henning Sievers

Theoretische Kernphysik, Universität Hamburg
Luruper Chaussee 149, D - 22761 Hamburg

Hamburg, den 17. Dezember 1997

# Abstract

In 1923 Louis de Broglie (1892-1987) discovered the material waves and six years later received for this discovery the Nobel price. Apart these well known fact this French physicist nevertheless seems to have been forgotten. Details of his life are as unknown as his efforts to describe quantum mechanics in a deterministic and objective way. Especially the actual discussion concerning the interpretation of quantum mechanics seems to justify a deeper occupation with the scientific work of Louis de Broglie. In this context the important influence of Albert Einstein is of special interest; for his photons announce the existence of material waves and it may surprise that Einstein himself did not postulate them.

The basis of this short scientific biography are the publications of de Broglie (with a complete bibliography given at the end), some more or less short memories of his friends and pupils and some unpublished documents found in the *Archives de l'Académie des Sciences* at Paris. The original text of the correspondence between de Broglie and Einstein and some excerpts of his thesis are enclosed German translation in the appendix.

# Résumé

En 1923 Louis de Broglie (1892-1987) a découvert l'onde de matière et il a reçu le prix Nobel six ans plus tard. A part de ces faits généralement connus, ce physicien français semble être oublié. Des détails concernant sa vie, vouée entièrement à la physique, sont aussi peu connus que ses essais d'une description déterministe et objective de la méchanique quantique. Surtout le débat actuel autour d'une interprétation nouvelle de la méchanique quantique justifie l'intérêt pour l'oeuvre scientifique de Louis de Broglie. Dans ce contexte, l'influence d'Albert Einstein est spécialement importante, d'autant plus que ses photons semblent annoncer l'existence de l'onde de matière. Il est presque étonnant qu'Einstein ne les aie pas postulée lui-même.

La base de cette courte biographie scientifique sont les nombreuses publications de de Broglie (une bibliographie complète est donnée à la fin), quelques témoignages plus ou moins courtes des ses amis et élèves ainsi qu'un nombre de documents inédits trouvés dans les *Archives de l'Académie des Sciences* à Paris. Le texte original de la correspondance de Broglie-Einstein et des extraits de la thèse de doctorat sont joints en traduction allemande.

# Zusammenfassung

Louis de Broglie (1892-1987) entdeckte 1923 die Materiewellen und erhielt für diesen Beitrag zur Entwicklung der modernen Physik sechs Jahre später den Nobelpreis. Über diese weithin bekannten Fakten hinaus scheint der geniale französische Physiker jedoch in Vergessenheit geraten zu sein: Details seines ganz der Physik geweihten Lebens sind ebenso unbekannt wie seine lebenslangen Bemühungen um eine deterministische und objektive Darstellung der durch die Quantenmechanik beschriebenen Effekte. Gerade die in den letzten Jahren neu entfachte Debatte um die Interpretation der Quantenmechanik rechtfertigt jedoch eine eingehendere Beschäftigung mit dem wissenschaftlichen Werk de Broglies. Dabei ist der prägende Einfluß Albert Einsteins von besonderem Interesse; zumal dessen Photonenhypothese die Existenz der Materiewellen so nahelegte, daß es fast erstaunen mag, daß Einstein diese nicht selbst postulierte.

Grundlage dieser Arbeit sind die zahlreichen Veröffentlichungen de Broglies (eine vollständige Bibliographie findet sich am Ende), einige mehr oder weniger kurze Erinnerungen seiner Bekannten und Schüler sowie eine Anzahl unveröffentlichter Dokumente, die in den *Archives de l'Académie des Sciences* in Paris aufbewahrt werden. Der Originaltext der Korrespondenz de Broglie-Einstein und einiger Ausschnitte der Doktorarbeit ist in deutscher Übersetzung am Ende der Arbeit zu finden.

# Vorwort

Die vorliegende Arbeit entstand in Anschluß an ein Seminar „Methoden und Ziele der Physik" an der Universität Hamburg, das von Prof. Dr. H. V. von Geramb geleitet wurde. Ausgangspunkt für die Beschäftigung mit Louis de Broglie war die Fragestellung, warum A. Einstein, nach dem Aufstellen der Photonenhypothese, nicht selbst den doch so naheliegenden Teilchen-Welle-Dualismus postulierte. Sehr schnell zeigte sich zu unserem Erstaunen, daß Louis de Broglie weder in der physikalischen, noch in der wissenschaftsgeschichtlichen Fachliteratur besondere Beachtung findet. So entstand die Idee, mit dieser Arbeit einen kleinen Beitrag zur Aufarbeitung von Leben und Werk des französischen Physikers zu leisten. Die Beschäftigung mit seinen zahllosen Veröffentlichungen, mehrere Reisen nach Paris, Gespräche mit Schülern, unveröffentlichte Texte und persönliche Briefe halfen, das Bild dieser interessanten und schillernden Persönlichkeit abzurunden.

Hamburg im Dezember 1997.

# Inhaltsverzeichnis

# Einleitung

Louis de Broglie (1892-1987) entdeckte 1923 den Welle-Teilchen-Dualismus und legte damit einen Grundstein der modernen Physik. Als Nobelpreisträger von 1929 ist sein Name nach wie vor der physikalischen Fachwelt ein Begriff; über sein Leben und seine wissenschaftliche Tätigkeit, die bis in unsere unmittelbare Gegenwart hineinreicht, erfährt man jedoch selten mehr[1] - weder in der englischsprachigen, noch in der deutschsprachigen Literatur und, was noch mehr erstaunt, auch in Frankreich scheint de Broglie weitgehend vergessen zu sein.

Vor zehn Jahren starb der französische Physiker und so scheint dies ein guter Zeitpunkt zu sein, einmal nachzufragen, welche Persönlichkeit sich hinter der de Broglie-Wellenlänge verbirgt. Es soll aber neben der Erinnerung an den genialen Physiker in dieser Arbeit auch darum gehen, die enge Verbindung nachzuweisen, die zwischen dem Menschen Louis de Broglie und der Ausrichtung seiner Forschung bestand.[2] Darüber hinaus möchte ich die zeitliche Veränderung seiner Forschungsinteressen mit der Entstehung und Weiterentwicklung der modernen Physik in Beziehung setzen, um so die wissenschaftsgeschichtliche Dimension des Schaffens von Louis de Broglie hervorzuheben.

Erfreulicherweise steht für dieses Vorhaben hervorragendes, teilweise unbearbeitetes französischsprachiges Material zur Verfügung. Zu verdanken ist dies vor allem der *Fondation Louis de Broglie*, die 1973 anläßlich der 50. Kommemoration der Entdeckung der Materiewelle gegründet wurde. Ehemalige Mitarbeiter de Broglies haben sich hier zusammengeschlossen, um sein wissenschaftliches Erbe zu pflegen und um in seinem Sinne weiter zu forschen. Der *Fondation Louis de Broglie* ist es zu verdanken, daß handschriftliche Abhandlungen, autobiographische Aufsätze und Briefe de Broglies erhalten sind und in den *Archives de l'Académie des Sciences* in Paris aufbewahrt werden. Dieses Archiv, das Teil des *Institut de France* ist und zahllose, oft unveröffentlichte Schriften de Broglies beherbergt, ist auch eine wertvolle Quelle für diese Arbeit gewesen.

Für den biographischen Aspekt dieser Arbeit waren mir zudem die Festschriften zum 60. [286], zum 80. [288] und 90. [290] Geburtstag hilfreich sowie eine Biographie von 1966 [285] und eine von 1992 [283] sowie Erinnerungen [287], Artikel und Gedenkreden (z.B. [280]-[282]) von Zeitgenossen de Broglies, die anläßlich seines Todes veröffentlicht wurden. Wiederum hat sich vor allem die *Fondation Louis de Broglie*, insbesondere Georges Lochak, um die Veröffentlichung des biographischen Materials bemüht.

Für den wissenschaftlichen Werdegang de Broglies steht eine unüberschaubare Anzahl von Primärwerken zur Verfügung: de Broglie hat über 40 Bücher [150-192], hunderte von Aufsätzen, Vorlesungen, Vorträgen und Abhandlungen publiziert [1-149, 193-279].

---

[1] Die Erwähnung de Broglies in der Fachliteratur beschränkt sich auf kurze Darstellungen wie in [302, 303].

[2] Karl von Meyenn bezeichnet es als die Aufgabe wissenschaftlicher Biographien, nachzuweisen, daß Wissenschaft auch „charakteristische Züge ihrer Schöpfer trägt". [312, S. 7].

Zudem finden sich in den Festschriften zum 80., 90. und 100. [293][3] Geburtstag sowie in [294] zahlreiche Aufsätze zur wissenschaftlichen Arbeit de Broglies und deren Weiterentwicklung durch seine Schüler.

Die Beschäftigung mit diesem Material ist jedoch nicht nur im französischen Sprachraum lücken- und mangelhaft: in der deutschen und englischen Literatur findet man praktisch keine ernsthafte Auseinandersetzung mit der Biographie und der Forschung de Broglies. Tatsächlich reduziert sich hier das Interesse auf de Broglies Entdeckung der Materiewelle. Zu diesem Thema existiert eine längere Abhandlung von 1968 [291] und neben der obligatorischen Erwähnung seines Namens auf den ersten Seiten eines jeden Lehrbuchs zur Quantenmechanik gibt es einige kurze Zeitungsartikel anläßlich seines Todes.[4] Zudem wurden einige seiner physikalischen und wissenschaftsphilosophischen Abhandlungen, namentlich zur Quantenmechanik in der Kopenhagener Deutung, in deutsche, z.T. englische Sprache[5] übersetzt.

Dieser kurze Überblick über das reichhaltige, aber noch kaum ausgewertete Quellenmaterial dürfte zeigen, daß es ein durchaus lohnendes Unterfangen sein kann, sich auch außerhalb des französischen Sprachraumes einmal um ein umfassenderes Bild des Menschen und Physikers Louis de Broglies zu bemühen.

Zum Aufbau dieser Arbeit sei folgendes gesagt: De Broglies Leben läßt sich in deutlich unterscheidbare Abschnitte gliedern, die chronologisch dargestellt werden sollen. Als letzter Nachkomme eines alten französischen Adelsgeschlechtes wuchs de Broglie in einem heute kaum noch vorstellbaren sozialen Umfeld auf, das ihn entscheidend geformt hat. So wird es nötig sein, auf seine Abstammung, seine charakterlichen Eigenheiten und seine unzeitgemäß erscheinenden Umgangsformen einzugehen. Dies soll im ersten Kapitel „Kindheit und Jugend (1892-1919)" dargestellt werden. Dafür werde ich kurz auf die bis weit in die französische Geschichte zurückreichende Familiengeschichte eingehen und de Broglies aristokratisch geprägte Kinderjahre sowie seine zunächst von Hauslehrern gewährleistete Ausbildung beschreiben; zudem sollen der Militärdienst und sein prägender Einfluß erwähnt werden. Im Kapitel II „Jahre der Kreativität (1919-1927)" möchte ich auf die wohl kreativste Schaffensphase im Leben de Broglies eingehen und die Arbeiten beschreiben, die de Broglies gesellschaftlichen Erfolg und seine wissenschaftliche Anerkennung gesichert haben: die drei kurzen Artikel in den *Comptes Rendus de l'Académie des Sciences*[6] von 1923 [16-18], deren Grundideen zur Doktorarbeit (1924) [150a] ausgearbeitet wurden. Es sollen dann die Weiterentwicklung des Wellenkonzeptes (die „Führungswelle") und dessen Scheitern (1927) beschrieben werden. Da der Einfluß Albert Einsteins auf die Arbeit de Broglies von entscheidender Bedeutung war, wird am Ende des Kapitels das Verhältnis der beiden Physiker genauer beleuchtet werden. Im dritten Kapitel „Jahre der Anpassung (1927-1951)" steht eine Zeit im Vordergrund, die geprägt war vom Scheitern des „Führungswellenkonzeptes", von der Anpassung an die gängige Kopenhagener Deutung und der Konzentration auf gesellschaftliche Funktionen und die Lehrtätigkeit am *Institut Henri Poincaré*. Im letzten Kapitel „Isolation (1951-1987)" werden die Jahre beschrieben, in denen de Broglie zu seinen ursprünglichen Ideen zurückkehrte und sich damit in der *scientific community* isolierte. Ich werde die neuen Aspekte seiner Theorie sowie die Organisation seiner Arbeitsgruppe und das schrittweise Ausscheiden de Broglies aus der aktiven Forschungstätigkeit darstellen. So sollen mögliche Gründe für de Broglies Verschwinden aus dem öffentlichen Bewußtsein aufgezeigt werden. Da gerade in den letzten Jahren die durch de Broglie 1952 ausgelöste Debatte um die Interpretation der Quantenmechanik neue Impulse erhalten hat, erscheint es mir angemessen, am Ende des Kapitels diesbezügliche neuere Versuche und deren Rechtfertigung ansatzweise vorzustellen.

---

[3] Der Band zum 100. Geburtstag [293] ist in den *Foundations of Physics* 1992 in englischer Sprache erschienen.

[4] *Nature*, Vol. 427, 28. Mai 1987; *Spektrum Physik* 18 (1987); *Physikalische Blätter* 43, Nr. 6, 1987 etc.

[5] Im Literaturverzeichnis ist in Klammern angegeben, welche Werke in Übersetzungen erschienen sind [150-192].

[6] Die *Comptes Rendus de l'Académie des Sciences* werde ich im Folgenden mit *Comptes Rendus* abkürzen.

Am Ende dieser Arbeit befindet sich eine vollständige Literaturliste der Werke de Broglies sowie eine Aufstellung der im Zusammenhang mit de Broglies Leben und Werk interessanten Veröffentlichungen anderer Autoren. Um einen kleinen Einblick in die Ausdrucksweise, Argumentations- und Gedankengänge und die oft ungewöhnliche Terminologie de Broglies zu geben, sind im Anhang A.1 zwei der wichtigsten Abschnitte seiner Doktorarbeit in einer deutscher Übersetzung von Gerda Boderseher wiedergegeben. Zudem befinden sich im Anhang A.2 ein Anzahl Briefe de Broglies und Einsteins, die in den *Archives de l'Académie des Sciences* aufbewahrt werden und die eine Grundlage des Kapitels II. 5 bilden.

# I Kindheit und Jugend (1892-1919)

In diesem ersten Kapitel werde ich die Kindheit und Jugend de Broglies sowie seine Studienjahre nachzeichnen. Dabei soll seiner humanistischen Ausbildung ebenso Beachtung geschenkt werden wie der langsamen Annäherung an die Physik. Vor allem aber soll Louis de Broglie auch als Persönlichkeit mit seinen charakterlichen Eigenheiten ins Blickfeld geraten.

## I.1 Abstammung

Von entscheidendem Einfluß auf die Erziehung de Broglies war die gesellschaftliche Stellung seiner Familie, deren Wurzeln zurückzuverfolgen sind. Um die anachronistisch erscheinende Lebensweise der de Broglies richtig beurteilen zu können, soll an dieser Stelle in abgekürzter Form auf die Familiengeschichte eingegangen werden. Die Ursprünge der Familie finden sich in Italien, im Piemont, wo bereits im 12. Jahrhundert die ersten Mitglieder der Familie „Broglia" bekannt waren als Verwalter, Militärs und Klostergründer. Im 17. Jahrhundert siedelte Francesco Maria Broglia (1611-1656), bereits mit dem Titel eines *Conte*, nach Frankreich über. Seine militärischen Erfolge brachten ihm die Titel *Marquis* und, postum, *Maréchal de France* ein. Seine Nachkommen, Victor Maurice (1647-1727) und François Marie (1671-1745), die sich französisch de Broglie nannten, sicherten den Ruhm der Familie und deren Platz in der französischen Geschichte als geschickte Feldherren. So erhielt François Marie für seine Erfolge 1716 in der Normandie das Schloß *Chambrais*, in dem er den Familiensitz einrichtete und das er 1742 in „Broglie" umbenennen durfte, wie es auch heute noch heißt.

Einer der Verwalter des Schlosses im 18. Jahrhundert war übrigens der Großvater des berühmten Novellenautors Prosper Mérimée und dessen Cousin Augustin Fresnel. Tatsächlich wurde der spätere Begründer der Wellenoptik 1788 in Broglie geboren - eine Verknüpfung, der Louis de Broglie zeitlebens große Bedeutung beimaß, da er diesen großartigen Physiker immer sehr verehrte und sich mit seiner eigenen Wellentheorie in dessen Nachfolge sah.

Ab dem Jahr 1742 erhielt der jeweils älteste Sohn der Familie de Broglie den Titel *Duc* und 17 Jahre später erkämpfte Victor-François (1718-1804) in der Schlacht bei Bergen (unweit von Frankfurt a. Main) vom 13. April 1759 auch einen sächsischen Fürstentitel (französisch: *Prince*) - noch Louis de Broglie trug diese Titel. Die französische Revolution brachte, wie für alle französischen Adelshäuser der Zeit, einen Einschnitt in der Erfolgsgeschichte der Familie; Charles Louis Victor (1756-1794) wurde als Opfer der *Terreur* guillotiniert. Dennoch sympathisierten die freiheitlich denkenden de Broglies mit den Idealen der neuen Gesellschaft. Wohl aus diesem Grund blieb ihr Anteil an der Restauration eher gering, so daß erst die Julimonarchie die volle gesellschaftliche Bedeutung der Familie wiederherstellen konnte: Léonce Victor Charles (1785-1870), der 1816 Albertine de Staël[7] heiratete, war einer der bedeutenden Staatsmänner Frankreichs

---

[7] Albertine war die Tochter der Schriftstellerin, die als Mme. de Staël berühmt wurde und die somit eine angeheiratete Vorfahrin Louis de Broglies war.

von 1830-1848. Auch der Großvater Louis de Broglies, Charles Victor Albert (1821-1901), war engagiert und erfolgreich in der Politik tätig, mußte jedoch wegen starker Anfeindungen seine Karriere 1877 aufgeben. Gerade auch auf die Laufbahn des Vaters Victor (1846-1906) wirkte sich dies negativ aus, so daß dieser trotz seines Engagements in der Regionalpolitik auf nationaler Ebene nie wirklich erfolgreich war.

## I.2 Kinderjahre

Dieser kurze Abriß nur der wichtigsten Aspekte der Familiengeschichte dürfte vielleicht verständlich machen, daß das Gewicht dieser Tradition schwer auf den Eltern Louis de Broglies lastete. Um dem bedeutenden Namen gerecht zu werden, führten sie ihr Leben in einer Weise, die heute unvorstellbar und für das späte 19. Jahrhundert bereits anachronistisch erscheinen mag.

Die Familie wechselte mehrere Male im Jahr mit dem gesamten Hausrat und der Dienerschaft in einem eigenen Saloneisenbahnwagen den Wohnort: Den Winter verbrachte sie zumeist in ihrem Stadthaus in der *Rue de la Boétie* in Paris, einige Monate verlebte man auf einem Besitz in Anjou, Saint-Amadour, in dessen Bezirk sich der Vater politisch engagierte, und im Sommer zog man in das Landhaus in Dieppe, wo Louis de Broglie am 15. August 1892 geboren wurde. Seine drei Geschwister waren vier, siebzehn und zwanzig Jahre älter als er, so daß Louis in einer Erwachsenenwelt aufwuchs: Außer zu seiner Schwester Pauline hatte er kaum Kontakt zu Gleichaltrigen und, was vielleicht noch prägender war, seine wenigen Spielkameraden entstammten der gleichen adligen Schicht wie er. Deshalb blieb Louis viele Jahre der Blick über die engen Grenzen der heilen, elitären Welt, in die er hineingeboren war und deren Regeln und Gesetze er als normal und alltäglich erachten mußte, verwehrt. Seine Erziehung übernahm eine Amme, die ihn überall hin begleitete, während er seine Eltern nur wenige Augenblicke des Tages zu sehen bekam; seine Schwester traf er nur zu festgesetzten Zeiten des Tages. Zur Normalität seines Lebens gehörten nicht nur das zahlreiche Personal, das ihm jede Aufgabe des täglichen Lebens abnahm, sondern auch die sehr kostspielige Lebensführung: täglich fanden im Haus Festbankette mit geladenen Gästen in Abendgarderobe statt. Über diese Zeit schrieb de Broglie:

> J'ai passé mon enfance et ma toute première jeunesse dans un milieu assez fermé, mais où les préoccupations intellectuelles étaient grandes [...].[8]

Und tatsächlich bestand sein Hauptvergnügen und die wichtigste Beschäftigung schon früh im Lesen[9]. Ihm standen die immensen Bibliotheken seines Vaters zur Verfügung, und da er sich in ihnen frei bewegen konnte, gab er sich einem unkontrollierten, oft vielleicht verfrühten Konsum von Literatur hin. Darüber schrieb seine Schwester Pauline, die Comtesse de Pange, in ihren Memoiren:

---

[8] Übersetzung: „Ich habe meine Kindheit und früheste Jugend in einem recht geschlossenen Milieu verbracht, in dem jedoch intellektuelle Betätigungen eine große Rolle spielten [...].", [288, S. 383].

[9] 1972 schreibt de Broglie scherzhaft: „J'ai tellement lu pendant ma vie que je suis étonné d'avoir encore des yeux." Übersetzung: „Ich habe in meinem Leben soviel gelesen, daß ich erstaunt bin, noch Augen zu haben.", [288, S. 383].

> Elevé comme moi, dans une solitude relative, il avait beaucoup lu et vécu dans l'irréel.
> Il parlait tout seul pendant des heures en marchant de long en large, inventant des
> personnages et leur donnant les répliques.[10]

Auch gemeinsam mit seiner Schwester erfand er Rollenspiele, in die sich beide stundenlang verloren [284c, S. 170].

Erlebte Pauline ihren kleinen Bruder zunächst als „nervöses, zwischen Lachen und Weinen schwankendes Kind" [284c, S. 165], so schien sich mit den Jahren, trotz der isolierten, fast einsamen Lebensweise, den z.T. kalten und unfreundlichen Wohnräumen[11] und der strengen Etikette der „optimistisch heitere Grundzug seines Wesens" [284, S. 165] gefestigt zu haben. Die Comtesse de Pange beschreibt den Elfjährigen als aufgeweckten, mitteilungsbedürftigen Jungen, dessen Fröhlichkeit das ganze Haus erfüllte, der in „jungenhaftem Übermut" Stimmen und Gesten anderer Leute täuschend nachahmte und der in seiner heiteren Art gegen die Etikette verstieß [284c, S. 166ff].

## I.3 Erste Ausbildung

Nachdem die Amme und ein weiteres Kindermädchen Louis die ersten Jahre begleitet hatten, übernahmen zwei geistliche Haushofmeister, zunächst ein ehemaliger Missionar, danach, 1904, ein Abbé Chanet, die Erziehung und den ersten Unterricht von Louis [284, S. 167]. Natürlich war der Unterricht stark humanistisch ausgerichtet: Latein, Griechisch und auch Französisch nahmen einen breiten Raum ein. Eine besondere Prägung bekam de Broglies Ausbildung, als die Familie 1901 nach dem Tod des Großvaters das Schloß Broglie und dessen Verwaltung übernahm. Dort fand Louis nämlich eine Vielzahl angesammelter Familienerinnerungen in Form von Dokumenten, Büchern, Biographien und Kunstgegenständen, deren Studium ihn gefangen nahm und ein reges Interesse für Historisches weckte. Und so schreibt de Broglie rückblickend über die ersten Jahre seiner häuslichen Schulausbildung:

> [...] ces préoccupations étaient nullement scientifiques, mais plutôt littéraires et surtout historiques.[12]

Ein Ereignis, das Louis de Broglies Werdegang entscheidend beeinflussen sollte, war der frühe Tod des Vaters im Jahr 1906. Da sich der Vater nie um ein intimes Verhältnis zu seinen Kindern bemüht hatte [284c, S. 253], war die emotionale Teilnahme der Kinder vor allem auf die vielfältigen Veränderungen des Alltagsleben beschränkt. Für Louis' Entwicklung bedeutete dies Ereignis aber vor allem, daß sein viel älterer Bruder Maurice an Stelle des Vaters die Verantwortung für seine Erziehung und Zukunft übernahm.

Maurice fühlte sich der Familientradition weit weniger verpflichtet als sein Vater. Bereits seine Berufswahl war auf den verhaltenen Protest seiner Eltern gestoßen: Er hatte sich, ganz anders als seine Vorfahren, für eine Marinekarriere entschieden und damit eine Art Kompromiß gesucht zwischen der Familientradition und seinem eigentlichen Interesse, der Naturwissenschaft. Im Rahmen seiner Ausbildung lernte er Professor Brizard kennen, der ihn mit der neuen Physik der Elektronen und Röntgenstrahlen vertraut machte. So begann er, eigene Experimente zu entwickeln, und nahm fast heimlich ein Physikstudium auf. Langsam wuchs die Entscheidung, die militärische Laufbahn aufzugeben. Nach seiner Heirat richtete er sich 1906 in seiner Pariser Stadtresidenz ein privates Laboratorium ein und gab, sehr zum Leidwesen seiner Familie, endgültig die Marinekarriere auf. Unter heute nicht mehr vorstellbaren unzureichenden Sicherheitsvorkehrungen untersuchte er die Eigenschaften der Röntgenstrahlen sowie den Photoeffekt und errang in Fachkreisen damit in den folgenden Jahren große Anerkennung.

---

[10] Übersetzung: „Ebenso wie ich war er in relativer Einsamkeit erzogen worden, er hatte viel gelesen und in einer Phantasiewelt gelebt. Stundenlang führte er Selbstgespräche, wobei er auf und ab ging, Personen erfand und ihnen antwortete.", [284b, S. 55ff].

[11] Zumindest die Wohnräume des alten Schlosses dürften nach der Beschreibung der Comtesse de Pange [284c, S. 122ff] alles andere als wohnlich gewesen sein.

[12] Übersetzung: „[...] diese Beschäftigungen waren in keiner Weise naturwissenschaftlich, sondern eher literarisch und vor allem geschichtlich.", [288, S. 383].

Dieser junge Mann übernahm also die Verantwortung für den Werdegang seines Bruders. Bislang hatte Louis kein sehr enges Verhältnis zu Maurice gehabt und hatte dessen Forschungstätigkeit kaum zur Kenntnis genommen. Erst jetzt bekamen seine Interessen eine erste Orientierung auf die Naturwissenschaft. Denn Maurice bewirkte, daß Louis in das *Lycée Janson de Sailly* eintrat, an dem Professor Brizard unterrichtete. Außerdem konnte er Louis dazu bewegen, den Zweig Latein-Naturwissenschaften zu wählen. Louis war ein ausgezeichneter Schüler in den Fächern Literatur, Philosophie, Geschichte und Physik, nur mittelmäßig jedoch in Mathematik und Chemie. Nach drei Jahren erhielt Louis de Broglie sein *Baccalauréat* in Mathematik und in Philosophie.

Im Alter von 17 Jahren standen ihm also sehr verschiedenartige Perspektiven offen: am meisten reizten ihn Geschichte, Recht und Physik. Die Entscheidung fiel ihm denkbar schwer, sein Bruder ließ ihm jedoch die volle Freiheit zu wählen. Maurice schrieb:

> Ayant éprouvé moi-même les inconvénients d'une pression sur les études d'un jeune homme, je me gardai bien d'imprimer une direction rigide aux études de mon frère, quoique je fusse un peu inquiet de son flottement.[13]

De Broglie entschied sich zugunsten eines Geschichtsstudiums: Er belegte das Wahlfach Mittelalter und erreichte 1910 mit exzellenten Ergebnissen seine *Licence*. Trotzdem blieb er unentschlossen und verbrachte ein Jahr mit Sprachkursen und ersten Anfängen eines Jurastudiums. Gleichzeitig begann er, sich für die Geschichte der Naturwissenschaften zu interessieren, eine Fachrichtung, die seinen unterschiedlichen Interessen gerecht werden konnte. Er nahm hier zum ersten Mal Kenntnis von dem Um- und Aufbruch, in dem sich die Physik damals befand. Die Faszination, die davon ausging, bewegte ihn schließlich dazu, sich an der Sorbonne für ein Physikstudium einzuschreiben.

## I.4 Naturwissenschaftliche Ausbildung

War die Einschreibung noch aus dem Zögern um den richtigen Weg heraus geschehen, so ging die wirkliche Berufung, die Entscheidung, sich voll und ganz der Physik zu widmen, von dem ersten Solvay-Kongreß aus, der Ende Oktober 1911 in Brüssel stattfand. Die größten Physiker der Zeit, darunter Planck, Sommerfeld, Wien, Nernst, Rutherford, Perrin, Marie Curie und Einstein, trafen sich hier, um unter dem Titel „Theorie der Strahlung und Quanten" die Wege der modernen Physik zu diskutieren. Maurice de Broglie, der wohl seinen experimentellen Erfolgen die Ehre zu verdanken hatte, Sekretär dieses Kongresses zu sein, war damit beauftragt, die Ergebnisse der Diskussionen zu veröffentlichen. Auf diesem Wege bekam Louis die Protokolle noch vor deren Veröffentlichung zu lesen. Er, der es gelernt hatte, in alten Handschriften zu lesen,

---

[13] Übersetzung: „Da ich selbst die Nachteile starken Druckes auf das Studium eines jungen Mannes kennen gelernt hatte, hütete ich mich davor, dem Studium meines Bruders eine strenge Richtung aufzuprägen, obwohl ich etwas besorgt war wegen seiner Unentschlossenheit.", zitiert in [285, S. 17].

bekam nun auf über vierhundert Seiten handschriftliche Kommentare, Ideen und
Aufsätze der bekanntesten Physiker zu lesen, er erlebte, wie sich die Geschichte der
Physik neu schrieb und war davon fasziniert. Mit einem Schlag waren alle Zweifel
ausgelöscht. Für ihn stand fest, daß er Anteil haben wollte an den Entwicklungen der
modernen Physik: Er würde sein Leben der Physik weihen.

Tatsächlich hatte diese Entscheidung etwas Absolutes: André George drückte es aus mit
den Worten, de Broglie sei entschlossen gewesen „à épouser la science et elle seule"[14].
Und in der Tat löste er seine Verlobung, die arrangierte Verbindung mit einem Fräulein
aus gutem Hause, gab Vergnügungen wie Bridge- oder Schachspiel als
Zeitverschwendung auf und reduzierte seine Besuche und gesellschaftlichen
Verpflichtungen aufs Minimum. Auch seine wissenschaftliche Arbeit konzentrierte sich
ganz und gar auf die Naturwissenschaften, er verbannte alle Geschichts- und Jurabücher
aus seiner Bibliothek.

Die Annäherung an die moderne Physik war allerdings für de Broglie trotz des durch den
Bruder einigermaßen geebneten Weges recht mühsam. Die Professoren an der Sorbonne
waren alles andere als gut informiert und standen den neuen Theorien kritisch gegenüber.
So scheint der Einwand eines Professors gegen die Relativitätstheorie repräsentativ
gewesen zu sein für die skeptische Stimmung: Dieser glaubte nämlich nicht an die
Zeitdilatation, da er selbst sehr schnell mit einem Karren in seinem Garten gefahren sei
und keine Zeitverzögerung habe feststellen können. De Broglie war nicht sehr
aufgeschlossen für diese Art der Polemik und behielt sein Leben lang eine Abneigung
gegen den Skeptizismus seiner Professoren. Besonders rückständig war aber gerade die
Mathematikausbildung:

> Il y avait certainement des lacunes dans cet enseignement. On ne parlait guère des
> intégrales de Fourier, ni des importantes fonctions particulières comme les fonctions
> de Bessel. Il n'était aucunement question de la théorie des fonctions propres et des
> valeurs propres [...].[15]

Auch Léon Brillouin äußerte sich ähnlich über den mathematischen Unterricht an der
Sorbonne:

> The situation in France was serious. There was no regular course of applied
> mathematics when I entered the Sorbonne - nothing at all. I didn't know about the
> Bessel function, I didn't know about any computations with expansions and different
> kinds of orthogonal functions [...]. There were really few people working in theoretical
> physics in France at that time. Theoretical physics was really at a low level, when I
> was a student.[16]

So wurde für de Broglie der einzig wirklich zuverlässige Zugang zur Physik und zur
Mathematik die eigene, intensive Lektüre der Primärwerke, ein Vorgehen, das ohnehin
am ehesten seinem Wesen und seinen Talenten entsprach: Louis de Broglie studierte
Mechanik in den Werken von Appell und Poincaré, Optik in der Abhandlung von Drude,
die gerade ins Französische übersetzt worden war. Gerade die Lektüre dieser drei
Autoren führte ihm die Analogie zwischen der Optik und der Mechanik vor Augen, eben
jene Analogie, die ihn zu seinem Wellenkonzept führen sollte. Weitere Autoren, die de
Broglie las, waren Lorentz, Boltzmann, Gibbs, Planck und Einstein. Nach nur zwei

---

[14] Übersetzung: „[...] die Wissenschaft und nur sie zu heiraten [...].", [288, S. 4].
[15] Übersetzung: „Es gab hier sicherlich Lücken in diesem Unterricht. Es wurde kaum von den
Fourierintegralen gesprochen, noch von speziellen wichtigen Funktionen, wie etwa den
Besselfunktionen. Es war nie die Rede von der Theorie der Eigenfunktionen und den Eigenwerten [...].",
de Broglie in einem Interview mit F. Kubli, zitiert in [291, S. 55].
[16] Léon Brillouin in einem Interview mit T.S. Kuhn 1968, zitiert in [291, S. 55].

Jahren, 1913, schloß de Broglie sein Physikstudium mit exzellenten Noten, gerade auch in den gefürchteten mathematischen Fächern, ab.

Obwohl ihn seine neuen Erkenntnisse begeisterten und er begierig war, mehr von den Quanteneffekten zu erfahren, konnte er während der folgenden sechs Jahre wegen des Militärdienstes und der allgemeinen Mobilmachung in dieser Richtung nicht weiter arbeiten.

# I.5 Militärdienst auf dem Eiffelturm

Aufgrund seiner besonderen Kenntnisse im Bereich der Radiotelegraphie und vielleicht auch aufgrund guter Beziehungen wurde de Broglie als Ingenieur bei den Funkern am Eiffelturm stationiert, wo er bis 1919 blieb. Diese Zeit ist von großem Interesse für de Broglies Werdegang: In gewisser Weise wurde hier seine Erziehung und Ausbildung komplettiert. Zwischenmenschlich lernte er eine sozial niedrigere, für ihn ganz unbekannte, niedrigere gesellschaftliche Schicht kennen: Zum ersten Mal hatte er direkten Kontakt zu ganz anders sozialisierten Menschen. Trotz der völlig veränderten Situation konnte er sich offensichtlich relativ gut an diese Lage gewöhnen und scheint ein herzliches Verhältnis zu seinen Kameraden gehabt zu haben. Mit einigen pflegte er noch lange Zeit nach dem Krieg Briefwechsel. Im Nachlaß de Broglies, in dem er nur die Briefe aufbewahrte, denen er einen historischen Wert zuordnete, fand sich unter all den Glückwunschschreiben zum Nobelpreis (darunter Briefe von Einstein, Schrödinger, Heisenberg, Pauli u.a.) auch die Karte eines Kameraden, die einen kleinen Einblick in die Zeit am Eiffelturm erlaubt:

> Mon cher Louis, Décidément, tu seras toujours aussi farceur. Mais cette fois tu as trouvé la meilleure, avec le prix Nobel.[17]

Bedenkt man, daß sich sogar die Brüder de Broglie siezten, so ist der vertrauliche Ton des Briefes ein deutlicher Hinweis darauf, daß de Broglie in seiner Militärzeit ganz wesentlich seinen sozialen Horizont erweitert hat.

Gleichzeitig lernte er experimentelle und praktische Aspekte der Physik kennen, für die er zeitlebens reges Interesse hegen sollte. Zu seiner Hauptaufgabe entwickelte sich die Reparatur defekter Geräte. Von prägendem Einfluß war aber wohl auch der alltägliche Umgang mit Wellenphänomenen, deren Realität seine Physikkonzeption entscheidend beeinflussen sollte, so daß er später sagte:

> Quand on s'est sali les mains pendant des jours et des nuits à faire démarrer les gros alternateurs qui servaient, à l'époque, aux émissions de radio, il n'est plus facile de croire qu'une onde ne puisse être qu'une probabilité de présence.[18]

Während des Militärdienstes war de Broglie fast vollständig von den neuesten Entwicklungen der Physik abgeschnitten. Nur ein Fachbuch, die *Thermodynamik* Plancks, hatte er mitgenommen, und so blieben ihm die neuesten Entdeckungen und Theorien bis zu seiner Demobilisierung 1919 weitgehend unbekannt.

---

[17] Übersetzung: „Mein lieber Louis, ganz sicher bist Du noch immer der gleiche Witzereißer. Aber diesmal hast Du mit dem Nobelpreis den besten gemacht.", zitiert in [283, S. 67].

[18] Übersetzung: „Wenn man sich Tag und Nacht die Hände damit dreckig gemacht hat, die großen Wechselstromgeneratoren anzuwerfen, die damals dazu dienten, Radiosignale auszusenden, fällt es nicht mehr leicht, daran zu glauben, daß eine Welle nur eine Aufenthaltswahrscheinlichkeit sei.", zitiert in [283, S. 69].

# I.6 Zusammenfassung

Die in diesem ersten Kapitel beschriebenen Lebensjahre waren von entscheidendem Einfluß für de Broglie. Da viele Aspekte seines bemerkenswerten Charakters, seiner ungewöhnlichen Lebensweise, seines historisch geprägten Weltbildes durch die besonderen Umstände seiner Kindheit, Jugend und Studienjahre wenn nicht erklärt, so doch plausibel gemacht werden, erscheint es mir angemessen, als eine Art Zusammenfassung des vorher Geschriebenen den Versuch zu unternehmen, ein möglichst genaues Bild des erwachsenen de Broglie zu entwerfen. Des Menschen also, der 1919 die Arbeit an dem neuartigen Wellenkonzept aufnahm und dessen Charakter im Laufe seines Lebens immer wieder in ähnlicher Weise beschrieben wurde. Hierbei stütze ich mich auch auf Erinnerungen seiner Verwandten, Bekannten und Zeitgenossen. Der Vollständigkeit halber werde ich an manchen Stellen etwas vorgreifen müssen.
In [282] beschreibt Jean Hamburger den eben Verstorbenen mit den Worten:

> [...] d'une élégance un peu surannée, courtois à l'extrême, mais volontiers silencieux, inspirant de soi la déférence et le respect [...].[19]

Diese Kombination aus distanzierter, formvollendeter Höflichkeit und respekteinflößender, in sich gekehrter Eleganz scheint für de Broglies Auftreten charakteristisch gewesen sein. - Ein Auftreten, das ihm sicherlich als arrogante Überheblichkeit ausgelegt werden konnte (und auch wurde), das tatsächlich jedoch das Resultat einer noch Idealen des 18. und 19. Jahrhunderts verpflichteten Erziehung war: Die unglaubliche Disziplin, das häusliche Zeremoniell und die hohen Anforderungen einer tief verwurzelten Familientradition, mit denen de Broglie aufwuchs, haben seine Umgangsformen so stark geprägt, daß er sie zeitlebens nicht ablegen konnte und wollte. Trotzdem scheint sein unzeitgemäßes Auftreten nicht unbedingt Spiegel seiner inneren Einstellung gewesen zu sein. Obwohl er stolz auf seine Familie war, schmückte er sich nicht mit den Titeln, die seine Vorfahren errungen hatten. Immer waren es die anderen, die in ihm den *Duc*, ja sogar den *Prince*[20] sahen. De Broglie ließ sich schlicht mit „Monsieur" ansprechen, auf seinen Visitenkarten und auf seinem Briefpapier fanden sich nur seine akademischen und universitären Titel; die Titel also, die er sich selbst verdient hatte. De Broglie ruhte sich offensichtlich nicht auf den familiären Lorbeeren aus, er spürte vielmehr die Verpflichtung, selbst Großes zu vollbringen, um seinem Namen gerecht zu werden.

> On ne choisit pas sa famille, [...], mais on choisit ce qu'on fait et on choisit ses amis.
> La famille, on ne peut pas s'en vanter, on peut seulement la mériter.[21]

Und selbst mit seinen Verdiensten prahlte er in keiner Weise: die de Broglie-Wellenlänge bezeichnete er in den Vorlesungen unter Vermeidung seines eigenen Namens ausschließlich als Materiewellenlänge. War de Broglie eitel, so war er dies in unaufdringlicher und zurückhaltender Weise.
Ganz offensichtlich hatten die relativ liberale Erziehung durch Maurice sowie die jahrelange Arbeit außerhalb des eigenen sozialen Milieus während des Krieges de Broglie lernen lassen, wie wenig aussagekräftig Herkunft und Titel sind: Diejenigen, die ihn näher kennenlernen konnten, bezeugen, daß de Broglie sich keinesfalls aufgrund seiner Herkunft den Mitmenschen überlegen fühlte: Er empfing Studenten mit der gleichen Höflichkeit wie Minister, half ausländischen Studenten mit der gleichen Selbstverständlichkeit wie französischen und verabschiedete seine Putzfrau vor einem Akademiemitglied, da sie die ältere Dame war.[22] In diesem Sinne ist es wohl auch zu verstehen, wenn Christian de Pange seinen Onkel Louis de Broglie als „moderne et un peu hors de temps"[23] charakterisiert. Modern war er insofern, als er die Menschen weder nach ihrer sozialen Stellung, Herkunft, Titel, noch nach ihrer Nationalität, Religion oder politischen Meinung, sondern nach moralischen Werten und persönlichen Verdiensten beurteilte [292, S. 13]. Und er war unmodern durch seine emotionale Verbundenheit mit der Familientradition und seinen anachronistischen Habitus. Daher scheint mir de Broglies traditionsbewußtes Festhalten am zukunftsweisenden Familienmotto „Pour l'avenir"[24] diesen scheinbaren Widerspruch zu versinnbildlichen.

---

[19] Übersetzung: „[...] von einer etwas veralteten Eleganz, von äußerster Höflichkeit, aber gerne schweigsam, flößte er aus sich heraus Achtung und Respekt ein.", [282].

[20] Georges Lochak berichtet [283, S. 16] von zahllosen Besuchern de Broglies, die ihn zuvor hilfesuchend befragten, welches wohl die geziemliche Anrede sei.

[21] Übersetzung: „Seine Familie sucht man sich nicht aus, man entscheidet sich jedoch, was man tut und wählt seine Freunde aus. Seiner Familie kann man sich nicht rühmen, man kann sie höchstens verdienen.", de Broglie zitiert in [292, S. 11].

Gerade dieser unmoderne Aspekt aber hat sicherlich zur Isolation und zum späteren Vergessen de Broglies beigetragen. Er war kein Mann mit Charisma, dem die Herzen der Studenten zuflogen, der in der Lage gewesen wäre, die Masse seiner Schüler anzusprechen und sie um sich zu scharen. Er war wenig medienwirksam [287, S. 105] und blieb damit einem breiten Publikum unbekannt.

Eben die etwas abgeschiedene Arbeitsweise, unbeobachtet den eigenen Ideen zu folgen, entsprach voll und ganz den Bedürfnissen und Neigungen de Broglies. Aufschlußreich ist die Erinnerung Léon Brillouins, der de Broglie regelmäßig traf, als dieser seine Doktorarbeit schrieb:

> [...] he wrote his Doctor's thesis. His brother, [...], Langevin and myself, I think we were the only three persons to whom he talked during the time of his thesis. [291, S. 31].

Die relative Einsamkeit seiner Kinderjahre und die Zustände an der Sorbonne hatten ihn gelehrt, sich Stoffe selbst zu erarbeiten, eigene Ideen zu entwickeln und auszuarbeiten. In Kombination mit der Eigenschaft, unmodern sein zu können, also der Fähigkeit, sich dem gesellschaftlichen Druck gängiger Moden zu entziehen, hatte er die Fähigkeit, wirklich Revolutionierendes zu schaffen. Tatsächlich scheint ein wesentlicher Zug seines Charakters die „liberté d'esprit"[25] gewesen zu sein, derer ein Wissenschaftler wohl bedarf, wenn er alte Konzepte in Frage stellen will. So sehr dieses Talent ausgeprägt war, hatte de Broglie jedoch nie gelernt, mit anderen zusammenzuarbeiten, und ihm waren damit die Möglichkeiten verbaut, die ein solcher konstruktiver Austausch bieten kann.[26]

Ganz wesentlich wurde de Broglies späteres Schaffen sicherlich auch durch die Art seiner Ausbildung beeinflußt. Durch die zunächst eher humanistische Ausbildung und die Begeisterung für Geschichte war de Broglies Blick auf die Physik ein anderer als der seiner Physikerkollegen. Interessierte ihn die Geschichte, die Interpretation und Anschauung, blieb ihm doch die Entwicklung stark mathematisierter Modelle verschlossen, da ihm hierfür das nötige Handwerkszeug fehlte. Möglicherweise war es

---

[22] Lochak in [292, S. 13].
[23] Übersetzung: „modern und ein bißchen außerhalb der Zeit", [287, S. 161].
[24] Übersetzung: „Für die Zukunft".
[25] Übersetzung: „Freiheit des Geistes", Lochak in [292, S. 12].
[26] Am Anfang seiner Karriere hatte er einige Artikel z.B. [3-6, 10, 11, 14, 15, 19, 20, 26] und eine Einführung in die Lehre der Röntgenstrahlen [152] zusammen mit seinem Bruder und dessen Mitarbeitern geschrieben, sich aber aufgrund der negativen Erfahrungen der Zusammenarbeit gegen weitere Kollektivwerke entschieden.

aber eben genau dieser andere Blickwinkel, der ihn zur Entdeckung der Materiewelle befähigte.

Um eben diese Entdeckung soll es im nächsten Kapitel gehen, in dem ich wieder in die Chronologie der Ereignisse zurückkehren werde. Es sollen die Jahre 1919-1927 betrachtet werden, in denen de Broglie seine größten wissenschaftlichen Erfolge hatte.

# II Jahre der Kreativität (1919-1927)

In dem nun folgenden Kapitel soll der Sprung vom Menschen zum Physiker de Broglie vollzogen werden, sofern man diese Trennung überhaupt wirklich sinnvoll vornehmen kann. Bevor aber auf seine eigene Forschung eingegangen werden kann, erscheint es mir nützlich, in aller Kürze den Stand der Physik um 1920 zu resümieren, um den Bezug zu den Erkenntnissen de Broglies besser herstellen zu können. Hierfür wird es nötig sein, kurz die entscheidenden historischen Etappen zu beschreiben, die zur Aufstellung der Quantenhypothese und der Relativitätstheorie geführt haben; besonders interessant ist in diesem Zusammenhang die Entwicklung der Theorie des Lichtes. Da es bereits ausführliche Darstellungen zu diesem Thema gibt,[27] will ich mich auf die Entdeckungen und Theorien beschränken, auf die sich de Broglie berief, und die ihn zu seinem Materiewellenkonzept geführt haben.

## II.1 Kurzer historischer Überblick

In der Geschichte der Optik, die bei weitem nicht erst im 17. Jahrhundert beginnt, stehen sich zwei zunächst widersprüchlich erscheinende Konzepte gegenüber: Während das eine besagt, daß sich Licht in Form eines Stroms von Teilchen fortbewegt, führt das andere dessen Ausbreitung auf Lichtwellen zurück.

Das weitaus ältere Konzept ist das des Korpuskelbildes, welches sich bereits in der Antike finden läßt.[28] Bis ins erste Jahrtausend glaubte man fest an die Teilcheneigenschaften des Lichtes,[29] das sich in einem Medium ausbreitet, wobei die Diskussion um die Frage kreiste, ob das Auge das Licht aussende, oder ob dieses von den Gegenständen ausgehe.

Einen ersten Schritt zur einer Wellentheorie des Lichtes vollzog im 13. Jahrhundert der englische Bischof Robert Grosseteste[30], der an eine Wesensverwandtschaft von Wellen und Licht glaubte, die von seinen Schülern weiterverfolgt wurde. Auch Kepler[31] sah drei Jahrhunderte später in der Lichtausbreitung eine wellenartige Bewegung. Galileo Galilei[32] führte den Begriff der Frequenz ein, den Descartes[33] auf das Licht anwendete. Weitere Erkenntnisse lieferten Olaüs Rømer[34], der bei der Beobachtung der Verfinsterung der Jupitermonde fand, daß das Licht sich mit endlicher Geschwindigkeit ausbreiten muß, und Fermat[35], der das „Prinzip der kürzesten Zeit" postulierte, das besagt, daß das Licht immer den schnellsten Weg wählt. Einen ersten Ansatz, die

---

[27] Z.B. in [295, 296, 298, 302, 303].

[28] Anmerkungen zu dieser Theorie finden sich z.B. bei: Demokrit (460-370 v. J.Chr.), Platon (427-347 v.Chr.), Aristoteles (384-322 v.Chr.), Lukrez (98-55 v.Chr.), Seneca (4 v. Chr. -64 n.Chr.)

[29] Hinweise hierzu finden sich in den Arbeiten von: Ptolemäus (90-168), Galien (129-199), Ibn-al-Haitham (965-1093).

[30] Robert Grosseteste (1168-1253)

[31] Johannes Kepler (1571-1630).

[32] Galileo Galilei (1564-1642).

[33] René Descartes (1596-1650).

[34] Olaüs Rømer (1644-1710).

[35] Pierre Fermat (1601-1665).

Wellennatur des Lichtes zu beschreiben, entwarf Malebranche[36], der als Quelle des Lichtes Vibrationen ansah, die um eine Mittellage in Richtung der Ausbreitungsrichtung ausgeführt werden. Erst Huygens[37] jedoch entwickelte die erste vollständige Wellentheorie des Lichtes, die nicht nur das Fermatsche Prinzip bestätigte, sondern auch eine Vielzahl experimenteller Ergebnisse erklärte. Nach Huygens besteht das Licht aus im Raum ausgedehnten Wellen, die sich über elastische Stöße des Äthers ausbreiten.

So richtungsweisend und zutreffend auch immer diese Ansätze waren, konnten sie sich in ihrer Zeit doch nicht durchsetzten. Weder das Prinzip Fermats, noch Huygens Theorie waren zu Lebzeiten ihrer Begründer anerkannt; es war das Konzept Newtons[38], das den Zeitgeist traf, Beachtung fand und von seinen Schülern weitergetragen wurde. Zunächst hatte er an eine Mischform von Korpuskel- und Wellentheorie gedacht: Das Licht sollte aus Teilchen bestehen, die mit dem aus Korpuskeln bestehenden Äther wechselwirken und diesen in Vibration versetzen können, tatsächlich konnte er mit dieser Mischform die nach ihm benannten Interferenzringe erklären. Newton unterschied also eindeutig Teilchen und die von ihnen ausgelöste Welle - eine Idee, die sich bei de Broglie wiederfinden sollte. Im Zusammenhang mit seiner Gravitationstheorie lehnte Newton jedoch in der Folgezeit die Existenz des Äthers ab und sprach sich ausschließlich für den Teilchencharakter des Lichtes aus. Er verwarf die Wellentheorie mit erst viel später widerlegten Argumenten und konnte seine Sicht durchsetzen. So blieb die Wellentheorie während des gesamten 18. Jahrhunderts unbeachtet.

Erst die Entdeckung des Interferenzprinzips durch Young[39] leitete 1801 eine Wende ein: Die Deutung des Versuchs durch Fresnel[40], der das Huygensche Prinzip anwendete, und auch die Experimente Frauenhofers[41] ließen nach kurzen Kontroversen die Korpuskeltheorie vergessen. Fresnel begründete eine reine Wellenoptik, die das Licht als Schwingungsfortpflanzung in einem elastischen Medium (dem Äther) erklärte und die den wellenmechanischen Vorstellungen de Broglies in ihren Grundzügen nicht unähnlich ist. Die Wellenoptik sollte ihre Gültigkeit bis ins frühe 20. Jahrhundert behalten.

Um die Arbeiten de Broglies in ihren historischen Kontext einordnen zu können, müssen nach diesem sehr verkürzten Überblick über die historische Entwicklung der Optik und des Lichtkonzeptes, noch einige entscheidende Entdeckungen und Theorien aus den anderen Bereichen der Physik dargestellt werden.

Die Mechanik entwickelte sich im 19. Jahrhundert durch zahlreiche praktische Anwendungen und durch neue mathematische Konzepte[42] zu einem experimentell abgesicherten und theoretisch festgefügten Zweig der Physik. Gleichzeitig entstanden aber auch neue Gebiete: In den zwanziger Jahren des 19. Jahrhunderts nahm Ampère[43] Experimente Œrstedts[44] auf, die den Einfluß fließenden Stroms auf einen Magneten betrafen, und legte damit die Grundlage für die Faradaysche[45] Entdeckung des Feldbegriffes. Erst Maxwell[46] jedoch gelang in den 60er Jahren des letzten Jahrhunderts die bestechende und endgültige mathematische Formulierung. Aus Maxwells Theorie folgte gleichzeitig die Existenz elektromagnetischer Wellen, die tatsächlich 1888 von H. Hertz[47] entdeckt wurden, und Maxwell gelang es zudem, das Licht in seine Theorie einzuordnen, allerdings trafen diese Ideen auf nicht unwesentlichen Widerstand und wurden erst später voll akzeptiert. Trotzdem hatte das Werk dieses genialen Physikers starken Einfluß auf die Vorstellung der physikalischen Realität: Maxwell selbst sah zwei konkurrierende Theorien: die Theorie eines „vollen Universums" (also der Feldtheorie) gegen die Theorie von „Atomen im leeren Raum"[48] und bezeichnete deren Verhältnis als einen seit der Antike andauernden Streit über die Struktur der Materie. Einstein schrieb zum Stand der Diskussion im 19. Jahrhundert, daß die Prozesse in der Natur vor Maxwell in der Form materieller Punkte, nach Maxwell aber in der Form von Feldern begriffen wurden.[49]

---

[36] Nicolas de Malebranche (1638-1715).
[37] Christian Huygens (1629-1695).
[38] Isaac Newtons (1643-1727).
[39] Thomas Young (1773-1829).
[40] Augustin Jean Fresnel (1788-1827).
[41] Joseph Frauenhofer (1787-1826).
[42] So hatte schon im 18. Jahrhundert Pierre-Louis Moreau de Maupertuis (1698-1759) sein Prinzip der kleinsten Wirkung formuliert, im 19. Jahrhundert entstanden das Konzept der Hamilton-Jacobi Mechanik und die Lagrangesche Formulierung.
[43] André Marie Ampère (1775-1836).
[44] Hans Christian Œrstedts (1777-1851).
[45] Michael Faraday (1791-1867).
[46] James Clerk Maxwell (1831-1879).

Gleichzeitig mit dem Feldkonzept entwickelte sich aber auch die scheinbare Gegenbewegung: Gerade in der Chemie entdeckte man im 18. Jahrhundert die schon in der Antike entwickelte Idee der Atome wieder.[50] Nach ersten tastenden Versuchen im 18. Jahrhundert akzeptierte man dann am Anfang des 19. Jahrhunderts auch in der Physik die Nützlichkeit der Atomhypothese und es entstand die Thermodynamik auf der Grundlage molekularkinetischer Vorstellungen, die eng mit den Anschauungen einer atomistischen Struktur verbunden war. In der zweiten Hälfte des 19. Jahrhunderts erhoben sich allerdings zahlreiche Einwände gegen die Atomistik: Trotz der Erfolge der Atomidee sahen viele Wissenschaftler darin nicht mehr als ein nützliches Gedankenkonstrukt und glaubten nicht an die Realität der Atome. Bemerkenswerterweise waren es nicht unbedingt die Vertreter des Wellenkonzeptes, die sich gegen die Realität des Atoms wehrten: Maxwell glaubte fest daran; vielmehr waren es die Positivisten unter Führung Ernst Machs und Wilhelm Ostwalds, aber auch Mitbegründer der Thermodynamik wie Rudolf Clausius[51] und selbst Planck, die die Atome als „Gedankending" abtaten.

In diese Stimmung der Skepsis fielen von 1895-1897 drei große Entdeckungen: die Entdeckung der Röntgenstrahlen,[52] die der Radioaktivität[53] und der experimentelle und theoretische Nachweis des Elektrons.[54] Diese neuen Erkenntnisse mußten nun auch die

---

[47] Heinrich Hertz (1857-1894).

[48] Zitiert in [283, S. 50].

[49] Zitiert in [283, S. 48].

[50] Die etwa im 5.Jh.v.Chr. entwickelte Idee (erste Quellen dazu finden sich z.B. bei Leukipp und Demokrit), es gebe unteilbare kleinste Einheiten, aus denen die Welt zusammengesetzt sei, wurde später zunächst von Epikur, dann von Lukrez modifiziert und überliefert und in der Renaissance von Francis Bacon, Giordano Bruno und Galileo Gallilei, Pierre Gassendi wiederentdeckt. Im 18. Jahrhundert griff Newton die Ideen Epikurs auf. Besonders in der Chemie akzeptierte man das Atomkonzept wegen seiner großen Erfolge früh. John Dalton gilt als der eigentliche Begründer der naturwissenschaftlichen Atomistik, Ant. Laur. Lavoisier, Wiliam Prout, und Amedeo Avogadro verhalfen der Atomidee zu neuer Modernität. Den größten Beitrag lieferte dann in der zweiten Hälfte des 19. Jahrhunderts Mandeleev mit seinem Periodensystem.

[51] Rudolf Clausius (1822-1888)

[52] 1895 durch W.C. Röntgen (1845-1923).

[53] Henri Becquerel fand 1896 heraus, daß Uranmineralien auch ohne Lichteinwirkung eine Photoschicht verändern. Gerhard Carl Schmidt, Marie und Pierre Curie fanden in den Folgejahren noch weitere strahlende Elemente.

[54] Hendrik Antoon Lorentz (1853-1928) hatte bereits um 1890 eine Elektronentheorie entwickelt, die eine Verbindung von Atomismus und Elektrizitätslehre darstellte. Erst Joseph John Thomson (1856-1940) gelang es jedoch 1897, den endgültigen Nachweis der Elektronen zu erbringen. Er griff Untersuchungen der Kathodenstrahlung von Hertz, Jean Baptiste Perrin (1870-1942), Sir William Crookes (1832-1919) u.a. auf, berichtigte Irrtümer und verbesserte Experimente. Schließlich konnte er

Skeptiker von der Existenz der Atome überzeugen. Plötzlich war es möglich und anerkannt, die makroskopischen Effekte der Thermodynamik und der Elektrizitätslehre durch bisher versteckte mikroskopische Parameter zu erklären. - Verborgene Parameter sollten auch im Werk de Broglies noch eine wichtige Rolle spielen.

In der Folgezeit bestimmten zwei wesentliche Probleme die Forschung: Zum einen wollte man die innere Struktur der Atome, die ihre Unteilbarkeit verloren hatten, kennen lernen, zum anderen fehlte eine schlüssige Verbindung zwischen Feldtheorie und Atom.

Die Entwicklung von Atommodellen gründete sich auf folgenden Wissensstand: Um 1900 war bekannt, daß ein Atom aus einem positiv geladenen Anteil und aus einer noch unbekannten Anzahl Elektronen besteht. Elektronen kannte man vom Kathodenstrahl und man vermutete, daß die seit Mitte des 19. Jahrhunderts bekannten Spektren der Elemente von der Bewegung der Elektronen im Atom herrühren mußten. Völlig unbekannt war jedoch ihr positiver Anteil. Es wurden verschiedene Modelle vorgeschlagen: Die einen vermuteten einen punktförmigen positiven Kern, so etwa Perrin mit seiner „structure nucléo-planétaire" (1901) und der Japaner Hantaro Nagaoka[55] mit dem „Saturnien System" (1904); andere, wie z.B. Thomson[56], nahmen stetig verteilte positive Ladung an, in die Elektronen eingebettet wären. Erst Rutherfords[57] Interpretation des Streuversuches von Alphateilchen an Metallfolien, den Geiger[58] und Marsden[59] durchgeführt hatten, konnte 1911 zeigen, daß die Hauptmasse des Atoms auf sehr kleinem Raum zusammengedrängt ist. Über die Bewegung der Elektronen entstanden bald Theorien, die stabile Zustände annahmen (etwa 1903 durch J.J. Thomson), die jedoch noch auf rein klassischer Argumentation basierten. Auch in diesem Bereich war das Modell Rutherfords richtungweisend: danach umkreisen die Elektronen den Kern wie Planeten ihr Zentralgestirn. Es setzte eine Suche nach völlig neuen Atommodellen ein, in deren Zug der Österreicher Arthur Erich Haas (1884-1941) 1910 als erster das Plancksche Wirkungsquantum auf den Atombau anwendete. Ein Jahr später griff Sommerfeld[60] diese Ideen auf, und Bohr[61] entwickelte, angeregt durch Rutherford, sein berühmtes Atommodell, das bereits eine Vielzahl von Phänomenen und Beobachtungen[62] erklären konnte. Bohr postulierte für die Elektronen gequantelte Energiezustände, in denen sie sich auf stabilen Bahnen, ohne Energie abzustrahlen, um den Kern bewegen können. Sommerfeld entwickelte das Modell weiter, indem er 1915/16 durch relativistische Betrachtungen die Feinstruktur der Spektren erklärte. Trotz seiner großen Erfolge konnte sich das Modell nur langsam durchsetzten, stellte dann aber viele Jahre hindurch einen unbestrittenen und wesentlichen Bestandteil der älteren, noch anschaulichen Quantenphysik dar.

Neben der Frage nach dem Aufbau der Atome stand eine andere große Frage im Mittelpunkt des Interesses der Forschung um 1900: welche Verbindung besteht zwischen

---

das Verhältnis e/m und die Geschwindigkeit v für die vermuteten Teilchen bestimmen. Die ermittelte Größe e/m wurde im gleichen Jahr von Pieter Zeeman (1865-1943) bestätigt. Der nach ihm benannte Effekt, der den Einfluß eines Magnetfeldes auf Spektrallinien nachwies, wurde von Lorentz' richtig interpretiert und bestätigte die Elektronentheorie.

[55]  Hantaro Nagaoka (1865-1950).
[56]  William Thomson (1824-1907).
[57]  Ernest Rutherford (1871-1937).
[58]  Hans Geiger (1882-1945).
[59]  Ernest Marsden (1889-1970).
[60]  Arnold Sommerfeld (1868-1951).
[61]  Niels Bohr (1885-1962).
[62]  So z.B. die Balmer- und die Paschenserie im Wasserstoffspektrum. Auch in den Folgejahren wurden zahlreiche Experimente, namentlich durch Moseley, Franck, Hertz durchgeführt, die die Richtigkeit des Modells bestätigten.

der Feldtheorie, die durch die Maxwell-Hertzsche Elektrodynamik beschrieben wurde, und dem Atom, das der klassischen Mechanik des Punktes zugeordnet wurde. Aufgeworfen war dieses Problem namentlich durch die Lorentzsche Elektronentheorie, die zum ersten Mal eine Verbindung zwischen den beiden bisher konkurrierenden Konzeptionen herstellte: Denn Lorentz führte die Interaktion von Feld und Atom, überhaupt alle elektrischen und magnetischen Erscheinungen auf das Verhalten elektrischer Ladungsträger zurück. Dieses Problem führte zu den konkreten Fragen, wie eine Elektrodynamik bewegter Körper aussehen könnte und wie Strahlung emittiert und absorbiert wird. Die erste Frage führte zur speziellen Relativitätstheorie, die zweite zur Entdeckung der Quanten.

Ein wesentliches Problem für die damalige Konzeption der Ausbreitung von elektromagnetischen Wellen stellte nach Fizeaus[63] Experiment der Versuch Michelsons[64] dar, den er 1887 mit Morley[65] wiederholte. War man bisher fest von der Existenz eines ruhenden Äthers als absolutes Bezugssystem überzeugt, zeigte das Experiment, daß es keinen Ätherwind gibt und daß die Lichtgeschwindigkeit im Vakuum unabhängig vom Bewegungszustand der Lichtquelle und des Beobachters immer den gleichen Betrag hat. Diese Ergebnisse beeinflußten die Suche nach einer Elektrodynamik bewegter Körper stark, mit der sich Hertz, Lorentz, Fitzgerald[66] und Poincaré[67] befaßten. In seinem Bestreben, die Hypothese eines ruhenden Äthers aufrecht zu erhalten, übernahm Lorentz die von Fitzgerald formulierte Kontraktionshypothese. Er fand 1904, daß die Maxwell Gleichungen gegenüber einer Galilei-Transformation nicht invariant sind und entwarf die nach ihm benannten Transformationsgleichungen, an denen auch Poincaré arbeitete. Erst Einstein, der erstaunlicherweise eher lückenhafte Kenntnis von Lorentz Arbeiten, Michelsons u.a. hatte, gelang es jedoch, mit seiner speziellen Relativitätstheorie ein schlüssiges theoretisches Konzept zu entwerfen: Einstein erhob die Konstanz der Lichtgeschwindigkeit zu einem neuen physikalischen Prinzip und widerlegte damit endgültig die Existenz des Äthers. Dieses geniale Werk, in dem der damals noch unbekannte junge Physiker die grundlegendsten Prinzipien der klassischen Physik revidierte, fand nicht sofort die breite Zustimmung aller Wissenschaftler; zu weiten Teilen ist es dem Einfluß und der Fürsprache Plancks zu verdanken, daß sich die Theorie in den Folgejahren durchsetzte.

Auf die Frage nach der Emission und Absorption von Strahlung fand Planck im Jahr 1900 den Schlüssel zu einer Antwort: Die sogenannte schwarze Strahlung stellte ein Problem dar, das nicht nur Wien[68], Rayleigh[69] und Jeans[70], sondern auch viele andere Physiker beschäftigte. Man hatte es nicht geschafft, auf der Grundlage der klassischen Gesetze ihre spektrale Intensitätsverteilung zu beschreiben. Einen verhältnismäßig guten Ansatz für die korrekte Beschreibung bei hohen Frequenzen stellte das Wiensche Strahlungsgesetz dar, das jedoch für niedrige Frequenzen weniger brauchbar war; in diesem Bereich stimmte aber das rein auf der klassischen Physik beruhende Rayleigh-Jeans-Gesetz mit der Realität überein, welches jedoch für hohe Frequenzen völlig verkehrte Ergebnisse lieferte, die sogenannte Ultraviolettkatastrophe. Planck gelang es, diese beiden Gesetze so zu interpolieren, daß sie das gesamte Spektrum perfekt beschrieben. Die Interpretation seines berühmten Strahlungsgesetzes brachte Planck auf ein ungewollt revolutionäres Ergebnis: die Energie der Oszillatoren des schwarzen Körpers ist gequantelt. Man schenkte jedoch, Planck eingeschlossen, der Quantentheorie kaum Beachtung, da sie allen Gesetzen der klassischen Mechanik und Elektrodynamik widersprach und dies obwohl man allgemein die Richtigkeit des Strahlungsgesetzes anerkannte. In einem kleinen Artikel revolutionierte Albert Einstein 1905 die Physik: Zu diesem Zeitpunkt gab es keinerlei Diskussion über die Natur des Lichtes. Selbst Planck schien nicht daran zu zweifeln, daß Licht aus elektromagnetischen Maxwell-Hertzschen Wellen bestehe. Einstein aber schreckte nicht wie Planck vor dem unglaublich Neuen der Quantenhypothese zurück, sondern dachte, vom Wienschen Gesetz ausgehend, konsequent in der Richtung Plancks weiter: Er postulierte die Welle-Teilchen Dualität des Lichtes und zeigte, daß im Bereich hoher Frequenzen und Temperaturen nicht mehr die übliche Wellentheorie anzuwenden sei, sondern die Vorstellung unabhängiger Lichtquanten angemessen sei. Er löste damit die Jahrhunderte dauernde Kontroverse um die Natur des Lichtes mit einem zu dieser Zeit noch undenkbaren Kompromiß. Diese Theorie wirkte auf die Zeitgenossen wie das Werk eines leichtsinnigen Jugendlichen, der ohne zu zögern die heiligsten Grundlagen der Physik verwarf (zumal er im selben Jahr noch die Newtonschen Raum- und Zeitbegriffe infrage stellen würde).

---

[63] Hippolyte Fizeau (1819-1896).
[64] Albert Abraham Michelson (1852-1931).
[65] Edward Williams Morley
[66] George Francis Fitzgerald (1851-1901).
[67] Jules Henri Poincaré (1854-1912).
[68] Wilhelm Wien (1864-1928).
[69] John Rayleigh (1842-1919).
[70] James Jeans (1877-1946).

## II.2 Wiederaufnahme der wissenschaftlichen Tätigkeit

Um 1920 waren entscheidende Schritte zu einer neuen Sicht auf die Physik getan, und man begann, die nicht klassisch zu beschreibende Welt der Atome zu verstehen. Die Relativitäts- und die Quantentheorie leisteten ihre Dienste, Bohrs Atommodell war akzeptiert. Trotzdem stieß die ältere Quantenmechanik an ihre Grenzen: Die Ergebnisse, die das Bohr-Sommerfeldsche Modell für wasserstoffähnliche Atome mit einem Valenzelektron im Magnetfeld lieferte, stimmten nicht mit dem Experiment überein, dazu ließ sich das Modell nicht auf Atome mit mehreren Leuchtelektronen ausweiten. Vor allem konnte man aber nur schwer rechtfertigen, daß sich das Modell in seiner Grundlage und Denkart deutlich an die klassische Physik anlehnte, aber trotzdem deren wesentlichste Gesetze verletzte. Es fehlte tatsächlich zu diesem Zeitpunkt ein wirklich neues physikalisches Weltbild, eine Theorie, die die zahlreichen neuen Erkenntnisse in einem neuen Formalismus zusammenführte.

An diesem Punkt halte ich es für angebracht, den historischen Rückblick zu beenden und zu Louis de Broglie zurückzukehren: 1919 befand sich de Broglie in der glücklichen Lage, den Krieg unverwundet und um viele praktische Erfahrungen reicher überstanden zu haben und in das zivile Leben zurückkehren zu können. Im Gegensatz zu vielen anderen jungen und talentierten Physikern hatte de Broglie nur Zeit verloren; immerhin wurde er im Jahr seiner Demobilisierung bereits 27 Jahre alt und hatte damit einen wichtigen Lebensabschnitt übersprungen, in dem er seine Studien wesentlich hätte vertiefen können.

So setzte er in den folgenden Jahren alles daran, den Rückstand aufzuholen. Ein schwieriges Unterfangen, wenn man bedenkt, wie rasant sich die Physik in diesen Jahren entwickelte. Neben seiner privaten Lektüre folgt er den Vorlesungen und Seminaren Paul Langevins (1872-1946) über die Relativitäts- und Quantentheorie am *Collège de France*, die in ihrer Form in Frankreich völlig neu waren und den Blick auf die aktuelle Arbeit der deutschen Physiker eröffneten. Gleichzeitig bekam de Broglie einen engen Kontakt zur Experimentalphysik in dem inzwischen international anerkannten Labor seines Bruders, wo er sich mit den eher theoretischen Aspekten der Forschung beschäftigte. De Broglies Ausbildung zu dieser Zeit war also vielseitig und von höchster Qualität.

Gemeinsam mit seinem Bruder und dessen Mitarbeitern veröffentlichte er erste Artikel zu den atomaren Spektren [3-6, 15], zum Photoeffekt [14] und zu den Röntgenstrahlen [10, 11, 20]. In diesen Arbeiten entwickelte de Broglie Formeln, die zwar die experimentellen Ergebnisse reproduzierten, aber theoretisch auf keinem sicheren Fundament standen, was ihm z.T. massive Kritik einbrachte: So schrieb Hendrik Kramers, Schüler Bohrs, über de Broglies Betrachtungen,

> [...] elles ne semblent pas s'accorder avec la manière dont la théorie quantique est aujourd'hui appliquée aux problèmes atomiques.[71]

Dieses Selbstbewußtsein, die Unbekümmertheit im Umgang mit der gerade erst akzeptierten modernen Physik, der mangelnde Respekt vor den Theorien der großen Physiker scheinen mir für de Broglie typisch zu sein.

Gerade die praxisorientierte Arbeit im Labor lenkte de Broglies Blick auf ein Arbeitsgebiet, das ihn sein Leben lang beschäftigen sollte. Mit den Experimenten zum Photoeffekt erneuerte Maurice de Broglie die Debatte um die Natur des Lichtes sowie der Röntgenstrahlung und folgerte aus den Ergebnissen, daß man die Teilchennatur nicht mehr übergehen könne.[72] Anfang der 20er Jahre war eine solche Interpretation, obwohl man das paradoxe Verhalten des Lichtes kannte, gerade auch in Frankreich noch nicht anerkannt.

## II.3 Die drei Artikel von 1923 und die Doktorarbeit

Stark geprägt durch den Einfluß des Bruders, fasziniert von den Arbeiten Einsteins und entschlossen, die wahre Natur der von Planck eingeführten Quanten zu verstehen, fing de Broglie Anfang der 20er Jahre an, sich eigene Forschungsziele zu setzen. Mit dieser Arbeit begann eine außerordentlich produktive und kreative Phase im Schaffen de Broglies. In den drei kurzen Artikeln vom September/ Oktober 1923 *Ondes et Quanta* [16], *Quanta de lumière, diffraction et interférences* [17] und *Les quanta, la théorie cinétique des gaz et le principe de Fermat* [18], veröffentlichte er erstmals die Ergebnisse, die die Physik revolutionieren und seine wissenschaftliche Anerkennung sichern sollten. Ein Jahr später, 1924, arbeitete er diese neuen Erkenntnisse aus und publizierte sie in seiner Doktorarbeit *Recherches sur la théorie des quanta* [150a]. Im nun folgenden sollen die wichtigsten Argumente und Ideen seiner Theorie dargestellt werden.[73]

Bereits vor dem Erscheinen der drei entscheidenden Artikel von 1923 hatte sich de Broglie 1922 in zwei Notizen in den *Comptes Rendus* [12, 13] als erster Physiker nach Einstein mit den Eigenschaften der postulierten Lichtquanten beschäftigt. Er ging darin noch um einiges über Einsteins Theorien hinaus: Die Lichtquanten sah er als echte

---

[71] Übersetzung: „[...] sie scheinen nicht mit der Art und Weise übereinzustimmen, in der heute die Quantentheorie auf atomare Probleme angewendet wird.", zitiert in „La découverte des ondes de matière", Actes de colloques, Académie des Sciences, Technique et documentation 1994, S. 43.

[72] Er betont dies etwa in einem Aufsatz von 1921: „Les phénomènes pour les rayons X et les spectres corpusculaires des éléments", Journal de Physique, 2 (1921), S. 265-287.

[73] Die Argumentationsgänge sollen hier in der von de Broglie verwendeten Terminologie zusammengefaßt werden. Da viele der von ihm verwendeten physikalischen Begriffe oft seinem ganz persönlichen Gebrauch entsprechen und nicht unbedingt zu gebräuchlichen Fachtermini geworden sind, werden sie in Anführungszeichen gesetzt. Ohne stets wörtlich zu zitieren, werde ich angeben, in welchem Artikel und an welcher Stelle der Doktorarbeit die kurz skizzierten Ideen und Theorien in ausführlicher Ausarbeitung zu finden sind.

Teilchen mit einem Spin und einer von Null verschiedenen Ruhemasse an.[74] 1923 griff er diese Ideen auf und erweiterte sie in folgender Weise:[75] De Broglie glaubte fest an die Energiebeziehungen Plancks und Einsteins: $E = h\nu_0$, bzw. $E = m_0c^2$. Deshalb postulierte er in Bezug auf die Lichtquanten die Gültigkeit einer Kombination der beiden Beziehungen: $h\nu_0 = m_0c^2$. Folglich konnte er bei einer vorgegebenen Frequenz auf die Ruhemasse des jeweiligen Photons schließen. De Broglie hatte nun den genialen Einfall, daß es genauso möglich sein sollte, von einer gegebenen Masse auf eine Frequenz zu schließen und erweiterte damit seine Annahme auf alle materiellen Teilchen, indem er ihnen eine Frequenz zuordnete. Diese im September 1923 zuerst veröffentlichte Hypothese wurde durch kein bis dahin bekanntes Experiment gestützt und war in ihren fast phantastisch anmutenden Postulaten nur aus dem Glauben an die Richtigkeit der Energiebeziehungen Plancks und Einsteins entstanden. Dies war die Geburtsstunde des Welle-Teilchen-Dualismus. In den folgenden Monaten gelang es de Broglie, seine Hypothese auf ein solides theoretisches Gerüst zu stellen.

Ein entscheidendes Problem stellte die relativistische Betrachtung der Beziehung $h\nu_0 = m_0c^2$ dar: Für einen relativ zum System ruhenden Beobachter wird die Energie des betrachteten Teichens um den Faktor

$$\left(1 - \beta^2\right)^{-1/2}$$

größer und die sich daraus ergebende Frequenz wäre:

$$\nu = \frac{1}{h}\frac{m_0c^2}{\sqrt{1-\beta^2}}.$$

Gleichzeitig verringert sich jedoch die „innere Frequenz" $\nu_0$, einer bewegten Uhr vergleichbar, für den ruhenden Betrachter um den Faktor

$$\sqrt{1-\beta^2},$$

so daß sich eine von der aus der Energiebeziehung abgeleiteten Frequenz $\nu$ verschiedene Frequenz ergeben würde:

$$\nu_1 = \nu_0\sqrt{1-\beta^2} = \frac{m_0c^2}{h}\sqrt{1-\beta^2}.$$

Diesen scheinbaren Widerspruch zweier unterschiedicher Frequenzen löste de Broglie, indem er die Frequenz $\nu$ nur auf die „periodische Bewegung" im Ruhesystem des Teilchens bezog, so daß dort gilt: $h\nu_0 = m_0c^2$. Diesem „periodischen Phänomen" ordnete de Broglie eine über den gesamten Raum ausgedehnte Schwingung $\nu_1$ zu, die ein ruhender Beobachter als eine Welle sieht, die sich in Richtung der Teilchenbewegung ausbreitet und deren Frequenz mit wachsender Geschwindigkeit zunimmt. Die

---

[74] De Broglies unorthodoxer Geist zeigt sich besonders schön am Beispiel seiner Lichttheorie: Er bezeichnete die erst seit 1926 als Photonen bezeichneten Lichtquanten als „Lichtatome" und gab ihre Masse mit $< 10^{-50}$ g an. Bis zu seinem Lebensende war er davon überzeugt, daß sie sich mit etwas geringerer Geschwindigkeit als $c$ bewegen und bezeichnete daher $c$ als „Grenzgeschwindigkeit des Lichtes" [150a, S.31]. Solche Betrachtungen zum Photon bilden auch den Inhalt des fünften Kapitels der Doktorarbeit [150a, S. 76ff].

[75] Die dargestellten Überlegungen bilden den Inhalt des ersten Artikels von 1923 [16] und des ersten Kapitels der Doktorarbeit [150a, S.31ff], (deutsche Übersetzung im Anhang A.1).

Ausbreitungsgeschwindigkeit der Welle, die de Broglie als „Phasengeschwindigkeit"
bezeichnete,

$$V = \frac{c}{\beta}$$

ist nicht nur größer als die des Teilchens, sondern größer als *c*. Das Verbindungsglied der
beiden Frequenzen stellt nach de Broglie die Phase dar: Für den ruhenden Beobachter
erscheinen ν und $ν_1$ immer in Phase. Diese wichtige Beziehung bezeichnete de Broglie
als das „Gesetz von der Phasengleichheit"[76].

Bemerkenswerterweise ist heute kaum noch bekannt, daß de Broglie auf dem eben
beschriebenen Weg, nämlich über die Relativitätstheorie, die Welleneigenschaften der
Materie und die nach ihm benannte Wellenlänge gefunden hat. Für ihn selbst war das
Auffinden der Phasenkonkordanz weitaus wichtiger als die Herleitung der Länge der
Materiewellen, und er bezeichnete später die Entdeckung des „Gesetzes von der
Phasengleichheit" als die größte Leistung seines Schaffens. Achtzigjährig sollte
erfeststellen:

> Si j'ai eu une grande idée c'est sans doute celle que j'ai exposée dans le premier
> chapitre de ma thèse.[77]

Mit diesem neuen theoretischen Gerüst gelang es de Broglie, die stabilen Bahnen des
Bohr-Sommerfeldschen Atommodells zu erklären: Er stellte sich vor, daß sich die
„interne Frequenz" eines Elektrons, das sich auf einem stabilen Orbit befindet, bei jeder
Umdrehung um ein ganzzahliges Vielfaches der Periode ändere, daß also solche Bahnen
stabil seien, die Resonanzen der „Phasenwelle" erlauben. Diese Betrachtungen finden
sich ebenfalls in dem Artikel vom 10. September [16] und bilden das dritte Kapitel der
Doktorarbeit [150a, S. 62ff].

Nur vierzehn Tage später veröffentlichte de Broglie den zweiten wichtigen Artikel [17],
in dem er die Eigenschaften der „Phasenwelle", wie sie hier erstmals genannt wurde,
noch weiter spezifizierte. Dies gelang de Broglie, indem er die Gesetze der
Elektrodynamik und der klassischen Mechanik auf die Ausbreitung der Welle und die
Bewegung des Teilchens anwendete. Ohne deren wirkliche physikalische Natur zu
erklären, unterschied er die „Phasengeschwindigkeit" der Welle von der
„Gruppengeschwindigkeit" und gab an, daß die Geschwindigkeit des Teilchens gleich der
„Gruppengeschwindigkeit" sein müsse. Über den Zusammenhang der Bewegung der
Welle und der des Teilchens stellte de Broglie folgendes Postulat auf:

> En chaque point de sa trajectoire, un mobile libre suit d'un mouvement uniforme le
> rayon de son onde de phase, c'est à dire (dans un milieu isotrope) la normale aux
> surfaces d'égale phase.[78]

---

[76] Die detaillierte Darstellung und Herleitung dieses Gesetzes findet man in [150a, S. 33ff],
(deutsche Übersetzung im Anhang A.1).

[77] Übersetzung: „Wenn ich in meinem Leben eine große Idee hatte, dann ist es sicherlich
diejenige, die ich im ersten Kapitel meiner Doktorarbeit dargestellt habe.", zitiert in [281, S. 580]. Da de
Broglie selbst den ersten Abschnitt seiner Dorktorarbeit so entscheidend fand, ist dieser im Anhang A.1
dieser Arbeit in deutscher Übersetzung wiedergegeben.

[78] Übersetzung: „In jedem Punkt seiner Bahn folgt ein freies Teilchen mit einer gleichförmigen
Bewegung dem Strahl seiner Phasenwelle, d.h. (in einem isotropen Medium) der Normalen auf den
Flächen gleicher Phase.", [17, S. 548].

Somit folgt das freie Teilchen einer der Bahnen, die sich durch die Anwendung des Fermatschen Prinzips auf die „Phasenwelle" ergeben. Die „Phasenwelle" bestimmt also die Bewegung des Teilchens, führt es gewissermaßen.

> Nous concevons donc l'onde de phase comme guidant les déplacements de l'énergie, et c'est ce qui peut permettre la synthèse des ondulations et des quanta [...].[79]

So entsprechen die Bahnen, die sich aus der Annahme einer „Phasenwelle" ergeben, denen, die man für die Bewegung eines materiellen Teilchens aus dem Maupertuischen Prinzip erhalten würde. Im fünften Abschnitt des zweiten Kapitels[80] schreibt de Broglie:

> Le principe de Fermat appliqué à l'onde de phase est identique au principe de Maupertuis appliqué au mobile; les trajectoires dynamiques possibles du mobile sont identiques aux rayons possibles de l'onde.[81]

Mit diesem Konzept konnte de Broglie in seinem zweiten Artikel [17] Interferenz- und Beugungserscheinungen von Licht, und im zweiten Kapitel seiner Doktorarbeit [150a, S. 44ff] die nicht-lineare Bewegung beschleunigter Teilchen auf Eigenschaften der „Phasenwelle" zurückführen.

De Broglie gelang es also, eine Analogie zwischen der Optik und der Mechanik festzustellen, so daß nach Jahrhunderten der Kontroverse nicht nur eine Antwort auf die Frage nach der wahren Natur des Lichtes gegeben werden konnte, sondern de Broglie auch das Verständnis von der Materie revolutionierte. Und obwohl die experimentelle Bestätigung der Materiewellen erst einige Jahre später erfolgen sollte, schlug de Broglie schon 1923 in seinem zweiten Artikel vor:

> De plus, un mobile quelconque pourrait dans certains cas diffracter. Un flot d'électrons traversant une ouverture assez petite présenterait des phénomènes de diffraction. C'est de ce côté qu'il faudrait chercher des confirmations de nos idées.[82]

Interessanterweise taucht dieser Vorschlag, den de Broglie erst Ende 1924 vor der Prüfungskommission wiederholte, nicht in der Doktorarbeit selbst auf.

Neben diesen fundamentalen Erkenntnissen, die die Physik einschneidend verändert haben, finden sich in den Arbeiten weitere richtungsweisende Ansätze, die hier erwähnt werden sollen. Bereits im zweiten Artikel [17, S. 549f] und im Kapitel V der Doktorarbeit [150a, S. 88] thematisierte Broglie erstmals das Problem der kohärenten Strahlung. In dem dritten Artikel [18, S. 630ff] und im siebten Kapitel [150a, S. 104ff] der Dissertation zieht de Broglie dann Betrachtungen zur statistischen Mechanik. De Broglie postulierte, daß sich ein Gas in einem abgeschlossenen Raum im Gleichgewicht befindet, wenn die „Phasenwellen", die die Atome begleiten, ein stationäres System bilden. Er verwendete also die gleiche Annahme, die ihn zu den stabilen Bahnen des Bohr-

---

[79] Übersetzung: „Wir stellen uns die Phasenwellen als die Ortsveränderungen der Energie leitend vor, und dies ist es, was es uns erlaubt, eine Synthese herzustellen zwischen Wellenbewegungen und Quanten.", [17]. Anmerkung: Für de Broglie bezeichneten die Begriffe „Masse" und „Energie" dieselbe physikalische Realität [150a, S.32], er benutze sie demnach synonym.

[80] Eine vollständige Übersetzung dieses wichtigen Abschnittes findet sich ebenfalls im Anhang A.1.

[81] Übersetzung: „Das Prinzip von Fermat angewendet auf die Phasenwelle ist identisch mit dem Prinzip von Maupertuis angewendet auf einen Massenpunkt; die dynamisch möglichen Teilchenbahnen sind mit den möglichen Wellenstrahlen identisch.", [150a, S. 56].

[82] Übersetzung: „Außerdem könnte man ein beliebiges Teilchen in gewissen Fällen beugen. Ein Strom von Elektronen, der durch eine Öffnung geht, die genügend klein ist, würde Beugungserscheinungen zeigen. In dieser Richtung müßte man wohl Bestätigungen unserer Ideen suchen.", [17, S. 549].

Sommerfeld-Atommodells geführt hatte, und gab in diesem Zusammenhang auch erstmals die berühmte Formel an

$$\lambda = \frac{h}{p}.$$

Die Forderung nach stationären Wellen führte zwangsweise zu diskreten Wellenlängen, woraus de Broglie auf diskrete Geschwindigkeiten der Atome schloß - eine schon länger bekannte Hypothese,[83] die bislang noch nicht erklärbar gewesen war. Er machte aber auch darauf aufmerksam, daß der Mechanismus, der diese Prozesse regelt, noch nicht - und das bis heute nicht - erklärbar sei. Mit seinen Betrachtungen konnte de Broglie zwar die statistischen Postulate Plancks, Nernsts und Maxwells bestätigen, die allgemeine Formulierung gelang jedoch erst ein Jahr später mit der Bose-Einstein-Statistik.

Es ist bemerkenswert, daß die drei kurzen Artikel von 1923 bereits alle wirklich neuen und kreativen Ideen de Broglies enthalten. Sie sind in ihrer Dichte, in ihrem Ideenreichtum und in ihrer Neuartigkeit so beeindruckend, daß sie unter die größten wissenschaftlichen Leistungen dieses Jahrhunderts gezählt werden müssen. Die berühmt gewordene Doktorarbeit stellt dagegen nur die gründliche Darlegung und Ausformulierung der in den Artikeln angerissenen Thesen dar. Die wenigen Wochen, in denen diese Artikel entstanden, waren sicherlich der produktivste Abschnitt im Leben de Broglies; so beschreibt er selbst den Prozeß der Erkenntnis als etwas Einmaliges, das ihn überkam, und das, fast einer göttlichen Eingebung vergleichbar, im Leben eines noch so genialen Physikers nur einmal eintreten dürfte, und er schrieb später über diesen entscheidenden Moment:

> Une grande lumière se fit alors soudain dans mon esprit.[84]

> Brusquement, à la fin de l'été 1923, toutes ces idées se mirent à cristalliser dans mon esprit.[85]

Zugespitzt ausgedrückt könnte man sagen, daß sich de Broglie sein gesamtes Leben lang damit beschäftigt hat, diese in kurzer Zeit entworfenen Ideen zu erklären, auszuarbeiten und darzustellen. Denn auch die Grundideen seiner späteren, weniger erfolgreichen Theorien, wie die „doppelte Lösung", die Photonentheorie und die „verborgene Thermodynamik"[86], sind in den Artikeln bereits angedeutet. Angesichts der immensen Bedeutung, die die Entdeckung der Materiewelle für die moderne Physik hatte und der immer noch damit verbundenen Fragen, kann man die kreativen Leistungen de Broglies im Jahr 1923 gar nicht überbewerten.

Doch Ende 1924 war man noch eher skeptisch. Die Prüfungskommission erkannte zwar die Kreativität de Broglies, so daß Jean Perrin, von Maurice de Broglie nach seiner Meinung befragt, antwortete:

> Tout ce que je puis dire c'est que votre frère est bien intelligent.[87]

---

Dieser vorsichtige, noch recht wenig schmeichelhafte Kommentar, zeigt die relative Hilflosigkeit der Kommission vor der Neuheit der Ideen. Niemand glaubte ernsthaft an die reale Existenz der Materiewellen. Eines der Kommissionsmitglieder, Charles Mauguin, erinnert sich später:

> Pourtant je dois faire un aveu. Je n'ai pas cru, lors de la soutenance de la thèse, à la réalité des ondes associées aux grains de matière.[88]

Auch im Umfeld Langevins erkannte man die Bedeutung der Arbeit, stand ihren Ergebnissen jedoch eher skeptisch gegenüber. Léon Briouillin sagte 1963 in einem Interview mit T. S. Kuhn:

> We in the group were immensely interested, but rather sceptical [...]. Altogether, Langevin was extremely interested, but he was never really convinced of the existence of the waves. [291, S. 27].

Trotzdem bescheinigt Langevin in seinem Bericht zur Dissertation:

> M. de Broglie a poursuivi ici, avec une cohérence et une maîtrise remarquables, un effort qui devait être tenté pour vaincre les difficultés au milieu desquelles nous sommes. L'originalité et la profondeur des idées, la coordination remarquable qu'elles permettent, justifient largement l'acceptation de son travail comme thèse de Doctorat ès Sciences.[89]

Mit der ausgezeichneten Beurteilung („très honorable") bewiesen Langevin und Perrin jedoch trotz ihrer Skepsis Weitblick.

## II.4 Weiterentwicklung der Ideen und deren Scheitern

Wenn de Broglie mit seinen Arbeiten die Physik seiner Zeit nachhaltig veränderte, so suchte er damit jedoch keinesfalls den Bruch mit den alten, bewährten Konzepten der Elektrodynamik, der klassischen Mechanik, der Relativitätstheorie und der Planckschen Quantentheorie; vielmehr führte ihn deren konsequente Anwendung, Verbindung und Ausweitung zum Erfolg. Er selbst sprach nur von einer Synthese und ordnete seine Erkenntnisse in die früheren Theorien ein:

> La nouvelle dynamique du point matériel libre est à l'ancienne dynamique (y compris celle d'Einstein) ce que l'optique ondulatoire est à l'optique géométrique. En y réfléchissant on verra que la synthèse proposée paraît être le couronnement logique du développement comparé de la dynamique et de l'optique depuis le XVIIᵉ siècle.[90]

---

[88] Übersetzung: „Dennoch muß ich ein Geständnis ablegen: Während der Verteidigung der Doktorarbeit habe ich nicht an die Realität der mit den Materieteilchen verbundenen Wellen geglaubt.", [291, S. 27].

[89] Übersetzung: „M. de Broglie hat hier mit bemerkenswerter Geschlossenheit und Meisterschaft eine Anstrengung unternommen, die versucht werden mußte, um die Probleme zu lösen, in deren Mitte wir uns befinden. Die Originalität und die Tiefe der Ideen, die erstaunliche Zusammenschau, die sie erlauben, rechtfertigen es vollauf, seine Arbeit als *Thèse de Doctorat ès Sciences* zu akzeptieren." Diese Beurteilung befindet sich im Original in den *Archives de l'Académie des Schiences* und ist in in [287, S. xxxviii] abgedruckt.

[90] Übersetzung: „Die neue Dynamik des freien materiellen Punktes verhält sich zur alten Dynamik (die Einsteins mit eingeschlossen), wie die Wellenoptik zur geometrischen Optik. Wenn man darüber nachdenkt, so wird man erkennen, daß die vorgeschlagene Synthese die logische Krönung der mit einander verglichenen Entwicklungen der Dynamik und der Optik seit dem 17. Jahrhundert ist.", [17, S. 549].

Die Suche nach Analogien und Symmetrien, der Wille, die bewährten Theorien zu erhalten, zu erweitern und zusammenzuführen sowie die feste Verbundenheit mit den tradierten Konzepten charakterisieren in besonderer Weise de Broglies Arbeit und die Ausrichtung seiner Forschung. Er glaubte an eine langsame Annäherung der Wissenschaft an die zu beschreibende Wirklichkeit, an große Synthesen, die das bisher Bekannte verbinden, nicht aber an die Notwendigkeit großer Brüche und Revolutionen, die alles bisher Dagewesene umstürzen. Über den Rhythmus wissenschaftlichen Fortschritts schrieb er 1949:

> Par moments, grâce à un vigoureux effort intellectuel, les savants parviennent à donner une vue d'ensemble cohérente de presque tous les faits connus dans le domaine dont ils s'occupent [...] une grande synthèse est réalisée, une ère d'unification de nos connaissances semble s'ouvrir.[91]

Seit seiner frühen Kindheit hatte sich de Broglie mit Geschichtlichem beschäftigt, auch in der Ausrichtung seiner wissenschaftlichen Arbeit verlor er die Vergangenheit nicht aus dem Blick. Tief verbunden mit dem klassischen Weltbild, glaubte er an eine physikalische Realität, die unabhängig vom Menschen und seinen unvollkommenen Wahrnehmungs- und Beschreibungsmöglichkeiten existiert. Für ihn mußte eine Theorie physikalischer Phänomene auf klaren Konzeptionen und präzisen Bildern in Zeit und Raum basieren. Er selbst schrieb später über die Orientierung seiner Forschung, sie bestünde darin,

> [...] à envisager les problèmes plutôt sous forme des images physiques intuitives que sous celle des formalismes mathématiques, [...].[92]

Dies unterschied ihn jedoch wesentlich von der Physikergeneration, die im Aufstreben begriffen war: In den folgenden Jahren sollten sich ja gerade die mathematisch begabten Wissenschaftler mit ihren eher formalen und abstrakten Theorien durchsetzen.
Werner Heisenberg (1901-1976) war einer dieser jungen aufstrebenden Physiker. Im Jahr 1925 entwarf er seine Matrizenmechanik, die beachtliche Erfolge feierte: Heisenberg errechnete molekulare Schwingungen des Wasserstoffs und des Sauerstoffs und fand dabei neue, mit dem Experiment übereinstimmende Ergebnisse; Wolfgang Pauli (1900-1958) gelang die Berechnung des Wasserstoffatoms, und gemeinsam mit Max Born (1882-1970) und Pascual Jordan (1902-1980) legte Heisenberg mit den Gleichungen, denen die Matrizen gehorchen sollten, die Basis für eine echte Mechanik im atomaren Bereich. Vorrangiges Ziel dieser neuen Quantenmechanik war es, die experimentellen Ergebnisse in einem mathematischen Formalismus zu beschreiben. Im Vordergrund stand dabei nicht, Anschaulichkeit oder physikalische Bilder zu liefern, wichtig war nur, daß die z.T. sehr komplizierten Rechnungen korrekte Ergebnisse lieferten. - Eine Entwicklung, die nicht nur de Broglie nicht gefiel: Einstein bezeichnete den Formalismus als echte Hexenrechnerei, die durch ihre große Komplexität ausreichend geschützt sei vor jedem Beweis ihrer Falschheit [283, S. 119].
Einer der wenigen, die neben Einstein die Ideen der Doktorarbeit de Broglies sehr ernst nahmen, war Erwin Schrödinger (1887-1961). Im Jahr 1926 veröffentlichte er eine Serie von Artikeln, in denen er die Idee der Materiewelle mathematisch konsequent zu einer kompletten Wellentheorie, der Wellenmechanik, ausarbeitete. Die Erfolge dieses neuen Formalismus waren überwältigend: In mathematisch übersichtlicher Form konnte

---

[91] Übersetzung: „Mitunter gelingt es den Wissenschaftlern, dank einer kraftvollen intellektuellen Anstrengung einen kohärenten Überblick fast aller bekannten Fakten in dem Bereich, mit dem sie sich beschäftigen, zu geben [...] eine große Synthese ist dann geschaffen und eine Ära der Vereinheitlichung unserer Kenntnisse scheint sich zu öffnen.", [188, S. 362f].

[92] Übersetzung: „[...] die Probleme eher über intuitive physikalische Bilder als über mathematische Formalismen anzugehen [...].", [189, S. 180].

Schrödinger mit großer Genauigkeit die unterschiedlichsten Probleme berechnen und fand dieselben Ergebnisse wie Heisenberg. Allerdings zog die Theorie Schrödingers Folgerungen und Einschränkungen nach sich, die de Broglie nicht vorgesehen hatte: Die Wellenmechanik war nicht-relativistisch, bezog sich ausschließlich auf Felder, gab also das Teilchen auf und fand eine elegante Beschreibung im Konfigurationsraum. Der Verlust an Anschaulichkeit entsprach aber ganz und gar nicht de Broglies (und auch nicht Schrödingers) Bedürfnis nach klaren physikalischen Bildern in Zeit und Raum. Nach solchen suchte er jedoch und orientierte seine Forschung in den Jahre 1925-1927 in diese Richtung: Er wollte die wahre physikalische Natur der Materiewellen erkennen und darstellen. Hatte er bereits in dem zweiten Artikel von 1923 davon gesprochen, daß die „Phasenwellen" die „Ortsveränderung der Energie" führe[93], suchte er nun in dieser Richtung eine Lösung des Problems. In einem kurzen Artikel von 1924 schrieb de Broglie:

> Cette propriété permet de considérer le point matériel comme une singularité du groupe d'ondes. [...]. Les rayons prévus par les théories ondulatoires seraient donc dans tous les cas les trajectoires possibles du quantum.[94]

Zum ersten Mal taucht hier die Idee auf, das Teilchen könne als eine Singularität der Welle identifiziert werden. Diese Idee weitete de Broglie zu seiner Theorie der „doppelten Lösung" aus, die er 1927 erstmals veröffentlichte [34]. Die Grundidee dieser Theorie ist, daß es neben der regulären Lösung der Schrödingergleichung Ψ noch eine weitere Lösung $u$ geben müsse, die dort eine Singularität besitzt, wo sich das Teilchen befindet. Dann würde die Schrödingerwelle nicht nur die ihr durch Born zugeordnete statistische Bedeutung haben, sondern auch alle theoretisch denkbaren Bahnen des Teilchens repräsentieren,[95] während die Lösung $u$ die individuelle Bewegung eines Teilchens darstellen würde. De Broglie postulierte, daß beide Wellen in Phase sein müßten und daß sich die Singularität in der Richtung ausbreite, in der die Phase am schnellsten anwachse. Diese Beziehung nannte er „Gesetz der Führung". Für die Geschwindigkeit des „geführten" Teilchens im nicht-relativistischen Bereich gab de Broglie die sogenannte „Führungsformel" an:

$$\mathbf{v} = -1/m \ \mathbf{grad} \ \varphi.$$

Worin φ die reelle Phase der Welle $u$ und $m$ die Masse des Teilchens ist. Um tatsächlich die Dynamik des Teilchens beschreiben zu können, führte de Broglie ein neues sogenanntes „Quantenpotential $q$" ein:

$$q = q_0 = -\frac{\hbar^2}{2m_0} \frac{\Delta a}{a},$$

wobei $a$ die von Ort und Zeit abhängige Amplitude der „führenden" Welle $u$ ist. Dieses Potential beschreibt den Einfluß der ausgedehnten Wellen auf die Bahn der „Region hoher Konzentration". Denn die Bewegung des „geführten" Teilchens ist dem Einfluß aller Hindernisse ausgesetzt, die die freie Ausbreitung der Welle behindern, also auch solchen Kräften, die klassisch betrachtet keinen Einfluß haben könnten. So konnte de Broglie anhand des „Quantenpotentials" auch Interferenz und Beugungserscheinungen von Materieteilchen erklären.

---

[93] Vergleiche Fußnote 79 des vorhergehenden Kapitels.

[94] Übersetzung: „Diese Eigenschaft erlaubt es, den materiellen Punkt als eine Singularität der Wellengruppe anzusehen. [...]. Die durch die Wellentheorien vorgesehenen Strahlen wären also in allen Fällen die möglichen Bahnen des Quants." [24].

[95] De Broglie spricht davon, das Teilchen folge einer der „lignes de courant envisagées par l'image hydrodynamiques de l'onde Ψ.", (Übersetzung: „Stromlinien, die durch das hydrodynamische Bild der Welle Ψ ins Auge gefaßt werden."), [189, S. 149].

Die Reaktionen auf den Artikel waren eher verhalten, man erkannte die Originalität der Idee an und konnte ihr keine schlagenden Argumente entgegensetzen. Wolfgang Pauli schrieb am 6. August 1927 an Niels Bohr:

> Aber immerhin, selbst wenn diese de Brogliesche Arbeit ein Fehlschlag ist (und das hoffe ich eigentlich), so ist sie doch sehr ideenreich und witzig und auf einem viel höheren Niveau wie die kindischen Arbeiten von Schrödinger, der heute noch glaubt, er könne der statistischen Deutung seiner Funktion entgehen [...]. [299a, S. 404].

Für Pauli stellte die „doppelte Lösung" anscheinend eine ernstzunehmende Konkurrenz für die sich andeutende Kopenhagener Deutung dar.

> [...] denn diese Arbeit gehört in die Kategorie der „interessanten und geistreichen Versuche" (wie Sie immer zu sagen pflegen, wenn Sie etwas nicht gerne haben, aber doch loben wollen). [299a].

Es ist bemerkenswert, daß Einstein im selben Jahr im Zusammenhang mit seiner allgemeinen Relativitätstheorie eine ähnliche Theorie aufgestellt hat: Ein materielles Teilchen sei eine Singularität im Gravitationsfeld und folge darin dem kürzesten Weg.

Im Oktober 1927 fand in Brüssel der fünfte Solvay-Kongreß statt: Alle Größen der modernen Physik fanden sich hier ein. Dieses Treffen kann als die Geburtsstunde der Quantenmechanik und als Triumph der Kopenhagener Schule angesehen werden. Bisher standen die unterschiedlichen Formalismen und Interpretationen unverbunden nebeneinander, alles deutete jedoch auf eine Synthese im Sinne der Kopenhagener Deutung hin: Mit den Theorien Schrödingers, Heisenbergs und Diracs waren drei komplexe Formalismen geschaffen, die die experimentellen Ergebnisse übereinstimmend und mit großer Genauigkeit wiedergaben. 1926 hatten Schrödinger und Pauli erkannt, daß die verschiedenen Formalismen nichts als unterschiedliche Darstellungen derselben Physik waren. Mit der Wahrscheinlichkeitsinterpretation, dem Komplementaritätsprinzip und der Unschärferelation standen physikalische Interpretationen zur Verfügung, es fehlte letztlich nur an der Synthese und an der Durchsetzung des gesamten Konzeptes.

Einige Physiker der älteren Generation wandten sich gegen den Verlust an Objektivität, der sich ankündigte: Planck, Lorentz, Einstein, Schrödinger u.a. De Broglie versuchte jedoch als einziger, ein konkurrierendes Konzept vorzustellen: In seinem Vortrag bot er die „doppelte Lösung" als eine mögliche Alternative an. Er hatte jedoch keine ausgereifte Theorie vorzuweisen: Seine mathematischen Fähigkeiten entsprachen nicht denen der neuen Physikergeneration und die von ihm anvisierte Theorie der „doppelten Lösung" wies zahlreiche formale Schwierigkeiten auf. So entschied er sich, nur eine abgeschwächte Form vorzustellen, höchstens in innovativen Bildern sprechend Wege aufzuzeigen, wie bisher unter der Oberfläche verborgene Parameter der statistischen Beschreibung der Schrödingerwelle darzustellen seien. Er stellte eine durch die vorgenommenen Vereinfachungen physikalisch wenig schlüssige Theorie vor, die er später als „Bastard" bezeichnete, geboren aus der Anpassung an die gängige Physik. De Broglie verzichtete auf die „innere Frequenz" des Teilchens, das „Gesetz der Phasengleichheit" und die Singularitäten. Übrig blieb nur die reguläre Schrödingerwelle, die das Teilchen nach diesem Konzept auf seiner Bahn leitet. - Mit dem „Führungswellenkonzept" machte er es seinen Kritikern leicht: Natürlich konnte eine Theorie, nach der eine Wahrscheinlichkeitswelle ein physikalisches Teilchen führt, nicht bestehen vor den größten Physikern des Jahrhunderts. Pauli antwortete ewas herzlos auf de Broglies Vortrag:

Ich habe Ihren Artikel im *Journal de France* gelesen, er ist sehr interessant, aber falsch.[96]

Nach de Broglies eigener Einschätzung [288, S. 343] wurde der wenig überzeugende Eindruck seines abgesehen von einigen kleinen Höhepunkten inhaltlich eher mittelmäßigen Vortrags, sicherlich noch verstärkt durch sein unsicheres, an öffentliche Vorträge nicht gewöhntes Auftreten.

Es bleibt jedoch festzuhalten, daß der Mißerfolg des unausgereiften „Führungswellenkonzeptes" nicht allein das Ergebnis wissenschaftlicher Auseinandersetzung war. Der Vortrag de Broglies traf in keiner Weise den Zeitgeist. In seinem Bedürfnis nach Anschaulichkeit stand de Broglie ziemlich isoliert einer wortführenden Gruppe junger, brillanter Physiker um Bohr gegenüber, die hochmotiviert war, die Leistungsfähigkeit ihrer Theorien auszutesten. Zum ersten Mal wurden die Ereignisse im atomaren Bereich faßbar, und es erschien realistisch, die Möglichkeiten praktischer Anwendung auszuloten. Zu diesem Zeitpunkt, da sich mit der Quantenmechanik in ihrer Kopenhagener Deutung eine großartige Theorie ankündigte, die alle Fragen zu beantworten schien, interessierte sich, nach Jahren des Suchens, niemand ernsthaft für eine weitere Theorie, die vielleicht anschaulicher, aber wenig ausgearbeitet war. Es herrschte so etwas wie Aufbruchstimmung, in der die Einwände und Ideen de Broglies nur als störende Bremse eines altmodischen klassischen Physikers angesehen wurden. Nur Einstein versuchte die Euphorie seiner jungen Kollegen zu bremsen, indem auch er die Wichtigkeit einer Lokalisierung des Teilchens unterstrich und hervorhob, daß de Broglie auf dem richtigen Weg sei (siehe Kapitel II.5).

Diese Kontroverse zwischen der Kopenhagener Schule und deren Gegnern war jedoch nicht nur eine Auseinandersetzung zwischen Fortschritt und Stillstand, sie wurde in Teilen wie ein Glaubenskrieg zwischen klassischem und indeterministischem Weltbild geführt.[97] Vor allem erstaunt die weit über die Argumente der Physik hinausgehende, fast aggressive sprachliche und rhetorische Form der Auseinandersetzung. Es wäre sicherlich eine lohnende Arbeit, die Vorträge der Vertreter der Kopenhagener Schule, namentlich die Reden Heisenbergs und Bohrs,[98] einmal genau unter diesem Gesichtspunkt zu analysieren: So könnte man leicht nachweisen, wie siegesgewiß und selbstbewußt sie auftraten, und wie sie dies in einer autoritären, machtergreifenden Sprache transportierten; Gegenargumente wurden durch sprachliche Konventionen und Postulate abgeschmettert und die Theorien Schrödingers und de Broglies respektlos verwässert, das physikalisch Brauchbare an ihnen jedoch einfach annektiert.[99]

Tatsächlich hatte es die Vertreter der Kopenhagener Schule leicht: Mit der Wahrscheinlichkeitsinterpretation, dem Komplementaritätsprinzip und der Unschärferelation gaben sie den mathematischen Formalismen eine physikalische, philosophische und erkenntnistheoretisch neue Motivation und postulierten die Vollständigkeit dieser neuen Quantenmechanik. Die Richtigkeit der Ergebnisse, die Anwendbarkeit der Theorie und die autoritäre, keinen Widerspruch duldende Argumentation Bohrs ließen die Gegner dieser Interpretation verstummen.

Gerade im Leben de Broglies bedeutete der Kongreß einen tiefen Einschnitt, setzte dieser doch seiner kreativen Schaffensphase ein Ende. Tief getroffen durch den erlittenen Mißerfolg kehrte er im Wissen um sein eigenes Versagen nach Paris zurück, wo er, entmutigt durch die unüberwindbaren Schwierigkeiten seiner „doppelten Lösung", für einige Jahre seine persönlichen Forschungsprojekte aufgab.

---

[96] Erinnerung de Broglies in französisch zitiert in [291, S. 40].

[97] Zu diesem Themenkomplex sei verwiesen auf [305].

[98] Veröffentlicht sind sie in dem Band [36a].

[99] Zu diesem Themenkomplex, der leider in diesem Rahmen nicht ausführlich besprochen werden kann, sei z.B. auf folgende Werke verwiesen: [283 (S. 134ff), 299, 312].

## II.5 De Broglie und Einstein

Obwohl de Broglie Einstein nur ein einziges Mal traf, nämlich 1927 während des fünften Solvay-Kongresses, beeinflußte dieser de Broglies Werk entscheidend. Da der Einfluß keines anderen Zeitgenossen so groß war, scheint es angemessen, im folgenden dieses Verhältnis etwas genauer zu beleuchten. Eine wertvolle Quelle dafür ist die Korrespondenz der beiden großen Physiker, die in den *Archives de l'Académie des sciences* aufbewahrt wird und die sich als Photokopien im Anhang dieser Arbeit findet. Es handelt sich dabei um acht noch unveröffentlichte Briefe, die in der Zeit von 1929 bis 1954 geschrieben wurden.

Bereits 1911, als de Broglie begann, sich für physikalische Probleme zu interessieren, waren es die Ideen Einsteins, auf die er sein besonderes Augenmerk richtete. In der Zusammenfassung der Ergebnisse des ersten Solvay-Kongresses nahm de Broglie zum ersten Mal von den Vorstellungen Einsteins Kenntnis, die zu dieser Zeit noch nicht voll anerkannt waren. Von den revolutionären Thesen begeistert, begann de Broglie, die Relativitätstheorie sowie Einsteins Theorie des Lichtes zu studieren. Der dauerhafte Einfluß dieser Lektüren auf de Broglies weitere Forschung ist bekannt: Denn die Entdeckung der Materiewelle fußt natürlich auf der Verallgemeinerung der Lichtquantentheorie. De Broglie hat diese Tatsache immer wieder betont[100] und damit seine Dankbarkeit für den Mann zum Ausdruck gebracht, den er sein Leben lang als sein Vorbild und einen der größten Denker unserer Zeit verehrte. In einem Brief vom 14. November 1929, in dem sich de Broglie für Einsteins Glückwünsche zur Verleihung des Nobelpreises bedankt, schreibt er:

> Ces félicitations sont de toutes celles que j'ai reçues celles qui m'ont le plus touché à cause de la grande admiration que j'ai pour vous depuis longtemps.[101]

Über die reine Bewunderung hinaus war de Broglie Einstein dankbar für dessen positive Einflußnahme zu Gunsten der Idee der Materiewelle. Einstein hatte nämlich in dem berühmten Brief vom 16. Dezember 1924 an Langevin über die Doktorarbeit de Broglies geschrieben:

> Die Arbeit von de Broglie hat großen Eindruck auf mich gemacht. Er hat einen Zipfel des großen Vorhangs gelüftet.[102]

Und in einem Brief an Lorentz vom 25. November 1924 schrieb Einstein:

> Ein jüngerer Bruder des uns bekannten De Broglie hat einen interessanten Versuch zur Deutung der Bohr-Sommerfeld Quantenregel unternommen (Pariser Dissertation 1924). Ich glaube, das ist der erste Strahl zur Erhellung dieses schlimmsten unserer physikalischen Rätsel. [283, S. 109]

Im Januar 1925 verwies Einstein dann in einem Artikel zur Bose-Einstein-Statistik auf die Dissertation de Broglies[103] und lenkte damit zum ersten Mal den Blick einer breiten Öffentlichkeit auf die Arbeit de Broglies. Schrödinger schrieb darüber:

---

[100] Bereits in [16-18, 150] beruft er sich ausdrücklich auf Einsteins Vorleistungen.
[101] Übersetzung: „Diese Glückwünsche waren von allen, die ich erhielt, diejenigen, die mich, aufgrund der großen Bewunderung, die ich für Sie seit langem hege, am meisten berührt haben.", Anhang A.2.1.
[102] Zitiert z.B. in [283, S. 109].

Übrigens wäre die ganze Sache sicherlich nicht jetzt und vielleicht nie entstanden (ich meine, nicht von meiner Seite), wenn mir nicht durch Ihre zweite Gasentartungsarbeit auf die Wichtigkeit der de Broglie'schen Idee die Nase gestoßen wäre.[104]

Sicherlich hat die Unterstützung des Nobelpreisträgers von 1921 wesentlich zur Beachtung und Weiterentwicklung der Ideen de Broglies beigetragen, wofür ihm de Broglie immer wieder dankte. In einem Brief vom 21. Juni 1951 schrieb de Broglie zum Beispiel:

En dehors de la profonde admiration que je ressens, comme tous les physiciens, pour votre oeuvre qui a renouvelé toute la science de notre temps, j'ai conservé pour vous une grande reconnaissance personnelle pour la grande bienveillance dont vous avez fait à mon égard au début de ma carrière scientifique [...].[105]

Einstein war aber gerade in späteren Jahren nicht nur Lehrer und bewunderter Meister de Broglies, es fand auch ein kollegialer Austausch statt, in dem immer wieder deutlich wurde, wie ähnlich sich die Anschauungen und die Ausrichtung der Forschung der beiden Physiker waren. Eine interessante Frage, die im folgenden diskutiert werden soll, kann die enge Verbindung der Arbeiten und das physikalische Weltbild de Broglies und Einsteins verdeutlichen: Betrachtet man das geniale Werk Albert Einsteins, so fällt auf, daß der Schritt von der Photonenidee zur Materiewelle ein für Einsteins Verhältnisse kleiner Schritt war, man mag sich darüber wundern, daß nicht er selbst diesen Schritt gegangen ist. Und tatsächlich gibt es Hinweise darauf, daß Einstein schon vor der Anerkennung der Dissertation de Broglies die Idee der Materiewelle gehabt hat: Pauli erinnerte sich, daß Einstein ihm im September 1924, also noch zwei Monate vor der Verleihung der Doktorwürde an de Broglie, Experimente vorgeschlagen hatte, die Interferenz- und Beugungsphänomen mit Molekularstrahlen nachweisen sollten [300, S. 47ff]. Um herauszufinden, ob Einstein zu diesem Zeitpunkt von der Arbeit de Broglies gewußt hat oder ob er unabhängig davon bereits ähnliche Theorien entwickelt hatte, muß man folgende Indizien betrachten:

Einen Hinweis dafür, daß Einstein die Arbeit schon im Frühjahr 1924 zugesandt bekommen haben könnte, liefert uns de Broglie selbst: Er war sein Leben lang davon überzeugt, daß erst Einsteins positive Reaktion auf die Dissertation die Prüfungskommission dazu bewogen habe, die Arbeit anzuerkennen. A. Pais hatte 1978 einen kurzen Briefwechsel mit Louis de Broglie, in dem der damals schon 86jährige seiner Überzeugung Ausdruck verlieh, Einstein habe bereits zu Beginn des Jahres 1924 ein Exemplar der Arbeit erhalten.[106] Etwas verwirrend ist jedoch, daß de Broglie als Beleg für die Einflußnahme Einsteins immer wieder den weiter oben erwähnten Brief vom 16. Dezember 1924 an Langevin zitiert.[107] Dieser ist aber erst einen Monat nach der Anerkennung der Doktorarbeit geschrieben worden und wäre damit zu spät gekommen, um noch irgendeinen Einfluß auf die Prüfer ausüben zu können. Zudem kündigt Einstein in dem Brief an Langevin an, im Kolloquium über die Ideen de Broglies referieren zu wollen. Bedenkt man wie aktuell die von de Broglie aufgestellten Thesen waren, fragt man sich, warum Einstein nicht schon früher darüber berichtet hat, wenn er die Arbeit schon im Sommer gelesen hätte. G. Lochak weist außerdem darauf hin, daß der spontane, begeisterte Tonfall, in dem Einstein sowohl in dem Brief an Langevin als auch in dem an Lorentz (25. November) über die Dissertation de Broglies urteilt, eher dafür spricht, daß Einstein die Arbeit erst Ende 1924 intensiv gelesen hat [283, S. 109].[108] Und, um die Verwirrung auf den Höhepunkt zu treiben, finden wir 1956 eine Aussage de Broglies, die genau dies bestätigt:

> Or, vers la fin de 1924, Einstein avait eu connaissance de ma thèse.[109]

Die widersprüchlichen Aussagen de Broglies legen die Annahme nahe, er habe selbst einen Einfluß Einsteins auf die Prüfungskommission nur vermutet und - da dies seinen eigenen Wünschen entsprach - im Nachhinein plausibel zu machen versucht. Einen schriftlichen Beleg hierfür gibt es jedenfalls nicht.

Sollte Einstein die Arbeit jedoch tatsächlich bereits im Frühjahr zugesandt bekommen haben, sprechen einige Argumente dafür, daß er sie zu diesem Zeitpunkt nicht wirklich aufmerksam gelesen hat: Im Jahr 1924 beschäftigte sich Einstein mit der Entwicklung der Bose-Einstein-Statistik. Er hatte einen Aufsatz von Bose übersetzt und veröffentlicht und kurze Zeit später, im Sommer 1924, eigene Überlegungen dazu publiziert, die in ihrer allgemeinen Form die Bose-Einstein Statistik begründeten; diese Überlegungen weitete Einstein zu dem bereits oben erwähnten Artikel aus, der im Januar 1925 veröffentlicht wurde. Während in der ersten Fassung des Artikels (Sommer 1924) de Broglies Arbeit unerwähnt bleibt, verweist Einstein 1925 ausdrücklich darauf. Da Einstein wohl kaum ein Physiker war, der sich mit fremden Lorbeeren schmückte, indem er die Arbeit anderer Wissenschaftler verschwieg, kann das Fehlen des Hinweises in der ersten Fassung so erklärt werden, daß Einstein bis dato die Dissertation de Broglies nicht richtig gelesen hatte.

Eine endgültige Klärung des genauen historischen Ablaufs scheint nicht mehr möglich zu sein, trotzdem stellen sich die Fakten etwa so dar: Sicher ist, daß de Broglie die Idee der Materiewelle als erster entwickelte und veröffentlichte, sein Verdienst, einen wirklich revolutionierenden Schritt gegangen zu sein, bleibt trotz all die vorangehenden Überlegungen ungeschmälert. Fest steht ebenfalls, daß Einstein sowohl die drei Artikel von 1923 in den *Comptes Rendus*, als auch die Doktorarbeit vor deren Veröffentlichung hätte lesen können, sicher ist aber auch, daß es hierfür bisher keine schriftlich gesicherten und eindeutigen Belege gibt. Es gibt aber sehr wohl Hinweise darauf, daß Einstein 1924 in eine ganz ähnliche Richtung wie de Broglie gedacht hat. Dies erscheint nicht unplausibel, wenn man bedenkt, wie nahe die Broglieschen Konzepte der Denkweise Einsteins standen. Es besteht also die Möglichkeit, daß Einstein unabhängig von de Broglies Darstellungen zu seinen früheren Ideen zurückgekehrt ist. Es ist wohl berechtigt, anzunehmen, daß Einstein das Dualismuskonzept 1905 nicht konsequent weiterverfolgt hat, da bereits seine Photonen nicht anerkannt wurden und die Möglichkeit, ein noch weitergehendes Konzept durchzusetzen, sehr gering war. Und vielleicht kam es Einstein 1924 sogar sehr gelegen, daß de Broglie in seinem Sinne weiter geforscht hatte und er selbst nur als Schirmherr der Materiewelle auftreten mußte, ohne mögliche öffentliche Anfeindung ertragen zu müssen.

---

[108] Dieses scheint jedoch angesichts der Tragweite der Erkenntnisse de Broglies kein sehr gewichtiges Argument zu sein: Selbst wenn Einstein die Arbeit schon Anfang des Jahres 1924 zur Kenntnis genommen hätte, würden sie einige Monate später noch sein hochlobende Begeisterung verdienen.

[109] Übersetzung: „Folglich wußte Einstein gegen Ende des Jahres 1924 um meine Dissertation."; zitiert in [189, S. 182].

Eine ganz ähnliche Position nahm Einstein auch gegenüber der de Brogllieschen „doppelten Lösung" ein. Er sympathisierte sehr mit den Bemühungen de Broglies um ein anschauliche Interpretation der Quanteneffekte und munterte ihn in bestmöglicher Weise zur Weiterarbeit auf. So war er der einzige, der de Broglie nach dem mißlungenen Solvay-Kongreß von 1927 unterstütze. Auf dem Bahnhof sagte er zu de Broglie:

> Continuez! C'est vous qui êtes dans la bonne voie [...] une théorie physique devrait pouvoir, en dehors de tout calcul, être illustrée par des images si simples qu'un enfant devrait pouvoir les comprendre.[110]

Für de Broglie war die Sympathie und Unterstützung Einsteins immer eine Quelle starker Motivation und Aufmunterung in seinem einsamen Kampf gegen die anerkannte Quantenmechanik.

> Elle [la lettre d'Einstein] m'a apporté un grand encouragement pour continuer à reprendre les idées, en les approfonissant, que j'avais entrevues en 1927.[111]

Wenn Einstein auch sehr mit seinem jüngeren französischen Kollegen sympathisierte, so war er doch nie wirklich versucht, dessen Konzepte aufzugreifen und selbst weiter zu entwickeln. Während des Solvay-Kongresses sagte er zu de Broglie:

> Ces problèmes de Physique quantique deviennent trop complexes. Je ne peux plus me mettre à étudier des questions aussi difficiles: je suis trop vieux.[112]

Obwohl ein solch pathetischer Ausspruch des 48jährigen, der nur drei Jahre zuvor die Bose-Einstein Statistik entworfen hatte, leicht überzogen wirkt, spiegelt er doch ganz gut Einsteins Resignation vor dem Eindruck wider, daß sich die Physik in eine falsche Richtung entwickle. Auch in späteren Jahren beschäftigte er sich kaum mit der Quantenmechanik und versteckte sich hinter den immensen Problemen, die es zu überwinden galt, wollte man der Kopenhagener Deutung eine andere Theorie gegenüberstellen. In einem Brief vom 8. Februar 1954 schrieb Einstein:

> Ich muß nämlich erscheinen wie der Wüsten-Vogel Strauss, der seinen Kopf dauernd in den relativistischen Sand steckt, damit er den bösen Quanten nicht ins Auge sehen muss.[113]

Aber immerhin hoffte auch Einstein, daß man eines Tages beweisen könne, daß die anerkannte Quantenmechanik unvollständig sei:

> In Wahrheit bin ich genau wie Sie davon überzeugt, dass man nach einer Substruktur suchen muss, welche Notwendigkeit die jetzige Quantentheorie durch Anwendung der statistischen Form kunstvoll verbirgt.[114]

Einstein war überzeugt, daß der Schlüssel zur Entdeckung einer Substruktur nur in einfachen physikalischen Vorstellungen gefunden werde könne, in dem Vertrauen auf

> [...] die denkbar grösste Einfachheit der Naturgesetze.[115]

---

[110] Übersetzung: „Machen Sie weiter! Sie sind es, der sich auf dem richtigen Weg befindet [...] eine physikalische Theorie sollte, abgesehen von jeglicher Berechnung, mit so einfachen Bildern erklärbar sein, daß sie ein Kind verstehen können müßte.", zitiert in [283, S. 141].

[111] Übersetzung: „Er [der Brief Einsteins] hat mich sehr darin bestätigt, darin forzufahren, die Ideen wieder aufzunehmen und zu vertiefen, die ich 1927 geahnt hatte.", Anhang A.2.9.

[112] Übersetzung: „Diese quantenmechanischen Probleme werden zu komplex. Ich kann mich nicht mehr daran machen, solch schwierige Fragen zu studieren: Ich bin zu alt.", zitiert in [189, S. 196].

[113] Anhang A.2.8.

[114] Anhang A.2.8.

[115] Anhang A.2.8.

Einstein bezeichnete sich selbst als „fanatischen Gläubigen der Methode der 'logischen Einfachheit'".[116] Er wußte, daß er in de Broglie einen Verbündeten hatte in seinem Widerstand gegen eine Physik, die mehr und mehr ihre Anschaulichkeit verlor. Einstein schreibt im Mai 1953 an de Broglie:

> Gemeinsam aber ist uns die Überzeugung, dass wir an der Idee der Möglichkeit einer vollständigen objektiven Darstellung einer physikalischen Wirklichkeit festhalten sollen [...].[117]

Zusammenfassend kann man sagen, daß Albert Einstein und Louis de Broglie eine große Sympathie und gegenseitiger Respekt verband. Diese Verbindung war allerdings weniger aus einer persönlicher Beziehung heraus entstanden, sie war vielmehr das Resultat weltanschaulicher und wissenschaftlicher Verwandtschaft. War de Broglie zu Beginn seiner wissenschaftlichen Karriere stark von Einstein beeinflußt und behielt er für Einstein zeitlebens die Dankbarkeit und Bewunderung eines Schülers, entwickelten sich doch seine Forschungsgebiete in eine andere Richtung. Trotzdem blieb die Ausrichtung der Arbeit, das übergeordnete Ziel das gleiche: die Suche nach einer anschaulichen, die physikalische Realität beschreibenden Wissenschaft. In eben jenem unmodernen und einsamen Streben, das sie als Außenseiter auswies, unterstützen sich diese beiden hervorragenden Physiker so gut es ging.

---

[116] Anhang A.2.8.
[117] Anhang A.2.7.

# III Jahre der Anpassung (1927-1951)

Für Louis de Broglie begann im Jahr 1927 ein neuer Lebensabschnitt. Die Zeit von 1927-1951 war geprägt durch innere Widersprüche: Sie brachte de Broglie zwar die lange verwehrte wissenschaftliche und gesellschaftliche Anerkennung, sie zwang ihn aber auch zur Aufgabe der eigenen Forschungsideale. Da diese Etappe im Leben de Broglies weder für ihn persönlich noch wissenschaftlich von herausragender Bedeutung war, sollen die wichtigsten Aspekte in aller Kürze behandelt werden.

## III.1 Aktivitäten und Ehrungen

Nach dem fünften Solvay-Kongreß sah de Broglie keine Möglichkeit, seine ursprüngliche Wellenvorstellung weiter zu verfolgen. Im Frühjahr 1928 erhielt er in einem vom 15.2.1928 datierten Brief[118] eine Einladung Paulis zu einem Gastvortrag in Hamburg. Dieser Vortrag ist insofern bedeutsam, als de Broglie hier erstmals seine Anpassung an die Kopenhagener Deutung öffentlich bekanntgab:

> Invité à faire des conférences à l'université de Hambourg au printemps de 1928, j'y donnais pour la première fois en public mon adhésion formelle aux idées nouvelles.[119]

Obwohl dieser Schritt die Aufgabe seiner inneren Überzeugungen bedeutete, folgte auf ihn die Phase gesellschaftlichen und wissenschaftlichen Ruhms. Paradoxerweise stand damit am Anfang der gesellschaftlichen und wissenschaftlichen Anerkennung sein persönliches Scheitern.

Seit dem experimentellen Nachweis der Elektronenbeugung durch Davidson und Germer sowie Thompson und seit der Einladung zum Solvay-Kongreß galt de Broglie auch in Frankreich als einer der führenden Physiker, so daß man ihm für den November 1928 einen Lehrstuhl am neu eingerichteten *Institut Henri Poincaré* für theoretiche Physik anbot. Letzten Endes hat ihn wohl auch die Verantwortung für seine Studenten dazu getrieben, die sich durchsetzende moderne Physik zu lehren:

> Une autre raison de mon renoncement, plus curieuse peut-être du point de vue psychologique, a été la suivante: au moment même où j'apercevais, en 1928, la difficulté de poursuivre mon effort d'interprétation, j'entrai dans l'enseignement supérieur, [...]. Or l'enseignement [...] exige [...] que l'on tienne ses élèves au courant des plus récents progrès et des résultats les plus instructifs.[120]

Im selben Jahr starb de Broglies Mutter, bei der er bis zuletzt am *Square de Messine* gewohnt und um die er sich sehr gekümmert hatte. Das Haus wurde verkauft und de

---

[118] Der Brief befindet sich in den *Archives de l'Académie des Sciences*.

[119] Übersetzung: „Im Frühjahr 1928 war ich eingeladen, an der Hamburger Universität Vorträge zu halten, dort gab ich zum ersten Mal öffentlich mein formelles Einverständnis zu den neuen Ideen.", [283, S. 144].

[120] Übersetzung: „Ein anderer Grund für mein Nachgeben, der vom psychologischen Standpunkt aus vielleicht noch erstaunlicher ist, war der folgende: In dem Moment, als ich 1928 die Schwierigkeiten erkannte, meinen Interpretationsversuch weiter zu verfolgen, begann ich meine universitäre Lehrtätigkeit. Nun verlangt die Lehre, daß man seine Schüler über die jüngsten Fortschritt und instruktivsten Ergebnisse auf dem laufenden hält.", zitiert in [283, S. 145].

Broglie änderte seinen Lebensstil grundsätzlich. Er kaufte ein fast ländlich gelegenes Haus in Neuilly und richtete sich dort ein geradezu bürgerliches Leben ein: An den alten Glanz seiner Familie erinnerten nur noch die geerbten Einrichtungsgegenstände und die große Bibliothek. Von der früher großen Dienerschaft nahm de Broglie nur einen Altgedienten mit, der in späteren Jahren durch ein Ehepaar ersetzt wurde, das de Broglie bis ins hohe Alter respektvoll umsorgte. Dies war die letzte einschneidende Veränderung in seinem Lebensstil, denn bis zu seinem Lebensende sollte er in diesem Haus ein gut organisiertes, pünktliches, ja fast klösterliches Leben führen.

Seine Mutter, die große Erwartungen in ihren jüngsten Sohn gesetzt hatte, erlebte dessen größten Triumph nicht mehr: 1929 erhielt Louis de Broglie den Nobelpreis für die Vorhersage der Elektronenbeugung. Sehr charakteristisch für de Broglie scheint die Art zu sein, mit der er dieses Ereignis seinen Studenten im Seminar mitgeteilt hat: Er könne nächste Woche nicht kommen, da er nach Stockholm reisen müsse. G. Lochak schreibt dazu:

> Je l'imagine bredouillant un peu, baissant la voix et regardant ses chaussures comme chaque fois qu'il était obligé de parler de lui.[121]

De Broglies Auftreten vor den Studenten war wohl in der Tat nicht sehr versiert. Insgesamt war sein Verhältnis zu ihnen geprägt von einer distanzierten Sympathie. Als Vorgesetzter ließ er seinen Mitarbeitern und Doktoranden die größtmögliche Freiheit, akzeptierte und veröffentlichte Arbeiten, die in sich schlüssig waren, auch dann, wenn die darin formulierten Theorien nicht seinen Ansichten entsprachen. Obwohl er immer für konstruktive Kritik, kompetente Beratung und Hilfeleistungen[122] zur Verfügung stand, ging von ihm nicht unbedingt eine aktive Betreuung aus:[123] Kontakt mit de Broglie kam nur dann zustande, wenn die Studenten auf ihn zugingen. Sein oft abwesendes, wenig aufgeschlossenes Auftreten war somit, wie M.-A. Tonnelat schreibt [285, S. 46f], eher dazu angetan, gerade Anfänger abzuschrecken. Aber auch mit seinen Vorlesungen und Seminaren scheint de Broglie nicht viele Studenten erreicht zu haben:

> Louis de Broglie n'a eu que peu d'élèves. Je ne suis certainement pas le seul à avoir éprouvé quelques déceptions lors des conférences qu'il nous fit à l'École normale Supérieure.[124]

Nicht nur die Themen, auch die Vortragsart dürfte wenig anregend gewesen sein: Viel zu leise und mit eintönig hoher, fast schriller Stimme soll er seine vorbereiteten Texte in atemberaubenden Tempo vorgelesen haben, so daß seine Studenten ihm nachsagten, er sei schneller als das Licht [283, S. 207].

Dieses methodische Mißgeschick wurde jedoch teilweise aufgewogen durch die bewundernswerte Klarheit der Darstellung [285, S. 46], die auch seine Veröffentlichungen auszeichnet. Deren Zahl ist in diesen Jahren unüberschaubar: Neben eher philosophisch oder wissenschaftstheoretisch ausgerichteten Darstellungen der Quantenmechanik [184-187], entstanden Bücher, Artikel und Aufsätze zu aktuellen Fragestellungen [31-93, 152-172, 214-261]. Im Rahmen seiner Tätigkeit als *Secrétaire*

---

[121] Übersetzung: „Ich stelle ihn mir vor, wie er leicht stotternd die Stimme senkte und seine Schuhe betrachtete, wie jedes Mal, wenn er gezwungen war, von sich zu reden.", [283, S. 151].

[122] Zahlreiche frühere Studenten de Broglies (z.B. Michel Hulin u.a.) haben seine freundliche Hilfsbereitschaft in ihren Beiträgen in [287] beschrieben.

[123] Dies berichtete mir auch Marcel Coz von der Universität Kentucky, der die damaligen Verhälnisse in Paris als Student erlebt hat.

[124] Übersetzung: „Louis de Broglie hatte nur wenige Schüler. Ich bin sicherlich nicht der einzige, der etwas enttäuscht war von den Vorträgen, die er uns in der *École Normale Supérieure* hielt.", Paul Germain in [281, S. 589].

*perpétuel de l'Académie des Sciences* entstanden zudem zahlreiche historische Wissenschaftlerporträts [193-211].

In einer einfachen, aber unglaublich präzisen Sprache weiß de Broglie darin den Leser an die Hand zu nehmen und ihm äußerst komplexe Gedanken verständlich zu machen. Auch Einstein hatte Freude an der Darstellungsweise de Broglies und schrieb ihm am 8. Februar 1954:

> Es ist eine drollige Sache, wie alles plastischer und lebhafter wirkt, wenn es in der altgewohnten Sprache erscheint.[125]

Es ist sicherlich kein Zufall, daß die heute noch gelesenen Bücher de Broglies diejenigen sind, die er in der Zeit bis 1951 zur orthodoxen Interpretation der Quantenmechanik schrieb.[126] Die gedankliche Klarheit dieser Werke macht sie zu Klassikern dieses Gebietes. Übrigens hat de Broglie die Neuauflage dieser Werke untersagt, ein Verbot, das die *Fondation Louis de Broglie*, die das wissenschaftliche Erbe de Broglies angetreten hat, aufrecht erhält.[127]

An dieser Stelle alle Aktivitäten, Auszeichnung und Ehrungen de Broglies während dieser Epoche erwähnen zu wollen, wäre ein langwieriges Unterfangen. Nur die wichtigsten seien erwähnt: 1933 wurde er in die *Académie des Sciences* gewählt, er hielt Vorträge im In- und Ausland, lehrte an der *École Supérieure d'Électricité* und an der *École Normale Supérieure*. 1938 erhielt er die Max-Planck-Medaille und den ehrenvollen Posten des *Secrétaire perpétuel de l'Académie des Sciences* und wurde 1944 in die *Académie Française* gewählt. Sein Rat war gefragt, in theoretischen Fragen aber auch in praktischen Anwendungen; so betreute er zahllose Doktorarbeiten, und sein 1931 gegründetes Seminar am *Institut Henri Poincaré* entwickelte sich zum wichtigsten Informationszentrum für theoretische Physik in Frankreich.[128] Kurz, de Broglie gelang es endlich, auf dem von ihm gewählten Weg dem Familiennamen alle Ehre zu machen.

## III.2 Forschungsrichtungen und Zweifel

Der Preis für diese Anerkennung war allerdings hoch: De Broglie mußte seine eigene Forschungstätigkeit auf ein Minimum reduzieren: Die gesellschaftlichen Verpflichtungen ließen kaum noch Zeit für persönliche Arbeit [285, S. 49]. Nur in der vorlesungsfreien Sommerzeit konnte er sich seinen eigenen Interessen widmen. Allerdings erschien ihm nach dem Solvay-Kongreß auch keine Aufgabe wirklich reizvoll, so daß seine Forschung vier Jahre lang fast vollständig ruhte. Erst 1932 begann er mit der Entwicklung einer neuen Theorie des Lichtes [43ff, 161, 162, 165, 170]. Er betrachtete darin das Photon nicht als Elementarteilchen; ihm zufolge setzt es sich aus zwei Dirac-Spin ½-Teilchen zusammen und besitzt einen Spin 1. Diese Spin ½-Teilchen identifizierte er mit den von Pauli postulierten Neutrinos, die zur damaligen Zeit noch nicht nachgewiesen waren und denen de Broglie eine Masse ungleich Null zuordnete. Daraus folgte für ihn, daß auch das Photon eine endliche, aber sehr kleine Masse besitzen müsse.[129] De Broglies Theorie reintegrierte damit das Photon in den Rahmen der wellenmechanischen Beschreibung, stellte eine Synthese her zwischen Materie und Licht und konnte außerdem die Polarisation des Lichtes als experimentell nachweisbares Phänomen des Spins des Photons erklären. Diese Theorie war immerhin Gegenstand reger Diskussionen in der Fachwelt: Pauli bezeichnet de Broglies Veröffentlichung dazu in einer Postkarte an Heisenberg vom 19. Januar 1934 als eine „nicht uninteressante Note" [299b, S. 253]. Während der nächsten acht Jahre taucht die Theorie de Broglies in Paulis Korrespondenz immer wieder auf, einzelne Aspekte werden besprochen, und es wird deutlich, daß Pauli die Idee mit Ausnahme der Masse des Photons gefiel, ihre Ausführung jedoch weniger. Diese heute beinahe völlig vergessene Theorie weitete de Broglie zu einer Theorie der Spinteilchen aus [167], die von seinen Schülern als sein zweites großes Werk bezeichnet wird. Im Zuge der Erarbeitung dieser Theorie prägte de Broglie übrigens den Begriff des „Antiteilchens", der auf ihn zurückgeht.

---

[125] Anhang A.2.8.

[126] Es handelt sich dabei vor allem um die Werke [184-187].

[127] Am Rande sei erwähnt, daß die *Fondation Louis de Broglie* tatsächlich nur das wissenschaftliche Erbe angetreten hat. Denn neben den Urheberrechten an den Werken de Broglies hat sie zwar ein Landhaus bekommen, das jedoch nicht aus de Broglies Familienbesitz stammte, sondern mit dem Geld für den Nobelpreis erstanden worden war.

[128] Fast alle großen Physiker der Zeit haben hier Vorträge gehalten.

[129] Vgl. Fußnote 74.

Während des zweiten Weltkrieges mußte de Broglies Forschungstätigkeit abermals längere Zeit ruhen. Die Jahre nach der Normalisierung des Universitätsbetriebes sind bestimmt durch die Suche nach neuen Betätigungsfeldern: Er arbeitete an einer Theorie des Atomkerns. Sein dreibändiges (1943-46) Lehrwerk [168] bringt keine wirklich neuen Ergebnisse, resümiert aber in der de Broglie so eigenen klaren Art den bisherigen Forschungsstand. Daneben beschäftigte er sich mit der Quantenfeldtheorie [170], zudem legte er vorsichtig die Grundlage für die Theorie, die er fünfzehn Jahre später als die „verborgene Thermodynamik" [73] (vgl. auch Kapitel IV.1 dieser Arbeit) bezeichnen sollte. Gleichzeitig begann er aber bereits, sich wieder verstärkt für die grundlegenden Fragen der Quantenmechanik zu interessieren: Nach zwanzig Jahren Unterbrechung beleuchtete er erstmals wieder die Postulate der Kopenhagener Quantenmechanik kritisch: So wandte er sich in [172] gegen die von Bohr und Pauli aufgestellte These, daß der Spin prinzipiell nicht direkt meßbar sei. Er veröffentlichte außerdem Betrachtungen zum Wahrscheinlichkeitsbegriff in der Quantenmechanik, die konsequent weitergedacht zur Frage nach verborgenen Parametern führen müßten [75]. Eine Einführungsveranstaltung in die Grundlagen der Quantenmechanik [183] und die Erscheinung der Jubiläumsschrift zum 70. Geburtsag Einsteins[130] nahm de Broglie ab 1949 zum Anlaß, die von Einstein und Schrödinger in den vergangenen zwanzig Jahren formulierten Einwände gegen die Kopenhagener Deutung durchzusehen. Hierbei fiel ihm auf, wie aktuell und zutreffend die Fragen und wie wenig befriedigend oft die Antworten der Kopenhagener Schule waren. Trotzdem war es 1946-1951 nicht de Broglies erklärtes Ziel, die Quantenmechanik in der Kopenhagener Deutung infrage zu stellen, wenngleich die neuerliche Beschäftigung mit ihren Grundlagen bereits auf aufkeimende Zweifel hinweist.

---

[130] *Albert Einstein Philosopher-Scientist*, P. A. Schilpp (Hrsg.), Library of Living Philosophers, Evanston III, 1949.

# IV Isolation (1951-1987)

Im Sommer 1951 erhielt de Broglie auf Einsteins Empfehlung hin eine Abhandlung des Amerikaners David Bohm [306], in der er seine „Führungswellentheorie" wiederfand. Vor dem Hintergrund der zahlreichen Zweifel, die er bereits seit einigen Jahren hegte, kann dieses Ereignis als Auslöser für de Broglies Abkehr von der Kopenhagener Deutung angesehen werden. Im Mittelpunkt des nun folgenden Kapitels soll die letzte Etappe im Leben de Broglies stehen, die bestimmt war durch intensive, von der übrigen *scintific community* abgeschiedene Arbeit auf der Suche nach einer neuartigen Beschreibung der Quanteneffekte.

## IV.1 Rückkehr zu den alten Ideen

Nach Lochak war es für ihn ein Schock, seine alte, aufgegebene Theorie von einem anderen Physiker enthusiastisch verteidigt zu sehen. Seine erste Reaktion war ablehnend, kannte er doch allzu gut die Schwächen dieses Konzeptes. Seine Kritik veröffentlichte er in einem Artikel in den *Comptes Rendus* [94] im September 1951; er gestand dennoch ein, daß das „Führungswellenkonzept" bei allen Schwierigkeiten dennoch in der Lage sei, das Einstein-Podolsky-Rosen-Paradoxon zu lösen und sah darin einen Gegenbeweis für das von-Neumann-Theorem[131] . Obwohl die Arbeit Bohms bei weitem nicht dem neueren Erkenntnisstand de Broglies entsprach, hatte sie entscheidenden Einfluß auf ihn: Er entschloß sich endgültig, seine angepaßte Haltung aufzugeben und nach Alternativen zur Kopenhagener Quantenmechanik zu suchen. Bestärkt wurde er in dieser Entscheidung durch den jungen französischen Physiker Jean-Pierre Vigier, der de Broglies Aufmerksamkeit auf Einsteins Idee der Singularitäten im Gravitationfeld lenkte. Erstaunlicherweise hatte de Broglie diese schon zwanzig Jahre alte Theorie noch nicht zur Kenntnis genommen, sah nun aber die frappante Analogie zu seiner „doppelten Lösung" und fühlte sich ermutigt, in dieser Richtung weiterzuarbeiten. In diesem Sinne kündigte de Broglie im Januar 1952 die Rückkehr zu seinen früheren Ideen an [96].

Dieser Artikel bedeutete einen Wendepunkt im Schaffen und im Leben de Broglies: er bestimmte nicht nur die Neuorientierung seiner Forschung, sondern trieb ihn auch in die Isolation. Denn de Broglies Abkehr von der Kopenhagener Deutung wurde ablehnend aufgenommen. So finden sich in dem Sammelband *Louis de Broglie - Physicien et Penseur* [286], der anläßlich seines 60. Geburtstages erschien, neben persönlichen Erinnerungen und wissenschaftlichen Beiträgen zahlreiche, manchmal versteckt ablehnende Stellungnahmen zu de Broglies Frontenwechsel.

Einzig die ewigen Kritiker der Kopenhagener Deutung, Einstein und Schrödinger, begrüßten de Broglie glücklich im Kreise der Außenseiter: Einstein etwa verlieh seinem Glauben an den „realen Zustand eines physikalischen Systems" [286b, S. 14] Ausdruck und stellte die Vollständigkeit der Quantenmechanik in Frage. Wenngleich Einstein die Lösung des Problems eher in der Relativitätstheorie zu finden hoffte, unterstützte er doch de Broglie mit seinem Beitrag indirekt in dessen Kritik an der Quantenmechanik. So auch Schrödinger, der eigensinnig und unwiderruflich an seinem schon 25 Jahre zuvor formulierten Wellenkonzept festhielt. Demgegenüber sprachen sich die Vertreter der Kopenhagener Deutung vehement gegen de Broglies neue Projekte aus. Erstaunlicherweise äußerten sich hierzu in besonders dogmatischer und kritischer Weise die Kopenhagen-Anhänger der zweiten Generation, also die Schüler Bohrs, wie Rosenfeld, aber auch Schüler de Broglies, wie Tonnelat und Destouches. Obwohl auch Pauli Argumente gegen verborgene Parameter vorbrachte, war in seinem Aufsatz der feindselige Unterton der zwanziger Jahre verschwunden. Auch Heisenbergs unpolemischer und sachlicher Beitrag zeigt, daß die großen Physiker der Gründungszeit der Quantenmechanik Frieden geschlossen hatten. Fast ist man geneigt, daraus zu schließen, daß diese genialen Physiker, die die revolutionierenden Ereignisse der zwanziger Jahre miterlebt hatten, sich durchaus vorstellen konnten, daß die inzwischen gealterte und orthodox gewordene Theorie noch verbesserungsfähig sei. Tatsächlich schreibt Pauli in seinem Beitrag, er sei

> Weit davon entfernt, den derzeitigen Stand der Quantenmechanik auf relativistischem Boden als endgültig zu betrachten [...].[286, S. 35].

---

[131] J. von Neumann, *Mathematische Grundlagen der Quantenmechanik*, Springer, Berlin 1932.

Auch von Dirac weiß man, daß er 1978 auf einer Konferenz in Australien sagte:

> Es kann letzten Endes passieren, daß Einstein doch Recht gehabt hat, denn die aktuelle Form der Quantenmechanik kann nicht als definitv angesehen werden. Es gibt große Schwierigkeiten [...]. Und es ist sehr gut möglich, daß es in der Zukunft eine perfektionierte Quantenmechanik geben wird, die eine Rückkehr zum Determinismus bedeuten wird und damit den Standpunkt Einsteins rechtfertigen wird.[132]

Demgegenüber machte sich eine neue Generation von vielleicht weniger erfahrenen, etwas weniger brillanten Physikern zum Anwalt der Kopenhagener Deutung; sie empfanden die Ansätze de Broglies als Sakrileg. Auch Anfang der fünfziger Jahre, in einer Zeit also, in der die theoretischen und praktischen Anwendungen der modernen Physik mit der Quantenfeldtheorie, den Möglichkeiten der Atom- und Kernphysik etc. beachtliche Erfolge feierten, wollte niemand etwas von de Broglies anschaulicher, aber unausgereifter Theorie hören.

Tatsächlich wurde die Situation um de Broglie herum schwieriger: In den Fluren des *Institut Henri Poincaré* sprach man von ihm, als sei er verrückt geworden, und dem einen oder anderen Studenten wurde nahegelegt, sich im eigenen Interesse anderen Professoren anzuschließen.[133] Die aus innerer Überzeugung heraus geschehene Rückkehr zu den alten Ideen wurde de Broglie von vielen Kollegen fast als Verrat angelastet. Sie fühlten sich überrumpelt und ausgeschlossen und sahen den von ihm eingeschlagenen Weg als rückständig, fruchtlos und falsch an.[134]

Nur wenige Mitarbeiter hielten zu de Broglie: Der eher mathematische Physiker Petiau, Destouches, der skeptisch war, sich aber für bestimmte Aspekte der Arbeit interessierte, Olivier Costa de Beauregard, Vigier und später G. Lochak, Andrade e Silva u.a. Natürlich arbeitete die Gruppe auch mit David Bohm zusammen. Im Vordergrund der Arbeit stand die Bemühung, der „doppelten Lösung" eine möglichst überzeugende Form zu geben. Neben kurzen Artikeln (vgl. [96ff]), entstand 1954 ein zwei Jahre später veröffentlichtes Werk [175], das einen guten Überblick über den Entwicklungsstand dieser Theorie gibt. De Broglie betrachtete darin die Welle $u$ außerhalb der „Region hoher Konzentration", er nennt sie „singuläre Region" [189, S. 154], als aus zwei Anteilen zusammengesetzte Funktion:

$$u = u_0 + \text{v}.$$

---

[132] In [283, S. 197] französisch zitiert aus P. A. M. Dirac, *Directions in Physics*, Wiley, New York 1978.
[133] Vgl. [283, S. 208].
[134] Vgl. [283, S. 212].

Hierin ist v eine reguläre, aus der Wellenmechanik bekannte Lösung, für die gilt $\Psi = C\text{v}$, ($C$ ist ein Normierungsfaktor); $u_0$ hingegegen ist eine Funktion in Phase mit v und mit sehr kleinen Werten außerhalb der „singulären Region", die jedoch in der Nähe der Region sehr schnell anwachsen. Für das Verhalten von $u$ im Bereich der „singulären Region" hatte de Broglie im Gegensatz zur früheren Formulierung erkannt, daß die Werte zwar sehr groß, nicht aber unendlich sein dürften. Zudem fand er in Analogie zu den Singularitäten im Einsteinschen Gravitationsfeld, daß in der „singulären Region" nach nicht-linearen Grundgleichungen zu suchen sei. Eben diese Nichtlinearität gewährleiste die Verbindung von $u_0$ und v außerhalb der „singulären Region". Somit konnte de Broglie sein Konzept der „doppelten Lösung" präzisieren: Die Funktion $u$ besitzt in der „singulären Region" sehr große, aber nicht unendliche Werte, und ist immer phasengleich mit v. Die sehr kleine „singuläre Region", einer bewegten Uhr gleichgesetzt, stellt das Teilchen dar, das entlang der „Stromlinien" der Welle v „geführt" wird. Die „Führung" des Teilchens wird durch die „Führungsformel" beschrieben. Es gelang de Broglie zudem, Erkenntnisse über die relativistische Betrachtung der „geführten Bewegung" des Teilchens zu gewinnen: Er erkannte, daß eine Beschreibung für die „geführte Bewegung" des Teilchens mit einer „Dynamik variabler Masse" möglich ist [137, S. 224f]. Die Ruhemasse des Teilchens $M_0$ entspricht dann nicht der üblichen und konstanten Ruhemasse $m_0$, sondern der variablen Masse

$$M_0 = m_0 + \frac{q_0}{c^2},$$

die von der Größe des sich zeitlich verändernden Quantenpotentials abhängt.

Neben diesen wichtigen Festellungen sah de Broglie die Notwendigkeit, ein zufallsbedingtes Element in die Theorie einzubauen. Ab 1960 entwarf er deshalb seine Theorie von der „verborgenen Thermodynamik", die bereits in früheren Artikeln angedeutet war.[135] Er veröffentlichte seine Ergebnisse in zahlreichen Artikeln [115ff] und insbesondere in den Büchern [179, 181]. Bisher war es de Broglie noch nicht gelungen, das „periodische Phänomen" $\text{v}_0$ im Ruhesystem des Teilchens physikalisch zu erklären und in sein Gesamtkonzept zu integrieren. Bereits 1945 (in [73]) entdeckte er jedoch, daß die von Planck und Laue angegebene relativistische Transformation einer Wärmemenge,[136] also

$$Q = Q_0 \sqrt{1 - \beta^2}$$

ganz analog zur Tranformation der „inneren Frequenz"

$$\text{v}_1 = \text{v}_0 \sqrt{1 - \beta^2}$$

funktioniert. Dies brachte ihn auf den Gedanken, daß die Energie im Ruhesystem des Teilchens $E_0 = h\text{v}_0 = M_0 c^2$ mit einer „inneren Wärme" $Q_0$ gleichzusetzen sei:

---

[135] Erste Hinweise finden sich schon 1921 in [7], weitergedacht hat de Broglie seine Ideen in [73], einem Artikel, der bereits 1945 geschrieben, aber erst 1948 veröffentlicht wurde. Erst ab 1960 arbeitete er jedoch wirklich konsequent an der Entwicklung einer neuen Theorie.

[136] Die Herleitung der Transformationsformel der Wärme durch Planck und Laue sowie seine eigenen Argumente dafür gibt de Broglie in [137, S. 210ff] wieder.

La particule nous apparaît alors comme un petit réservoir de chaleur en mouvement avec la vitesse βc.[137]

Die innere Energie nimmt demnach für einen relativ zum System ruhenden Betrachter die folgende Form an:

$$E_1 = h\nu_1 = h\nu_0 \sqrt{1-\beta^2} = M_0 c^2 \sqrt{1-\beta^2} = Q_0 \sqrt{1-\beta^2}$$

Mit dieser Beziehung fand de Broglie 1967 in [132] den Ausdruck für die Gesamtenergie eines relativ zum System ruhenden Beobachters:

$$E_{ges} = \frac{M_0 c^2}{\sqrt{1-\beta^2}} = E_1 + E_t = Q_0 \sqrt{1-\beta^2} + \frac{M_0 v^2}{\sqrt{1-\beta^2}}$$

$E_t$ bezeichnete de Broglie als „translatorischer Anteil" oder auch „pseudo-énergie cinétique".[138] Es gelang ihm nachzuweisen, daß eine Verbindung zwischen der „doppelten Lösung" und der „verborgenen Thermodynamik" über die Masse $M_0$ möglich ist. Denn auch im Rahmen der „verborgenen Thermodynamik" konnte de Broglie (z.B. in [137]) zeigen, daß nicht mit der sonst üblichen Ruhemasse $m_0$, sondern mit der variablen Masse $M_0$ zu rechnen sei. De Broglie erklärte die auch in der Theorie der „doppelten Lösung" auftretende Veränderung der Masse $M_0$ als einen zufälligen Wärmeaustausch mit einem „verborgenen Thermostaten". Die normale „geführte Bewegung" würde also überlagert durch eine Brownsche Bewegung, die in zufälliger Weise das Teilchen von einer der möglichen Bahn auf eine andere bringe. Dieser Austausch von Wärme sei verbunden mit den Veränderungen des Quantenpotentials und damit der Amplitude der Welle, so daß diese als Verbindung zwischen dem „verstecktem Thermostaten" und dem Teilchen diene [137, S. 230].

De Broglie kam weiterhin seinen Lehrverpflichtungen am *Institut Henri Poincaré* nach. Während sich das wöchentliche Seminar eher an den Themen der orthodoxen Physik orientierte, nutzte de Broglie seine Vorlesung als Forum für seine eigene Forschungstätigkeit: Oftmals bekamen seine Doktoranden hier Theorien zu hören, die denen komplett widersprachen, die sie in ihren Dissertationen verteidigen würden. Diese etwas paradoxe Situation endete 1962 mit de Broglies Emeritierung, nach der er das *Institut Henri Poincaré* verließ, ohne es jemals wieder zu betreten. Von den Verpflichtungen des Lehrstuhls befreit, konnte er sich ganz auf seine eigene Forschungsarbeit konzentrieren. Mit vier Mitarbeitern, Silva, Lochak, Fer und Thiounn, gründete er in der *Académie française* ein Seminar, das sich zum Ziel setzte, im Sinne der „doppelten Lösung" der „verborgenen Thermodynamik" eine konsistente Theorie zu finden. In der Zeit von 1963-1976 entstanden zahlreiche Veröffentlichungen [116-149, 178183, 189-192, 266-279] zu diesem Thema, und de Broglie bezeichnete diese Zeit als die vom intellektuellen Standpunkt aus betrachtet schönsten Jahre seines Lebens [288, S. 86].

---

## IV.2 Die letzten Lebensjahre

Mehr und mehr zog sich de Broglie aus dem öffentlichen Leben zurück und nahm aufgrund seines fortgeschrittenen Alters immer weniger aktiv an der Forschungstätigkeit seiner Gruppe teil. Die Veröffentlichungen von 1972-1976 [145-149, 182-183] sind zumeist nur noch Resümees seines wissenschaftlichen Lebenswerkes. 1973 weihte de Broglie die *Fondation Louis de Broglie* noch selbst ein. Im Oktober 1975 fand sein letzter öffentlicher Auftritt statt: Er hielt den Eröffnungsvortrag für das Seminar in der *Fondation Louis de Broglie* [274] und beendete damit seine Karriere und sein Leben in der Öffentlichkeit.

Die letzten Lebensjahre Louis de Broglies waren überschattet von altersbedingter Krankheit; er starb am 19.3.1987 in aller Stille. Er hatte sich eine einfache Beerdigung gewünscht, zu der neben den Familienmitgliedern nur einige enge Mitarbeiter kamen. Sein Tod rief in der Presse ein sehr verhaltenes Echo hervor: nur in einigen kurzen Notizen wurde seiner gedacht, wobei er nur als Entdecker der Materiewelle erwähnt wurde. Die späteren Bemühungen um eine neue Interpretation der Quantenmechanik, seine anderen unzähligen Aktivitäten wurden ebenso übergangen, wie sie auch heute noch, zehn Jahre nach de Broglies Ableben, namentlich in Frankreich totgeschwiegen werden. Wäre er vor 1950, auf dem Höhepunkt seiner Karriere, verschieden, so hätte man ihm wohl einen unbestrittenen Patz in der französischen Geschichte eingeräumt. De Broglie hatte jedoch die Forschung seiner letzten aktiven Jahre in eine Richtung orientiert, die ihm von vielen seiner Kollegen als Verrat an der französischen Physik angelastet wurde. Dies trübt nach wie vor die Erinnerung an einen Mann, der in seinem öffentlichkeitsscheuen, wenig medienwirksamen Auftreten zeitlebens eine umstrittene, dem breiten Publikum eher unbekannte Persönlichkeit geblieben war.

## IV.3 Die Debatte um die Quantenmechanik

Als eine Art Würdigung der zu Lebzeiten de Broglies nicht mit Erfolg gekrönten Versuche einer neuen Interpretation der Quantenmechanik möchte ich zum Abschluß dieser Arbeit auf die heute wieder aktuelle Debatte um die Quantenmechanik eingehen. Dies erscheint umso angebrachter, als diese durch de Broglie und Bohm Anfang der 50er Jahre erneut entfachte Kontroverse gerade in den letzten Jahren neue Impulse erhalten hat. Da die im Weiteren angesprochenen Forschungsrichtungen teilweise außerordentlich komplex sind und ihre ausführliche Darstellung den Rahmen dieser Arbeit sprengen würde, werde ich nur die wichtigsten Aspekte andeuten und auf weitergehende Literatur verweisen.

Niemand bestreitet heutzutage ernsthaft die Leistungsfähigkeit der Quantenmechanik, die durch Übereinstimmung von Experiment und den durch die Theorie vorhergesagten Resultaten bestens bestätigt ist: Kein anderer Formalismus hat in der Physik so exakte Ergebnisse geliefert wie die Quantenmechanik. Auch de Broglie hat dies nie bezweifelt. Während aber der mathematische Formalismus unbestritten richtig ist, bleibt die Diskussion um seine Interpretation noch immer aktuell. Schon Einstein erdachte, ebenso wie Schrödinger und de Broglie, eine Vielzahl von Gedankenexperimenten,[139] die unweigerlich zu Paradoxa führten oder der Relativitätstheorie widersprachen. Die Unfähigkeit der Kopenhagener Interpretation, anschauliche, klare und logische Bilder zu liefern sowie die Abhängigkeit von sprachlichen Konventionen boten Kritikern ausreichend Angriffspunkte. Die neue Debatte um die Quantenmechanik findet sowohl auf einer wissenschaftstheoretischen und als auch auf einer physikalischen Ebene statt. In aller Kürze sollen die wichtigsten Diskussionsbeiträge beider Richtungen resümiert werden.

---

[139] So etwa sein berühmtes mit Rosen und Podolsky 1935 veröffentlichtes Experiment [308].

Die wissenschaftstheoretische Diskussion um die Kopenhagener Quantenmechanik ist insofern in diesem Kontext interessant, als sie die physikalischen Bemühungen um eine neue Interpretation zusätzlich motiviert und anregt. Hat doch gerade die wissenschaftsgeschichtliche Forschung in den letzten Jahrzehnten die Zeit- und Milieuabhängigkeit der Entstehung wissenschaftlicher Theorien gezeigt und damit auch die Kritik an aktuellen Paradigmen der Gegenwart möglich gemacht.

Schon bei de Broglie findet sich ein Beitrag über den Rhythmus des wissenschaftlichen Fortschritts [188, S. 362ff], in dem er die wichtigsten Aspekte des Kuhnschen Paradigmabegriffes vorwegnimmt. T. S. Kuhn stellte Ende der 60er Jahre fest, daß wissenschaftliche Arbeit immer innerhalb eines disziplinären Systems, unter einem Paradigma, d.h. mit stillschweigend anerkannten experimentellen und theoretischen Standards stattfindet. Das Paradigma selbst ist solange nicht Gegenstand der Forschung, gilt als wahr und vollständig, bis verstärkt Phänomene auftreten, die innerhalb des Paradigmas nicht mehr erklärbar sind; dann kommt es in deren Folge zur Krise, zur wissenschaftlichen Revolution, zum Paradigmenwechsel, und mit der Etablierung des neuen Paradigmas wiederum zu einer Phase der Normalwissenschaft. Diese in der heutigen wissenschaftstheoretischen Diskussion weitgehend anerkannte Lehrmeinung läßt sich an der Geschichte der Naturwissenschaften belegen (vgl. etwa [302]).

Bezieht man es auf die Kopenhagener Quantenmechanik, so liefert das Konzept Kuhns der modernen Wissenschaft erstmals eine epistemologische Grundlage, sich selbst als Teil der Geschichte zu verstehen. Begreift man die Paradigmen früherer Jahrhunderte als inzwischen überholte Etappen auf dem Weg des wissenschaftlichen Fortschritts, wird man auch das eigene, gegenwärtig aktuelle Paradigma in seiner geschichtlichen Eingebundenheit verstehen. Dies könnte zu der Einsicht führen, daß auch die Kopenhagener Deutung als eine unvollständige historische Zwischenstufe auf dem Weg zu einer zukünftigen physikalischen Theorie anzusehen ist. Eben diese Einsicht würde neue revolutionierende Forschung außerhalb der durch das geltende Paradigma gesetzten Grenzen fördern. Diese Argumentation entspricht genau de Broglies Ansicht über den Nutzen und Sinn der Wissenschaftsgeschichtsschreibung:

> L'histoire des sciences nous montre la science en progrès constant, remaniant et révisant sans cesse les connaissances acquises et leur interprétation, elle nous montre le passé, malgré bien des insuffisances, préparant le présent. Mais nous ne devons jamais oublier que notre science n'est aussi qu'un stade provisoire du progrès scientifique, plein lui-même, sans aucun doute, d'insuffisances et d'erreurs et dont le rôle, de ce point de vue, est surtout de préparer l'avenir. C'est une erreur très grande et très facile à commettre, de croire que les conceptions actuelles de la science sont définitives.[140]

---

[140] Übersetzung: „Die Wissenschaftsgeschichte zeigt uns die Wissenschaft in konstantem Fortschritt, indem sie die erworbenen Kenntnisse und deren Interpretation ständig überarbeitet und überprüft, sie führt uns vor Augen, wie die Vergangenheit trotz ihrer Unzulänglichkeiten die Gegenwart vorbereitet. Aber wir sollten niemals vergessen, daß unsere Wissenschaft auch nur ein provisorisches Stadium des wissenschaftlichen Fortschritts ist, das seinerseits voller Schwächen und Fehler ist und dessen Rolle von dahergesehen vor allem darin besteht, die Zukunft vorzubereiten. Es ist ein großer und leicht zu begehender Fehler, zu glauben, daß die aktuellen wissenschaftlichen Konzepte definitiv seien.", [292].

Dieser von de Broglie angedeutete Fehler, das aktuelle Paradigma als das letzte und endgültige zu betrachten, findet sich sehr häufig in der Wissenschaftsgeschichte. Auch war die Paradigmenbildung in den letzten Jahrhunderten nie ganz frei von Weltanschauung und außerwissenschaftlichem Einfluß.[141] In der wissenschaftstheoretischen Diskussion der letzten Jahrzehnte scheint sich nun herauskristallisiert zu haben, daß auch bei der Entstehung und Durchsetzung der Kopenhagener Deutung neben den physikalischen und formalen Sachzwängen auch wissenschaftsexterne Faktoren eine Rolle gespielt haben.[142]

Bereits in den 20er und 30er Jahren hatten sich zahlreiche Wissenschaftler und Wissenschaftstheoretiker zur Entstehung wissenschaftlicher Theorien und zum Einfluß des sozialen und kulturellen Umfeldes auf die Ausrichtung der Arbeit eines Wissenschaftlers geäußert. So existiert z.B. von Schrödinger ein Aufsatz [317], in dem er auf den großen Einfluß hinweist, den gesellschaftliche Faktoren auf das Weltbild des Physikers und damit auf seine Forschung haben. Der Amerikaner Forman veröffentlichte 1971 einen viel diskutierten Ansatz [309], dessen Grundaussagen hier nur kurz angedeutet werden können. In seinen provokanten Thesen weist Forman einen Zusammenhang zwischen den Geistesströmungen in Deutschland zwischen 1919-1933 und den Grundkonzepten der Quantenmechanik nach. Gerade die Aufgabe der strengen Kausalität stellt Forman in den Kontext eines kulturpessimistischen geistigen Umfelds. Wenngleich seine Thesen in der dargestellten Form oft allzu zugespitzt erscheinen und trotz eines vielfach unsauberen Umgangs mit dem Begriff der Kausalität,[143] gelingt es Forman, den nicht zu vernachlässigenden Einfluß des geistig-kulturellen Milieus auf die Entstehung und Durchsetzung der Kopenhagener Deutung nachzuweisen.

Einen weiteren Hinweis auf externe Faktoren liefert uns die Untersuchung der Prozesse, die zur Durchsetzung der Kopenhagener Deutung geführt haben. Die glaubenskriegartige Auseinandersetzung, der Ehrgeiz aller Forscher, ihrem Weltbild zum Durchbruch zu verhelfen, die Attitüde der Wissenschaftler um Bohr mit ihrer absoluten, ja autoritären Sprache, die bestimmte Fragestellungen verbot, - all dies scheint die These zu stützen, daß neben wissenschaftsinternen Faktoren auch Ehrgeiz, Durchsetzungvermögen und Autorität zur Etablierung der Kopenhagener Quantenmechanik geführt haben.[144]

Natürlich ist Vorsicht geboten in der Anwendung der externalistischen Perspektiven auf die Physik. Bei einer derart gut fundierten Wissenschaft, deren methodische Grundlage die Überprüfung der Theorien durch experimentelle Ergebnisse ist, scheint der Spielraum gesellschaftlicher Einflüsse deutlich eingeschränkt zu sein. Obwohl öffentliches Interesse, Finanzierung und persönlicher Ehrgeiz zwar die Entdeckung neuer Phänomene vorantreiben oder bremsen können, bleibt das Experiment und sein Ergebnis selbst doch davon unbeeinflußt. Wie die laufende wissenschaftstheoretische Diskussion zeigt, kann einzig die Entstehung und Akzeptanz neuer Theoien wissenschaftsexternen Faktoren unterliegen, denn:

> Die gleichen experimentellen Daten können mit verschiedenen theoretischen Konzepten interpretiert werden.[145]

---

[141] Berühmtestes Beispiel ist sicherlich die Einflußnahme der mächtigen Kirche auf die Gestaltung des physikalischen Weltbildes in früheren Jahrhunderten.

[142] Zu diesem Themenkomplex sei auf [312] hingewiesen. Hier findet sich auch eine ausführliche Literaturliste (S. 347-405) zu den neueren wissenschaftstheoretischen Arbeiten und zur Entstehung der Kopenhagener Deutung, zum Kulturmilieu der Weimarer Republik.

[143] Vgl. den Aufsatz *Weimarer Kultur und Quantenkausalität*, in dem der Autor John Henry den Ansatz und das Vorgehen Formans kritisiert, abgedruckt in [312, S. 201ff].

[144] Vgl. die Fußnoten 97 und 99.

[145] W. Kuhn in [287, S. 129].

Die Entscheidung darüber, welches Konzept sich allgemein durchsetzt, scheint nicht die Wissenschaft allein zu fällen: Auch Geistesströmungen, kulturelles Umfeld, Durchsetzungsvermögen und Ehrgeiz können sie beeinflussen. Eben darum kreist die neue Debatte um die Quantenmechanik: die Neubewertung der Interpretation der allgemein als richtig geltenden experimentellen Daten. Die neueren Forschungsaktivitäten der Wissenschaftstheorie scheinen nahe zu legen, daß die Kopenhagener Deutung nicht die unumstößlich letzte mögliche Interpretation der Phänomene in den atomaren Größenordnungen sein muß, sondern daß sie diejenige Theorie war, die am ehesten dem damaligen Zeitgeist entgegenkam und sich somit am leichtesten durchsetzen konnte. Mit diesen Untersuchungen zeigt die Wissenschaftheorie die Sinnhaftigkeit der aktuellen physikalischen Bemühungen um eine neue Interpretation.

Diese Bemühungen haben spätestens in den 50er Jahren neue Bedeutung gewonnen und gerade in den letzten Jahren haben sich grundlegend neue Möglichkeiten eröffnet, wobei es sehr unterschiedliche Ansätze gibt: die Viele-Welten-Theorie, die modale Interpretation, die statistische Interpretation [316, S. 1]. Zwei sehr unterschiedliche Theorien seien hier noch exemplarisch erwähnt.

Ein eher anschaulich-physikalischer Ansatz beschäftigt sich mit der Vollständigkeit der Quantenmechanik und ist wesentlich in der Tradition de Broglies, Bohms (z.B. [306]) und des Engländers John Bell[146] zu sehen. Die zahlreichen Doktoranden und früheren Mitarbeiter de Broglies haben schon seit den frühen 50er Jahren auch unabhängig von ihm, jedoch durch ihn beeinflußt, gearbeitet.[147] Gerade unter der Schirmherrschaft der *Fondation Louis de Broglie* ist jedoch in den letzten zwanzig Jahren eine rege Forschungstätigkeit entstanden. Olivier Costa de Beauregard, Jean-Pierre Vigier, Georges Lochak u.a. entwickeln Experimente und Konzepte, die darauf abzielen, den Welle-Teilchen-Dualismus im Sinne de Broglies und im Gegensatz zum Komplementaritätprinzip als objektive physikalische Realität nachzuweisen.[148] Unbefriedigt durch die Paradoxa, die aus der Kopenhagener Deutung folgen, sucht man unter der als statistisch angesehenen Oberfläche der Schrödingerwelle nach verborgenen Parametern, die eine objektive, deterministische und logische Beschreibung der durch die Quantenmechanik beschriebenen Effekte erlauben sollen.[149]

Ein anderer, weitaus modernerer Weg ist der als „consistent histories" bekannt gewordene Ansatz, den namentlich R. Omnès,[150] R. B. Griffiths[151] und C. J. Isham[152] erdacht haben. Auch innerhalb dieser Theorie sucht man nach einer objektiven Beschreibung realer Phänomene im atomaren Bereich, die unabhängig von einem Beobachter existieren. Die Theorie steht der nichtrelativistischen Hilbertraum-Quantenmechanik sehr nahe: Sie übernimmt in weiten Teilen den mathematischen

---

[146] Für Verweise auf die Veröffentlichungen Bohms und Bells sei auf die Literaturverzeichnisse in [304, 320] verwiesen.

[147] An dieser Stelle auch nur ansatzweise einen Überblick über die Veröffentlichungen der Mitarbeiter de Broglies geben zu wollen würde den Rahmen dieser Arbeit sprengen; deshalb sei nur auf [290, 293] verwiesen, worin sich zahlreiche weiterführende Literaturhinweise finden lassen.

[148] Hierzu sei auf die wichtigen Veröffentlichungen der Fondation zum 90. und 100. Geburtstag de Broglies verwiesen: [290, 293].

[149] Zu diesem Themenkomplex sei auf [304, 320]. In diesen zwei Bänden werden die wichtigsten Argumente für und gegen verborgene Parameter genannt sowie Experimente dargestellt und vorgeschlagen. Zudem sind darin Originalschriften von Einstein, Heisenberg, Bohr, Schrödinger, Bohm, Bell u.a zu diesem Thema abgedruckt.

[150] Vgl. etwa [313].

[151] Vgl. etwa [310].

[152] Für eine tiefergehende Diskussion dieses Ansatzes und ausführliche Literaturverweise sei verwiesen auf [316].

Formalismus und zahlreiche ihrer Axiome, wie etwa den Indeterminismus. Eine wesentliche Neuerung ist jedoch der Geschichtenbegriff. Eine Geschichte im einfachsten Fall ist in der nichtrelativistischen Quantenmechanik die chronologische Abfolge von Ereignissen. Sind Wahrscheinlichkeiten in der orthodoxen Quantenmechanik an Ereignisse zu bestimmten Zeitpunkten gebunden, so werden sie in der neuen Theorie auf ganze Geschichten bezogen. Ganz wesentlich unterscheidet sich der „consistent histories"-Ansatz von der Kopenhagener Quantenmechanik durch den Umgang mit der klassischen Physik. Wie O. Rudolph ausführt, ist die orthodoxe Quantenmechanik auf die makroskopische Beschreibung der klassischen Physik genauso angewiesen, wie die klassische Physik auf die mikroskopische Beschreibung der Quantenmechanik. Erst beide Konzepte ergänzen sich zu einer vollständigen Beschreibung eines physikalischen Systems als Ganzes. Aus der Anwendung klassischer Bilder jedoch auf mikroskopische Effekte, der Vermischung der von einander unabhängigen Bereiche also, entstehen schwerwiegende logische Paradoxa, wie sie etwa von Einstein immer wieder angeführt wurden. Diese versucht namentlich R. Omnès zu überwinden, indem er die klassische Physik in ihren dynamischen und logischen Aspekten aus der Quantenmechanik herleitet. Dieser vielversprechende, mathemathisch und philosophisch komplexe Ansatz steckt jedoch, wie es O. Rudoph ausdrückt, noch in den Kinderschuhen und bedarf weiterer wesentlicher Ausarbeitung.

Trotz der bereits erzielten Erfolge ist die Suche nach einer neuen Interpretation der Quantenmechanik noch immer das Betätigungsfeld nur weniger Wissenschaftler. Es besteht jedoch die Hoffnung, daß ein breiteres Interesse die Entwicklung erheblich beschleunigen könnte.

# Fazit

Louis de Broglie hat ein ruhiges, fast mönchisches Leben geführt, das vollständig der Wissenschaft geweiht war. Mit allen Traditionen seiner adligen Herkunft brechend, aber vom Bruder unterstützt, suchte er am Anfang seiner wissenschaftlichen Laufbahn in einer Fachrichtung dem berühmten Familiennamen gerecht zu werden, die sehr wenig den Idealen seines gesellschaftlichen Umfeldes entsprach. Tatsächlich erreichte de Broglie in der Physik ähnlich große gesellschaftliche Erfolge, wie sie bereits den Ruhm seiner Vorfahren, der Feldherren, Staatsmänner und Politiker ausgemacht hatten. De Broglie stellt damit gewissenmaßen ein Bindeglied zwischen Vergangenheit und Gegenwart dar: Seine Lebensweise, sein Auftreten, seine Umgangsformen ließen ihn als hochadligen Vertreter einer vergangenen Zeit erscheinen, sein Interesse für die moderne Physik, für deren Weiterentwicklung und Anwendungen, wies ihn hingegen als modernen Menschen des 20. Jahrhunderts aus. Eben dieser Zwiespalt macht de Broglie zu einer der interessantesten und unverständlichsten Physikerpersönlichkeiten unserer Zeit.

Denn nicht nur de Broglies Persönlichkeit wurde durch diesen scheinbaren Widerspruch charakterisiert, auch sein Weltbild, seine Konzeption der Physik bekamen dadurch ihre besondere Ausrichtung. Er war modern in seinem Streben, die Welt der Quanten zu verstehen, um sie durch neue Konzepte faßbar zu machen, in seinem Glauben an eine physikalische Realität und deren Beschreibung in Zeit und Raum blieb er jedoch den klassischen Idealen verwachsen und suchte wissenschaftlichen Fortschritt auf der Grundlage der klassischen Theorien, nicht im Bruch mit ihnen.

Dieser Glaube an die Gültigkeit scheinbar vergangener Ideale und die gleichzeitige Suche nach revolutionär Neuem haben de Broglie zeitlebens zum Außenseiter gemacht, gesellschaftlich ebenso wie wissenschaftlich. Die ungewöhnliche Doktorarbeit wurde 1924 zwar als „intelligent und interessant" akzeptiert, de Broglies Leistung wurde in ihrer Modernität aber erst voll anerkannt, als man die Weiterentwicklung seiner Theorie, das „Führungswellenkonzept", bereits als unspektakulär und zu klassisch abtat. Fünfundzwanzig Jahre später nahm de Broglie die Suche nach einer kausalen und deterministischen Interpretation der Effekte im atomaren Bereich wieder auf, zu einer Zeit also, als die moderne Physik ungeahnte Erfolge feierte und die *scientific community* keine Veranlassung für einen Paradigmenwechsel sah.

Das Paradigma der Quantenmechanik in der Kopenhagener Deutung, in seiner unglaublichen Anwendbarkeit, gilt heute seit nunmehr siebzig Jahren und hat mit seinen zahlreichen Anwendungen und Weiterentwicklungen in bisher unbekanntem Ausmaß den Fortschritt vorangetrieben. De Broglie selbst hat jedoch in diesem Rahmen wenig zu den erreichten Erfolgen beigetragen. Die Jahre seiner Anpassung an die gängige Interpretation waren, vielleicht mit Ausnahme seiner Arbeiten zur Lichttheorie, eher unproduktiv und zeichnen sich durch eine Suche nach angemessenen Betätigungsfeldern aus. De Broglie, in seinem milieubedingten Außenseitertum, konnte sich nicht wirklich den ihm fremden Denkweisen der Kopenhagener Quantenmechanik anpassen, und vielleicht widerstebte es ihm auch, sich vollständig der Theorie unterzuordnen, die als am Ende triumphierender Konkurrent gegen seine eigene Theorie angetreten war.

Zudem war de Broglie kein Normalwissenschaftler im Kuhnschen Sinne, der im Rahmen eines bestehenden Paradigmas arbeitet. Für seine eigene Forschung akzeptierte er nicht, daß erst Phasen der Normalwissenschaft die Erkenntnisse und Arbeitsweisen eines Paradigmas für den Menschen nutzbar machen und diese damit für den Fortschritt ähnlich wichtig sind wie die Aufstellung des Paradigmas selbst. Er selbst sah sich als Vordenker, als Visionär, der in großen Linien dachte und dessen Ideen die Grundlage für weitergehende Forschungsbereiche der Physik legten. Somit war die Rückkehr zu seinen früheren Ideen Anfang der fünfziger Jahre nur allzu natürlich. Mit seinen Theorien von der „doppelten Lösung" und der „verborgenen Thermodynamik" konnte de Broglie in kreativer Weise Schwachpunkte der Kopenhagener Deutung aufzeigen, offene Fragen bloßlegen und in gewisser Weise scheint ihm die neuere Debatte um die Interpretation der Quantenmechanik in seiner Skepsis recht zu geben.

Tatsächlich erfüllte de Broglie mit diesem Infragestellen der Grundlagen der modernen Physik die wichtige Aufgabe einer Kontrollinstanz. Denn nach Popper ist jeder Versuch, eine etablierte Theorie zu falsifizieren, entscheidender Bestandteil wissenschaftlicher Arbeit.[153] Wie die Wissenschaftsgeschichte aufzeigt, kann gerade die Überwindung der Grenzen eines alten Paradigmas die Forschung entscheidend voranbringen. Und diese Erkenntnis sollte zu einem bewußten und konstruktiven Umgang mit Querdenkern ermutigen: Denn jede Epoche braucht geniale Denker, die den Mut und die Kompetenz haben, die Grundlagen der wissenschaftlichen Forschung zu überprüfen und zu hinterfragen. Aber auch heute noch werden diese Denker nicht als Teil der *scientific community* angesehen, sondern zu Außenseitern gemacht. In diesem Sinne kann de Broglie als Vorbild für jeden Forschen gelten, denn er hatte 1951, nach Jahren des Schweigens und der Anpassung, den Mut zum Zweifeln und Nachfragen, den unkonventionellen Geist wiedergefunden, der ihn in den 20er Jahren zur Entdeckung der Materiewelle geführt hatte. Wohl mit Recht kann man sagen, daß Louis de Broglie einer der bedeutenden Physiker dieses Jahrhunderts war, dessen Leben und Werk auch heute noch alle Beachtung verdient.

---

[153] Zitiert in [289, S. 1433].

# Bibliographie

## Veröffentlichungen de Broglies[154]

---

[154] Nach [183].

Wissenschaftsphilosophische Werke

[184]  *La physique nouvelle et les quanta, Bibliothèque de philosophie scientifique*, (geleitet von Paul GAULTIER), Flammarion, Paris 1937, (italienische Übersetzung).

[185]  *Matière et lumière*, Albin-Michel, Paris 1937, *Sciences d'aujourd'hui*, (geleitet von André GEORGE) (deutsche, englische, amerikanische, italienische, japaische, spanische und holländische Übersetzungen).

[186]  *Continu et discontinu en physique moderne*, Albin-Michel, Paris 1941, *Sciences d'aujourd'hui*, (geleitet von André GEORGE), (deutsche, holländische, italienische Übersetzungen).

[187]  *Physique et microphysique*, Albin-Michel, Paris 1947, *Sciences d'aujourd'hui*, (geleitet von André GEORGE) (deutsche, italienische, spanische Übersetzungen).

[188]  *Savants et découvertes*, Albin-Michel, Paris 1951, *Les savants et le monde*, (geleitet von André GEORGE), (spanische Übersetzung).

[189]  *Nouvelles perspectives en microphysique*, Albin-Michel, Paris 1956, *Sciences d'aujourd'hui*, (geleitet von André GEORGE), (englische Übersetzung, Basic Books, New York 1962).

[190]  *Sur les sentiers de la science*, Albin-Michel, Paris 1960, (italienische Übersetzung, Paolo Boringhieri, Turin 1962).

[191]  *Certitudes et incertitudes* de *la science*, Albin-Michel, Paris 1966.

[192]  *Recherches d'un demi-siècle*, Albin-Michel, Paris 1976.

Akademische Vorträge und Notizen

[193]  La vie et l'oeuvre d'Emile Picard lue à la séance publique du 21 décembre 1942.

[194]  La vie et l'oeuvre d'André Blondel lue à la séance publique du 18 décembre 1944.

[195]  Discours de réception à l'Académie française prononcé sous la Coupole le 31 mai l945 (édition de luxe, par Albin-Michel).

[196]  La réalité des molécules et l'oeuvre de Jean Perrin lue à la séance publique du 17 décembre 1945.

[197]  Rapports sur les prix vertus lue à la séance publique du 10 janvier 1946.

[198]  La vie et l'oeuvre de Charles Fabry lue à la séance publique du 16 décembre 1946.

[199]  La vie et l'oeuvre de Paul Langevin lue à la séance publique du 15 décembre 1947.

[200]  La physique contemporaine et l'oeuvre d'Albert Einstein lue à la séance publique du 19 décembre 1949.

[201] La vie et l'oeuvre de Hendrik Antoon Lorentz lue à la séance publique du 10 décembre 1951.

[202] La vie et l'oeuvre d'Aimé Cotton lue à la séance publique du 14 décembre 1953.

[203] Le dualisme des ondes et des corpuscules dans l'quevre d'Albert Einstein lue à la séance publique du 5 décembre 1955.

[204] Notice sur la vie et l'ouevre d'Émil Borel lue à la séance publique du 9 décembre 1957.

[205] Notice sur la vie et l'ouevre de Frédéric Joliot lue à la séance publique du 14 décembre 1960.

[206] Notice sur la vie et l'ouevre de Georges Darmois lue à la séance publique du 9 décembre 1962.

[207] Notice sur la vie et l'ouevre de Jean Bequerel lue à la séance publique du 9 décembre 1964.

[208] Notice sur la vie et l'ouevre de Camille Gutton lue à la séance publique du 11 décembre 1965.

[209] Notice sur la vie et l'ouevre d'Albert Pérard lue à la séance publique du 11 décembre 1967.

[210] Notice sur la vie et l'ouevre de Bernard Lyot lue à la séance publique du 8 décembre 1969.

[211] Notice sur la vie et l'ouevre d'André Danjon lue à la séance publique du 13 décembre 1971.

Vorträge und allgemeine Artikel

[212] La théorie des quanta, synthèse de la dynamique et de l'opitique, *Revue générale des Sciences*, 35. Jahrgang, Nr. 22, 1925, S. 629.

[213] Deux conceptions adverses de la lumière et leur synthès possible, *Scientia*, September 1926, S. 128.

[214] La physique moderne et l'oeuvre de Fresnel, *Revue de Métaphysique et de Morale* XXXIV, Nr. 4 1927 S. 421.

[215] L'oeuvre de Fresnel et l'évolution actuelle de la physique, (Vortrag anläßlich des hundersten Geburtstages von Fresnel am 29 Oktober 1927 gehalten), *Revue d'Optique théorique et expérimentale* 6, 1927, S. 493.

[216] Continuité et individualité dans la physique moderne, *Cahier de la nouvelle journée*, Nr. 15, continu et discontinu, S. 60.

[217] La crise récente de l'optique ondulatoire, (Vortrag gehalten im *Conservatoire des Arts et Métiers* am 17. April 1929), *Revue Scientifique*, 67. Jahrgang, Nr. 12, 1929, S. 353.

# Werke anderer Autoren

## Zur Biographie

## Zur Materiewelle

# Anhang

## A.1 Auschnitte der Doktorarbeit in deutscher Übersetzung

A.1.1 Kapitel I.1 der Doktorarbeit. Übersetzung: G. Bodenseher.
A.1.2 Kapitel II.5 der Doktorarbeit. Übersetzung: G. Bodenseher.

## A.2 Korrespondenz Louis de Broglie - Albert Einstein (1929-1954)

A.2.1   de Broglie an Einstein am 14. November 1929.
A.2.2   de Broglie an Einstein am 29. Januar 1929.
A.2.3   de Broglie an Einstein am 21. Juni 1951.
A.2.4   Einstein an de Broglie am 2. April 1953.
A.2.5   de Broglie an Einstein am 9. April 1953.
A.2.6   Einstein an de Broglie am 14. April 1953.
A.2.7   Einstein an de Broglie im Mai 1953.
A.2.8   Einstein an de Broglie am 8. Februar 1954.
A.2.9   de Broglie an Einstein am 8. März 1954.

## A.1.1 Kapitel I.1 der Doktorarbeit. Übersetzung: G. Bodenseher.

Kapitel I [150a, S. 31-37], Die Phasenwelle
1. Die Quantenbeziehung und die Relativitätstheorie

Eines der wichtigsten neuen Konzepte, die durch die Relativitätstheorie eingeführt worden sind, ist das von der Trägheit der Energie. Nach Einstein muß die Energie so betrachtet werden, als hätte sie Masse, und jede Masse stellt Energie dar. Masse und Energie sind immer miteinander durch die allgemeine Relation verknüpft:

<div align="center">Energie = Masse $c^2$</div>

worin c eine Konstante ist, die als „Lichtgeschwindigkeit" bezeichnet wird, für die wir aber die Bezeichnung „Grenzgeschwindigkeit der Energie" vorziehen. Die Gründe dafür werden wir später erläutern. Da immer Proportionalität zwischen Masse und Energie besteht, darf man Materie und Energie als zwei synonyme Ausdrücke ansehen, die die gleiche physikalische Realität bezeichnen.

Zuerst die Theorie des Atoms, sodann die Theorie des Elektrons haben uns gelehrt, Materie als ihrem Wesen nach diskontinuierlich zu sehen und dies führt uns dazu, anzunehmen, daß alle Formen von Energie, im Gegensatz zu den alten Vorstellungen über das Licht, wenn nicht gänzlich in kleinen Raumgebieten konzentriert, so doch mindestens um gewisse singuläre Punkte verdichtet sind.

Das Prinzip von der Trägheit der Energie ordnet einem Körper, dessen Ruhemasse (d.h. die von einem mit ihr verbundenen Beobachter gemessen ist) $m_0$ ist, eine Ruheenergie $m_0 c^2$ zu. Wenn sich der Körper in gleichförmiger Bewegung mit einer Geschwindigkeit $v = \beta c$ relativ zu einem Beobachter befindet, den wir der Einfachheit halber als festen Beobachter bezeichnen wollen, so hat seine Masse für ihn den Wert $\frac{m_0}{\sqrt{1-\beta^2}}$, in Übereinstimmung mit einem wohlbekannten Resultat aus der relativistischen Dynamik. Folglich ist seine Energie $\frac{m_0 c^2}{\sqrt{1-\beta^2}}$. Da die kinetische Energie eines Körpers als die Zunahme definiert werden kann, die die Energie eines Körpers für den fixen Beobachter erfährt, wenn dieser Körper vom Ruhezustand in eine Bewegung mit der Geschwindigkeit $v = \beta c$ übergeht, findet man für ihren Wert:

$$E_{kin} = \frac{m_0 c^2}{\sqrt{1-\beta^2}} - m_0 c^2 = m_0 c^2 \left( \frac{1}{\sqrt{1-\beta^2}} - 1 \right)$$

Für niedrige Werte von $\beta$ führt dies zur klassischen Form:

$$E_{kin} = \frac{1}{2} m_0 v^2$$

Nachdem wir das in Erinnerung gerufen haben, wollen wir untersuchen, in welcher Form wir Quanten in die relativistische Dynamik einführen können. Es scheint uns die grundlegende Idee der Quantentheorie zu sein, daß es unmöglich ist, eine isolierte Quantität an Energie ins Auge zu fassen, ohne ihr eine bestimmte Frequenz zuzuordnen. Diese Verbindung drückt sich durch die von mir als Quantenbeziehung genannte Relation aus:

<div align="center">Energie= $h$ * Frequenz</div>

$h$ Plancksches Wirkungsquantum. Die fortschreitende Entwicklung der Quantentheorie hat mehrfach die Bedeutung der mechanischen Wirkung hervorgehoben, und man hat des öfteren versucht, die Quantenbeziehung mit Hilfe der Wirkung anstatt der Energie zu formulieren. Gewiß, die Konstante $h$ besitzt die Dimension einer Wirkung, nämlich $ML^2T^{-1}$, und das ist kein Zufall, da uns die Relativitätstheorie lehrt, daß die Wirkung zu den fundamentalen „Invarianten" der Physik gehört. Aber die Wirkung ist eine sehr abstrakte Größe, und in der Folge zahlreicher Überlegungen zu den Lichtquanten und dem photoelektrischen Effekt sind wir dazu geführt worden, die Energie als Basis zu nehmen und sodann zu untersuchen, warum die Wirkung in zahlreichen Fragestellungen so eine große Rolle spielt.

Die Quantenbeziehung hätte ohne Zweifel nicht viel Sinn, wenn Energie auf kontinuierliche Weise im Raum verteilt sein könnte, aber wir haben gesehen, daß dem sicherlich nicht so ist. Man kann also annehmen, daß als Folge eines großen Gesetzes der Natur jedem Energiepaket der Masse $m_0$ ein periodisches Phänomen zugeordnet ist, so daß gilt:

$$h\nu_0 = m_0 c^2$$

wobei wohlgemerkt $\nu_0$ im Eigensystem des Energiequants gemessen wird. Diese Hypothese ist die Grundlage unseres Systems: Ihr Wert liegt, so wie bei allem Hypothesen, im Wert der Konsequenzen, die man daraus ableiten kann.

Dürfen wir das periodische Phänomen als im *Inneren* des Energiequants lokalisiert annehemen? Dies ist keinesfalls notwendig, und es wird sich im Paragraph III ergeben, daß es zweifelsohne in einem ausgedehnten Raumbereich vorhanden ist. Was ist im übrigen unter dem Inneren eines Energiequants zu verstehen? Das Elektron, welches wir vielleicht zu Unrecht am besten zu kennen glauben, stellt für uns den Prototyp eines isolierten Energiepakets dar. Nun ist aber nach überkommenen Vorstellungen die Energie des Elektrons im ganzen Raum verteilt, mit einer sehr starken Konzentration in einem Gebiet sehr kleiner Dimension, dessen Eigenschaften uns nahezu unbekannt sind. Das, was das Elektron als unteilbare Energiepartie charakterisiert, ist nicht das kleine Volumen, das es im Raum einnimmt, - ich wiederhole, es erfüllt ihn gänzlich - sondern die Tatsache, daß es eine *Einheit* ist.

Nachdem wir die Existenz einer mit dem Energiequant verknüpften Frequenz angenommen haben, wollen wir untersuchen, wie sie sich für den vorhin eingeführten fixen Beobachter manifestiert. Die Lorentztransformation für die Zeit ergibt, daß ein mit einem bewegten Körper verbundenes periodisches Phänomen für einen festen Beobachter im Verhältnis von 1 zu $\sqrt{1-\beta^2}$ verzögert erscheint, dies ist die berühmte Verlangsamung der Uhren. Die vom fixen Beobachter festgestellte Frequenz wird sein:

$$\nu_1 = \nu_0 \sqrt{1-\beta^2} = \frac{m_0 c^2}{h} \sqrt{1-\beta^2}$$

Da die Energie des bewegten Mobils andererseits für den gleichen Beobachter $\dfrac{m_0 c^2}{\sqrt{1-\beta^2}}$ ist, ist die entsprechende Frequenz nach der Quantenbeziehung $\nu = \dfrac{1}{h} \dfrac{m_0 c^2}{\sqrt{1-\beta^2}}$. Die beiden Frequenzen $\nu_1$ und $\nu$ sind wesentlich verschieden, da der Faktor $\sqrt{1-\beta^2}$ nicht in derselben Art und Weise eingeht. Dies stellt eine Schwierigkeit dar, die mich lange Zeit beschäftigt hat; es ist mir gelungen, sie zu beheben, indem ich das nun folgende Theorem, welches ich das Theorem von der Harmonie der Phase nennen werde:

„Das periodische Phänomen, das mit dem bewegten Körper verknüpft ist, und dessen Frequenz für den fixen Beobachter $\nu_1 = \dfrac{1}{h} m_0 c^2 \sqrt{1-\beta^2}$ ist, erscheint für diesen immer in

Phase mit einer Welle der Frequenz $\nu = \dfrac{1}{h} m_0 c^2 \dfrac{1}{\sqrt{1-\beta^2}}$, die sich in derselben Richtung

wie das Teilchen fortpflanzt und die Geschwindigkeit $V = \dfrac{c}{\beta}$ besitzt."

Der Nachweis ist sehr einfach. Nehmen wir an, zur Zeit $t = 0$ bestünde Phasengleichheit zwischen dem periodischen Phänomen und der definierten Welle. Zur Zeit $t$ hat das bewegte Objekt seit dem Startzeitpunkt eine Distanz von $x = \beta c t$ zurückgelegt, und die

Phase des periodischen Phänomens hat sich um $\nu_1 t = \dfrac{m_0 c^2}{h} \sqrt{1-\beta^2} \dfrac{x}{\beta c}$ geändert. Die

Phase dieses Bereichs der Welle, die das bewegte Objekt überdeckt, hat sich geändert um den Betrag:

$$\nu\left(t - \frac{\beta x}{c}\right) = \frac{m_0 c^2}{h} \frac{1}{\sqrt{1-\beta^2}}\left(\frac{x}{\beta c} - \frac{\beta x}{c}\right) = \frac{m_0 c^2}{h} \sqrt{1-\beta^2}\ \frac{x}{\beta c}$$

Wie wir es vorausgesagt haben, bleibt die Übereinstimmung der Phase aufrecht.

Es ist möglich, von diesem Theorem eine andere Ableitung zu geben, die im Grunde genommen mit der ersten identisch ist, aber vielleicht etwas anschaulicher. Wenn $t_0$ die Zeit für einen mit dem bewegten Objekt verbundenen Betrachter ist, ergibt die Lorentztransformation:

$$t_0 = \frac{1}{\sqrt{1-\beta^2}}\left(t - \frac{\beta x}{c}\right)$$

Das periodische Phänomen, das wir uns vorstellen, wird für den Beobachter durch eine Sinusfunktion dargestellt von $\nu_0 t_0$ dargestellt. Für den festen Beobachter wird es durch die gleiche Sinusfunktion von $\nu_0 \dfrac{1}{\sqrt{1-\beta^2}}\left(t - \dfrac{\beta x}{c}\right)$ dargestellt, einer Funkton, die eine

Welle von der Frequenz $\dfrac{\nu_0}{\sqrt{1-\beta^2}}$ darstellt, die sich mit der Geschwindigkeit $\dfrac{c}{\beta}$ in der selben Richtung wie das Teilchen fortbewegt.

Es ist nun unerläßlich, über die Natur der Welle, deren Existenz wir angenommen haben,

nachzudenken. Die Tatsache, daß die Geschwindigkeit $V = \dfrac{c}{\beta}$ notwendigerweise größer als $c$ ist ($\beta$ ist immer $<1$, sonst wäre die Masse unendlich oder imagnär), zeigt uns, daß es sich nicht um eine Energie transportierende Welle handeln kann. Unser Theorem lehrt uns im übrigen, daß sie die Verteilung eines Phänomens im Phasenraum darstellt, es ist eine „Phasenwelle". [...]

Die vorhergehenden Resultate scheinen uns von höchster Wichtigkeit zu sein, weil sie mit Hilfe einer Hypothese, die durch den Begriff des Quants förmlich nahegelegt wird, eine Verknüpfung herstellen zwischen der Bewegung eines Massenpunktes und der Fortpflanzung einer Welle, und auf diese Weise lassen sie die Möglichkeit einer Synthese der antagonistischen Theorien über die Natur der Strahlung erkennen. Schon können wir feststellen, daß die geradlinige Ausbreitung der Phasenwelle mit der geradlinigen Bewegung eines Massenpunktes verknüpft ist. Das Fermtsche Prinzip angewendet auf die Phasenwellen bestimmt die Form von diesen Strahlen, die geradlinig sind, wohingegen das Maupertuische Prinzip angewandt auf den bewegten Massenpunkt seine geradlinige Bahn bestimmt, die einem der Strahlen der Welle entspricht. Im Kapitel II werden wir versuchen, diese Übereinstimmung zu verallgemeinern.

# A.1.2 Kapitel II.5 der Doktorarbeit. Übersetzung: G. Bodenseher.

## Kapitel II.5 [150a, 55f], Die Erweiterung der Quantenbeziehung

Wir sind am Kulminationspunkt dieses Kapitels angelangt. Seit Beginn des Kapitels haben wir die folgende Frage gestellt: „Wenn sich ein Massenpunkt in einem Kraftfeld ungleichförmig bewegt, wie breitet sich dann seine Phasenwelle aus?" Anstatt, so wie ich es zuerst gemacht habe, durch tastende Versuche die Fortpflanzungsgeschwindigkeit in jedem Punkt und für jede Richtung zu bestimmen, werde ich eine eine vielleicht etwas hypothetische Erweiterung der Quantenbeziehung vornehmen, deren tiefe Übereinstimmung mit dem Geist der Relativitätstheorie aber unwidersprochen ist.

Wir haben stets $h\nu = \omega$ gesetzt, wobei $\omega$ die Gesamtenergie des Massepunktes und $\nu$ die Frequenz seiner Phasenwelle ist. Andererseits hat uns der vorhergehenden Paragraphen gelehrt, zwei Vierervektoren $J$ und $O$ zu definieren, die vollkommen symmetrische Rollen bei der Untersuchung der Bewegung eines Massenpunktes und in der Fortpflanzung einer Welle spielen.

Unter Benutzung dieser beiden Vektoren schreibt sich die Beziehung $h\nu = \omega$:

$$O_4 = \frac{1}{h}J_4$$

Die Tatsache, daß zwei Vektoren eine gemeinsame Komponente haben, beweist nicht, daß das auch für die anderen Komponenten gilt. Dennoch setzten wir mit einer ganz naheliegenden Verallgemeinerung:

$$O_i = \frac{1}{h}J_i \qquad (i = 1, 2, 3, 4)$$

Die Änderung $d\varphi$ bezüglich eines infinitesimalen Elements der Phasenwelle hat den Wert:

$$d\varphi = 2\pi O_i dx^i = \frac{2\pi}{h}J_i dx^i.$$

Das Fermatsche Prinzip wird also zu:

$$\delta \int_A^B \sum_1^3 J_i dx^i = \delta \int_A^B \sum_1^3 p_i dx^i = 0$$

Wir gelangen also zu folgender Feststellung:

„Das Prinzip von Fermat angewendet auf die Phasenwelle ist identisch mit dem Prinzip von Maupertuis angewendet auf einen Massenpunkt; die dynamisch möglichen Teilchenbahnen sind mit den möglichen Wellenstrahlen identisch."

Wir meinen, daß diese Idee einer tieferen Beziehung zwischen den beiden großen Prinzipien der geometrischen Optik und der Dynamik ein wertvoller Leitgedanke sein könnte, um eine Synthese zwischen Wellen und Quanten zu verwirklichen.

Die Hypothese von der Proportionalität der Vektoren J und O ist eine Art Erweiterung der Quantenhypothese, deren gegenwärtige Form offensichtlich unzureichend ist, da sie nur die Energie beinhaltet und nicht den ihr untrennbar zugeordneten Impuls. Die neue Formulierung ist bei weitem zufriedenstellender, weil sie durch die Gleichheit zweier Vektoren ausgedrückt wird.

## A.2.1 de Broglie an Einstein am 14. November 1929.

Übersetzung:[155]

Lieber Herr Einstein
Ich danke Ihnen von ganzem Herzen für die Grüße, die Sie mir anläßlich des Nobelpreises haben zukommen lassen. Diese Glückwünsche waren von allen, die ich erhielt, diejenigen, die mich am meisten berührt haben, wegen der großen Bewunderung, die ich für Sie seit langem hege.
Vielleicht hat meine Idee der Elektronenwellen einen relativ wichtigen Fortschritt bedeutet, aber ich weiß, wieviel es noch zu tun gibt, um dies alles wirklich zu verstehen und sich der Wahrheit etwas zu nähern. Wir müssen viel arbeiten, aber das ist nicht traurig, denn beim Arbeiten hat man die größten Freuden.
Ich danke Ihnen nochmals, lieber Herr Einstein [...].

## A.2.2 de Broglie an Einstein am 29. Januar 1929.

Übersetzung:

Lieber Herr Einstein,
ich muß mich zunächst vielmals dafür entschuldigen, daß ich Sie bei Ihren Arbeiten störe. Aber meine Neugierde ist lebhaft geweckt worden, da die französischen Zeitungen angekündigt haben, daß Sie kürzlich eine Mitteilung über eine sehr wichtige neue Entdeckung veröffentlicht hätten. - Falls dies richtig ist, wäre ich Ihnen unendlich dankbar, wenn Sie mir ein Exemplar Ihrer Mitteilung zusenden könnten, denn ich glaube, daß sie in den Berichten der Berliner Akademie erschienen ist, und ich habe große Mühe, die Berichte in Paris zu bekommen.
Ich bin immer noch nicht sehr zufrieden mit dem aktuellen Zustand der Wellentheorie und ich glaube seit langem, daß Sie es sein werden, der uns den richtigen Weg aufzeigen wird, den es zu verfolgen gilt, um sie zu verbessern. Deshalb war ich sehr gepackt von der Neuigkeit, die ich in den Zeitungen las.
Ich habe erfahren, daß Sie in letzter Zeit krank waren und dehalb nicht kommen konnten, um an unserem neuen Institut Henri Poincaré Vorträge zu halten. Ich hoffe, daß Sie jetzt wieder vollständig gesund sind, und wir alle am Institut Henri Poincaré wünschen uns, daß Sie in einem anderen Jahr zu uns kommen können.
Ich entschuldige mich nochmals, Sie gestört zu haben [...].

---

[155] Alle Briefe de Broglies enden mit den üblichen, sehr langen französischen Höflichkeitsformen, die aufgrund ihres förmlichen Charakters inhaltlich uninteressant sind und somit nicht übersetzt werden.

## A.2.3 de Broglie an Einstein am 21. Juni 1951.

Übersetzung

Lieber Herr Einstein

Ich möchte Ihnen sehr dafür danken, daß Sie mir durch Frau François Ihre Photographie mit Widmung haben zukommen lassen. Diese Photographie, die wirklich sehr schön ist, wird für mich eine wertvolle Erinnerung sein. Außer der tiefen Bewunderung, die ich, wie alle Physiker, für Ihr Werk hege, das die gesamte Wissenschaft unserer Zeit erneuert hat, habe ich ich eine große persönliche Dankbarkeit bewahrt für das große Wohlwollen, welches Sie mir zu Beginn meiner wissenschaftlichen Laufbahn haben zuteil werden lassen und für die so wertvollen Ermutigungen, die Sie mir damals machten.

Es hat mich sehr gefreut zu erfahren, daß eine meiner Schülerinnen, Frau Tonnelat, die wirklich eine bemerkenswerte Person ist, mit Ihnen bezüglich ihrer Arbeiten über die unitären Theorien Kontakt hatte und daß Sie sich für ihre Resultate interessiert haben. [...]

## A.2.4 Einstein an de Broglie am 2. April 1953.

Dear de Broglie.

On my advice Mr. Bohm sent your little paper of mine (which I wrote for Max Born Volume) together with his reply. This had to be done for it seemed to me not appropriate that Bohm's replica should appear together with my article without offering you the opportunity to do the same. I had not send you my article because it seemed to me not important enough to bother you with it. I did expect that Bohm would care to publish his reaction together with my paper. However, as Bohm insisted I could not prevent it that you were bothered.

I should be grateful to hear from you so that I can feel free to send Bohm's answer to the Editor of the Born Volume.

With cordial regards, yours, A. Einstein.

## A.2.5 de Broglie an Einstein am 9. April 1953.

Übersetzung:

Lieber Herr Einstein

Ich danke Ihnen für Ihren freundlichen Brief. Herr Bohm hat mir tatsächlich Ihren Artikel für das Buch von Herrn Born, ebenso wie seinen eigenen zugesandt. Ich habe die beiden Dokumente mit großem Interesse gelesen.

Wenn Sie glauben, daß es keine Schwierigkeiten bereitet, sende ich Ihnen gerne einen kurzen Artikel, der meinen aktuellen Standpunkt bezüglich der Interpretation der Wellenmechanik präzisiert. Dieser Artikel könnte Ihrem und dem Bohms hinzugefügt werden. Ich glaube, daß dies nicht unnütz wäre, da mein Standpunkt doch sehr verschieden ist von dem Bohms.

Aber der Artikel wird in französisch abgefaßt sein und wird ohne Zweifel ins Englische übersetzt werden müssen. Zudem werde ich einige Tage brauchen, um ihn zu schreiben, und ich glaube nicht, daß ich ihn Ihnen vor Ende April werde zusenden könne. Ich wäre Ihnen sehr dankbar, wenn Sie mir mitteilen würden, ob Sie glauben, daß ich Ihnen diesen Artikel für die Festschrift Borns zusenden soll. Wenn Ihre Antwort positiv sein sollte, würde ich mich bemühen, Ihnen meinen Text so schnell als möglich zuzusenden.

Ich danke Ihnen für das Interesse, das Sie für meine Arbeiten aufbringen, [...]

## A.2.6 Einstein an de Broglie am 14. April 1953.

Dear de Broglie.

I received your kind letter of April 9[th] and feel glad that you decided to add some remarks of your own to Bohm's paper. This will be a beautiful occasion for me to see how you think now about the interpretation of the basis of quantum theory. But at the same time I am sorry that indirectly - without my fault - I have caused you to be troubled.

You may send your contributing directly to the editor of the Born Volume (Dr. Robert Schlapp, The University, Edinburgh, Scotland). No English translation will be needed, for my contribution will also be published in German. I shall send Bohms contribution directly to Dr. Schlapp. I do not know, of course, whether he will accept such belated contributions and especially such which are connected with the contribution of somebody else, I am mentioning this so that you may not feel annoyed if it should happen that the editor sees no possibility to print your remarks. It would be cautious, in any case, to ask him before you do the work. But even if your and Bohm's remarks could not appear in the Born volume it seems to me that it would be highly desirable that they should appear together in some place, because I know that the interest for the question of principle is very vivid in the younger generation of physicists.

The whole affair reminds me a little of the biblical tale of the Tower of Babel: „And the lord said: Now nothing will restrain them to do what they have imagined. Let us go down and there confound their language, that they may not understand one another's speech!" But in our case the Lord not only confounded the language but also thoughts. So you see He has since improved his methods. With cordial regards, yours, A. Einstein.

## A.2.7 Einstein an de Broglie im Mai 1953.

Lieber de Broglie.

Der Standpunkt, den Sie in Ihrer Note eingenommen haben, scheint mir sehr einleuchtend. Sie glauben - wenn ich Sie richtig verstehe - nicht an die neuerdings von Herrn Bohm wieder vertretene Möglichkeit gemäss dem Programm

a) Lösung der Schrödinger-Gleichung durch ein $\Psi$-Feld

b) Zuordnung einer „Bahn", die mit der $\Psi$-Funktion vereinbar ist.

Stattdessen schwebt Ihnen eine Darstellung der physikalischen Realität („vollständige Beschreibung") vor, die von der Art ist

$$\Psi = \Psi\Omega$$

Dies ist eine Produktform, in welcher der Faktor der Partikel-Struktur, der andere der Wellen-Struktur entspricht. Dies würde in der Tat eine befriedigende Erklärung der Doppel-Struktur sein, wie sie sich uns empirisch darbietet. Dies würde eine wirklich neue Theorie sein, nicht eine Vervollständigung der gegenwärtigen.

Soweit ich sehen kann, denken Sie daran, dass das Produkt der ursprünglichen Schrödinger-Gleichung genügen soll. (?) Oder soll nur der „Wellen"-Faktor diese Eigenschaft haben? Oder beide Faktoren? Oder beide Faktoren und das Produkt?

Ihr Ziel wäre auch erreicht, wenn die gesuche Funktion als eine Summe solcher Produkte dargestellt wäre. Endlich scheint es dabei auch nicht nötig, dass das Ganze nur durch eine einzige Funktion (Komponente) dargestellt werden müsse, sondern vielleicht durch ein Gebilde mit mehreren Komponenten.

Sie meinen, daß die Freiheit das grösste Glück für den Theoretiker bedeutet. Diese Freiheit hat mich so entmutigt, dass ich krampfhaft nach einem formalen Prinzip gesucht habe, das diese Freiheit beschränkt. Nun habe ich zwar die Freiheit überwunden, aber vielleicht auf eine ganz naturfremde Weise.

Gemeinsam aber ist uns aber die Überzeugung, dass wir an der Idee der Möglichkeit einer vollständigen objektiven Darstellung einer physikalischen Wirklichkeit festhalten sollen.

Mit freundlichen Grüssen, Ihr A. Einstein.

## A.2.8 Einstein an de Broglie am 8. Februar 1954.

Lieber Louis de Broglie.

Gestern las ich einen mir schon bekannten Aufsatz von Ihnen über die Frage Quanten und Determinismus in deutscher Übersetzung und hatte grosse Freude an Ihren klaren Gedanken. Es ist eine drollige Sache, wie alles plastischer und lebhafter wirkt, wenn es in der altgewohnten Sprache erscheint.

Dass ich Ihnen schreibe, hat eine eigentümliche Ursache. Ich will Ihnen nämlich sagen, wie ich zu meiner Methodik getrieben worden bin, die von außen gesehen recht bizarr ist. Ich muß nämlich erscheinen wie der Wüsten-Vogel Strauss, der seinen Kopf dauernd in dem relativistischen Sand verbirgt, damit er den bösen Quanten nicht ins Auge sehen muss. In Wahrheit bin ich genauso wie Sie davon überzeugt, dass man nach einer Substruktur suchen muss, welche Notwendigkeit die jetzige Quantentheorie durch kunstvolle Anwendung der statistischen Form kunstvoll verbirgt.

Ich bin aber schon lange der Überzeugung, dass man diese Substruktur nicht auf konstruktivem Wege auf dem bekannten Wege aus dem bekannten empirischen Verhalten der physikalischen Dinge wird finden können, weil der nötige Gedankensprung zu gross wäre für die menschlichen Kräfte. Zu dieser Meinung kam ich nicht nur durch die Vergeblichkeit vieljähriger Bemühungen sondern auch durch die Erfahrungen bei der Gravitationstheorie. Die Gravitationsgleichungen waren nur auffindbar auf Grund eines rein formalen Prinzips (allgemeine Kovarianz), d.h. auf Grund des Vertrauens auf die denkbar grösste logische Einfachheit der Naturgesetze. Da es klar war, dass die Gravitationstheorie nur einen ersten Schritt zur Auffindung möglichst einfacher allgemeiner Feldgesetze darstellt, schien es mir, dass dieser logische Weg erst zu Ende gedacht werden muss, bevor man hoffen kann zu einer Lösung auch des Quantenproblems zu gelangen. So wurde ich zu einem fanatischen Gläubigen der Methode der „logischen Einfachheit".

Nun sind zwar die Physiker dieser Generation davon überzeugt, daß man auf solchem Wege nicht zu der Theorie der atomistischen und Quanten-Struktur gelangen kann. Vielleicht haben sie darin recht. Vielleicht gibt es keine Feldtheorie der Quanten. Dann kann meine Bemühung nicht die Lösung des Problems der Atomistik und der Quanten ergeben, vielleicht sogar nicht einmal uns einer Lösung näher bringen. Aber diese negative Überzeugung ist nur intuitiv, nicht objektiv begründet. Auch sehe ich keinen anderen klaren Weg zu einer logischen einfachen Theorie.

Dies zur Erklärung der Vogel-Strauss-Politik. Ich dachte, dies könnte Sie vom psychologischen Standpunkt interessieren, zumal Sie das Vertrauen in die Endgültigkeit der statistischen Methode wieder verloren haben.

Herzlich grüsst Sie Ihr Albert Einstein

# A.2.9 de Broglie an Einstein am 8. März 1954.

Übersetzung:

Lieber Herr Einstein

Ihr Brief war für mich sehr interessant zu lesen und zu überdenken. Er hat mich sehr darin bestärkt, darin fortzufahren, die Ideen wieder aufzunehmen und zu vertiefen, die ich 1927 geahnt hatte. Wie Sie wissen, arbeite ich mit einigen jungen Mitarbeitern daran, diese Konzeptionen zu präzisieren und auszuweiten, und ich habe so einige Ergebnisse erhalten, die mir ermunternd erscheinen.

Aber es bleiben, wie Sie richtig glauben, nennenswerte Probleme, die weit davon entfernt sind gelöst zu sein. Dennoch neige ich neuerlich dazu, zu glauben, daß die aktuell anerkannte statistische Interpretation „unvollständig" ist und daß man präzise Bilder in Zeit und Raum des Welle-Teilchen-Dualismus suchen muß, die es erlauben, die Erfolge der statistischen Gesetze der Quantenmechanik zu erklären.

Was Sie in ihrem Brief über Ihre Einstellung zu den Quantenproblemen und über Ihr Vertrauen auf die Methode der „logischen Einfachheit" sagen, hat meine Aufmerksamkeit sehr auf sich gezogen. Es erscheint mir in der Tat wahrscheinlich, daß die sehr allgemeinen Standpunkte logischer Kohärenz, die Sie zu den großartigen Ergebnissen in der allgemeinen Relativitätstheorie und den unitären Theorien geführt haben, dieselben sind, die es eines Tages erlauben werden, den wahren Sinn der Quanten und des Welle-Teilchen-Dualismus zu verstehen.

In meinen aktuellen Forschungen bin ich zu der Idee gelangt, daß man, um den Welle-Teilchen-Dualismus erklären zu können, eine Wellenmechanik entwickeln muß, die auf nicht-linearen Gleichungen beruht und deren lineare Gleichungen nichts anderes sind als näherungsweise Formen, die unter gewissen Voraussetzungen gelten. Aber um auf diesem Wege Fortschritte zu machen, müßte es gelingen, die Form dieser unbekannten nicht-linearen Gleichungen zu präzisieren. Dies ist ein sehr schwieriges Problem, und ich weiß nicht, wie man sie nur von den physikalischen Ergebnissen ausgehend finden soll. In Übereinstimmung mit Ihren Ideen kann dieses Problem sicherlich nur gelöst werden, indem man denselben Weg begeht, der zu den Gleichungen der allgemeinen Relativitätstheorie geführt hat, nämlich den Weg der logischen Einfachheit.

Frau Tonnelat, deren Arbeiten zu den unitären Theorien Sie gut kennen, interessiert sich mit Herrn Vigier und meinerselbst für diese Aspekte des Problems der Quanten, die selbstverständlich sehr schwierig sind.

Ich danke Ihnen nochmals herzlich für den Gewinn, den mir die Lektüre Ihres wertvollen Briefes und für die große Aufmunterung, die er mir für meine Arbeit gebracht hat [...].